\documentclass{aastex631}
\usepackage{graphicx}
\usepackage{color}
\usepackage{rotate}
\usepackage[maxfloats=256]{morefloats}
\newcommand{\gsim}{\raisebox{-.4ex}{$\stackrel{>}{\scriptstyle\sim}$}}
\newcommand{\lsim}{\raisebox{-.4ex}{$\stackrel{<}{\scriptstyle\sim}$}}

\newcommand{\s}{\mbox{$''$}}
\newcommand{\fbolunit}{\mbox{W m$^{-2}$}}

\newcommand{\my}{\mbox{$M_{\odot}$~yr$^{-1}$}}
\newcommand{\ls}{\mbox{$L_{\odot}$}}
\newcommand{\ms}{\mbox{$M_{\odot}$}}
\newcommand{\kms}{\mbox{km~s$^{-1}$}}

\begin{document}

\title{Common Envelope Mass Ejection in Evolved Stars: Modeling the Dust Emission from post-RGB stars in the LMC}

\email{geetanjali.sarkar@gmail.com}

\author{Geetanjali Sarkar}
\affil{Department of Physics,
Indian Institute of Technology, Kanpur \\
U.P., India}

\author{Raghvendra Sahai}
\affiliation{Jet Propulsion Laboratory \\
Pasadena, CA, USA\\}

\begin{abstract}
Common Envelope (CE) systems are the result of Roche lobe overflow in interacting binaries. The subsequent evolution of the CE, its ejection and the formation of dust in its ejecta  while the primary is on the Red Giant Branch (RGB), gives rise to a recently identified evolutionary class -- dusty post-RGB stars. Their spectral energy distributions (SEDs) suggest that their mass-ejecta are similar to dusty post-Asymptotic Giant Branch (post-AGB) stars. We have modeled the SEDs of a select sample of post-RGB and post-AGB stars in the Large Magellanic Cloud (LMC),  quantified the total dust mass (and gas mass assuming gas-to-dust ratio) in the disks and shells and  set constraints on the dust grain compositions and sizes.  We find that the shell masses in the post-RGBs are generally less than those in post-AGBs, with the caveat that substantial amount of mass in both types of objects may lie in cold, extended shells. Our models suggest that  circumstellar disks, when present, are geometrically thick structures with a substantial opening angle, consistent with numerical simulations of CE evolution (CEE). Comparison of our model dust masses with the predictions of dust production during CEE on the RGB suggest that CEE occurred near or at the tip of the RGB for our post-RGB sources. A surprising result is that some post-RGB stars harbor carbon-rich dust, believed to form when C/O $>$ 1, e.g. following triple-alpha nucleosynthesis and third dredge-up events in AGB stars. This anomaly strengthens the hypothesis that dusty post-RGBs are born in binary systems.
\end{abstract}


\section{Introduction} 

The demise of most stars in the Universe (i.e., in the 1--8\,\ms~range) has traditionally been thought to occur as a 
result of heavy mass-loss on the Asymptotic Giant 
Branch (AGB), with rates in the range $10^{-8}-10^{-4}$\,\my \citep{hofner2018}, when the stars are very luminous ($L\sim5000-10,000$\,\ls) and cool ($T_{eff} < 3000$\,K). This heavy 
mass-loss is believed to occur via a dusty, spherical, radiatively-driven wind, typically expanding at 
$\sim5-20$\,\kms. After the mass-loss has depleted most of the stellar envelope, the 
stars evolve to higher temperatures through the post-AGB phase at almost constant 
luminosity. A major unsolved problem in this evolution is that post-AGB stars are surrounded either by a disk 
(dpAGB objects\footnote{disk-prominent post-AGB objects: \citet{sahai2011}}), and/or an expanding dusty shell (pre-planetary 
nebulae or PPNe) that is generally highly aspherical -- objects with round morphology are absent during this evolutionary phase! HST imaging surveys have clearly revealed the aspherical morphologies of pre-planetary nebulae (PPNs), e.g. \citep{sahai2007, ueta2000}. 
The current consensus is that these dramatic changes in the geometry and kinematics of 
mass-loss during the AGB-to-PPN transition are a result of strong interaction with a binary 
companion, see e.g. \cite{sahai2018} and references therein.

Intriguingly, binarity may also be responsible for the rapid and unexpected evolution of Red Giant Branch (RGB) stars to the 
PPN stage (post-RGB stars).  The Boomerang Nebula -- the coldest object in the Universe \citep{sahai1997}, is the only 
known representative of a post-RGB star
that has evolved to the PPN phase \citep{sahai2017}, in our Galaxy. The Boomerang's luminosity is much lower 
($L\sim300$\,\ls) than possible for a post-AGB object, and \citet{sahai2017} show that merger with a binary companion most likely triggered 
extreme mass loss ($\sim$10$^{-3}$ M$_\odot$yr$^{-1}$) at a very high ejection velocity (165 \kms),
over a relatively short period (3500 yr) via common-envelope evolution (CEE).

A class of post-RGB objects similar to the Boomerang (i.e., with $L\lesssim1000$\,\ls) have recently been identified in the Large and Small Magellanic Clouds 
(LMC and SMC), using optical spectroscopy and $Spitzer$ photometry \citep{kamath2014, kamath2015, kamath2016}. The circumstellar dust in the ejecta heated by the central star results in these objects having large mid-IR excesses. The optical spectroscopy allows a determination of stellar parameters ($T_{\rm eff}$, log\,$g$, $[Fe/H]$ and $E(B-V)$) and integration of the spectral-energy-distribution (SED) constrains the bolometric flux. The known distance to the LMC ($\sim$ 50 kpc) then allows 
reliable luminosity estimates for these stars; the low luminosities ($<$ 2500 L$_\odot$) 
indicate that these stars have not yet reached the AGB phase. The post-RGB stars are thus separated from post-AGB ones by their low luminosities. 
Further, the spectroscopically estimated log\,$g$ values are compared with theoretical values that a star with similar luminosity and $T_{\mathrm eff}$ would have 
in the post-RGB and pre-Main Sequence phase. Stars in the latter phase 
would be
typically 15-20 times more massive and hence differ in log\,$g$ by $\sim$ 1.2
\citep{kamath2016}. However, for stars with poor
quality optical spectra, the estimated log\,$g$ and hence subsequent classification may not be accurate; 
e.g. J051845.47-690321.8, a post-AGB
star was wrongly classified as a Young Stellar Object (YSO): \citet{kamath2019}. 
In addition,
since the post-AGB/RGB stars are an old to intermediate age population, they are expected to be more metal poor 
than the YSOs which would belong to the young LMC population with a mean metallicity of $[Fe/H]$ $\simeq$ $-0.5$
\citep{kamath2015}.

In this paper, we investigate a 
sample of  this new class of post-RGB stars in the LMC, believed to be rapidly 
evolving to 
the PN stage. We also study a sample of post-AGB LMC stars discovered by 
\citet{kamath2015}: hereafter KWVW15\footnote{According to KWVW15, ``shell" sources are those that show a double-peaked SED in the near-to mid-IR region, whereas the ``disk" sources 
do not show two distinct flux peaks but display a near-IR excess.} for comparison, in order to investigate if the properties of the ejecta (e.g., mass, mass-loss rate, temperature, disk-to-shell mass ratio) are different in these two classes. In order to understand the influence of a binary companion on the mass loss history and rapid evolution 
of a post-RGB star to the PPN stage, we model the circumstellar dust distribution of these objects, using their 
near to far-infrared SEDs.  We chose objects classified as ``shell" or ``disk" sources from
KWVW15. The presence of a disk and/or shell and the ratio of the shell
to disk mass (when both can be identified) should help in understanding the physical processes contributing to the 
early evolution of these stars to the PN stage. 

The plan of this paper is as follows. \S\,\ref{targsel} describes the selection of sources for our modeling study,  \S\,\ref{modsed} describes our modeling methodology, and \S\,\ref{prgbmod} and \S\,\ref{pagbmod} provide details of the SED modeling for each of our post-RGB and post-AGB objects. In \S\,\ref{discuss}, we provide a discussion of our modeling results, and \S\,\ref{conclude} gives our conclusions.

\section{Target Selection}\label{targsel}

The post-RGB and post-AGB stars in this study are taken from KWVW15 -- these are listed in Table\,\ref{srcs} together with their photometry. 
The photometry has been compiled by \citet{kamath2015} 
and is available online from the Vizier database (https://vizier.u-strasbg.fr/viz-bin/VizieR).
The Vizier database was further searched to compile the error bars 
and upper (lower) limits if any on the photometric magnitudes. KWVW15 use the average of epoch 1 and 2 \citep{meixner2006} $Spitzer$ MIPS 24 $\micron$ photometry. Although these values are close enough and the use of an average is justified, we use the photometric data from both the epochs with their corresponding error bars from the NASA/IPAC Infrared Science Archive
(https://irsa.ipac.caltech.edu).

A total of eight post-RGB and post-AGB stars have been analysed. These are divided equally between ``shell'' and ``disk'' sources. KWVW15 introduced this classification based on a visual inspection of their dust SEDs. The shell sources show far-IR excess and  the peak of the SED lies beyond 10 $\micron$.  The disk sources show near-IR excess indicative of hot dust and the peak of the SED lies around 10 $\micron$ or sometimes shortwards. Further, on a color-color plot, see Fig.\,12, \citet{kamath2015}, the shell sources have [3.6]$-$[4.5] $<$ 0.5 and [8]$-$[24] $>$ 4.0. Disk sources, on the other hand, have [3.6]$-$[4.5] $>$ 0.5, or in some cases, [3.6]$-$[4.5] $<$ 0.5 in combination with [8]-[24] $<$ 3.0.

In order to ensure that for the post-RGBs in our study, CEE occured on the RGB, we selected objects with
L $<$ 1000 $L_{\odot}$. 
The post-AGB stars are common between KWVW15 
and \citet{aarle2011}, the only exception being J051906.86-694153.9. This star showed 
the BaII line at 4554.03\AA, (KWVW15).  The presence of BaII indicates a 
s-process enriched post-AGB object.

\section{Modelling the Spectral Energy Distribution: Methodology}
\label{modsed}

We have modeled the SEDs of the selected sources using the one-dimensional radiative transfer code, DUSTY \citep{ivezic2012}. Optical properties for six different grain types are included
in the DUSTY code. These are `warm' and `cold' silicates (Sil-Ow and Sil-Oc) from \citet{ossenkopf1992},
silicate and graphite grains (Sil-DL and grf-DL) from \citet{draine1984}, amorphous carbon (amC-Hn) from \citet{hanner1988} and silicon carbide (SiC-Pg) from \citet{pegourie1988}. 
For stars on the RGB, the circumstellar envelope is expected to be oxygen-rich.
The change from oxygen-rich to carbon-rich chemistry may occur as a result of
specific binary interactions (see \S\,\ref{pahs}) and/or once the star has undergone thermal pulses on the AGB. Silicates are 
expected to be the main dust species in oxygen-rich outflows, see e.g. \citet{gail2009}. 
Further, as the name suggests, `warm silicates' are expected 
to dominate in regions of high dust temperature and/or in fast moving circumstellar outflows \citep{ossenkopf1992}. 
In contrast, `cold silicates' are expected in the cool highly 
obscured circumstellar environment around OH/IR stars and in the interstellar medium (ISM). 
Hence, for our models, the dust grain composition was chosen to be `warm' (Sil-Ow) silicates. 
Only when Sil-Ow grain composition did not provide a satisfactory fit, we tried different 
grain compositions.  We used the Mathis, Rumpl, Nordsieck 
(MRN), \citet{mathis1977}
grain size (a) distribution function, n(a) $\propto$ a$^{-q}$ for $a_{\mathrm min}$ $\leq$ a $\leq$ $a_{\mathrm max}$. DUSTY allows one to use standard MRN parameters (q = 3.5, $a_{\mathrm min}$ = 0.005 $\micron$ and $a_{\mathrm max}$ = 0.25 $\micron$) or modified MRN parameters. The MRN grain size distribution, for a mixture of silicate and graphite grains, reproduces the Milky Way extinction curve. The Magellanic Cloud extinction curves also obey the same
power law for the grain size distribution albeit for different silicate and graphite abundance ratios \citep{pei1992}.
Whenever required, we used a modified MRN distribution by altering $a_{\mathrm min}$ and $a_{\mathrm max}$. The dust density was assumed to be proportional to r$^{-2}$, where r is the radial distance from the star. 

There are multiple input parameters associated with the dusty circumstellar
environment: the dust temperature at the inner shell boundary, $T_{\mathrm d}$(in); the relative
shell thickness (ratio of outer, $R_{\mathrm out}$ to inner radius, $R_{\mathrm in}$), $Y$ = $R_{\mathrm out}$/$R_{\mathrm in}$; the optical
depth at 0.55 $\micron$, $\tau$; choice of grain composition and the grain-size distribution.

Our strategy for exploring the parameter space is as
follows. We first attempt to fit the SED using a single shell (one-component) model,
varying the above input parameters. These model
numbers are given a suffix `s'. If systematic discrepancies
remain between the fit and the data, we then attempt a two-component fit, in which we add
an inner component, representative of a hot, compact disk, and vary its $T_{\mathrm d}$, Y, $\tau$,
and dust-grain properties. Specific wavelength ranges of the SED are relatively more sensitive to the shell and disk, and help us to constrain their properties in a non-degenerate manner. 

For the two-component fit (inner disk + outer shell), we approximated the inner disk by a spherical shell that intercepts a fraction of the direct starlight e.g., as in \citep{sahai2003, sahai2006}. Such a shell is thus roughly equivalent to an axially-symmetric wedge-shaped fraction of a sphere; this fraction is hereafter referred to as the "disk fraction" and is listed in Tables\,\ref{mod-tbl} and \ref{mod-tbl-pagb}. Thus, a disk with an opening angle of $\theta_{\rm d}$ is approximated by a shell that intercepts a
fraction, sin($\theta_{\rm d}$/2) of the radiation emitted within a 4$\pi$ solid angle, and the
corresponding "disk-fraction" is sin($\theta_{\rm d}$/2).  An illustration of the circumstellar geometry is provided in Fig.\,\ref{cartoon}.

A ``correctly illuminated" model of the outer shell is constructed assuming the shell to
be divided into two parts. The fraction of the shell that lies in the shadow of the disk
(= the disk-fraction) is illuminated by star light attenuated by the disk, together with
the sum of the scattered and thermal emission from the disk within that fraction. The
remaining fraction of the shell is illuminated by direct starlight plus the remaining
fraction of the sum of scattered and thermal emission from the disk. The DUSTY code is run
separately for each of the two parts of the outer shell and the outputs are added
proportionately to obtain the final SED. These correctly illuminated model numbers are given
a suffix 'c'.

In some post-AGB stars, the SEDs may be a result of an interaction between a slow-moving cold outer shell ejected during
the previous AGB phase and a fast-moving warm inner shell ejected during the post-AGB phase
similar to the case of IRAS 22036+5306 \citep{sahai2006}.  A fit to the SED is then obtained
assuming a pair of `nested shells', i.e., a warm inner shell covering 4$\pi$ solid angle and a cold outer shell.
In the correctly illuminated nested shells model, the 
cold shell is illuminated by the radiation emerging from the warm inner shell.

\subsection{Chi-square statistic}
\label{chisq}
We arrive at the best-fit models based on visual inspection of the observed and modeled
data. In doing so, we have given more importance to matching the photometry at longer
wavelengths ($\lambda$ $\gsim$ 2 $\micron$) because these are much less affected by the relatively
uncertain intervening interstellar absorption along the line-of-sight to each object, and
potential stellar variability in the optical and near-infrared. For the sake of completeness,
we also calculated the reduced chi-square statistic for each of the models. 
The reduced chi-square is given by:
\begin{equation}
\chi^{2} = \sum \left(\frac{O_{\rm i}-M_{\rm i}}{\sigma_{\rm i}}\right)^{2}/(N-p-1)
\end{equation}
where, O$_{\rm i}$ is the observed flux, M$_{\rm i}$ is the model flux, 
$\sigma_{\rm i}$ is the error in the observed flux and N$-$p$-$1 is the number 
of degrees of freedom, with N equal to the number of observed datapoints and p equal to the number of free parameters (= 5 for single shell models
and 10 for two-component models).

While estimating $\chi^{2}$ we omitted the WISE photometric data and used
the ALLWISE data due to its improved photometric sensitivity at bands 1 and 2 (\S\,\ref{wiseall}).
For our dataset then, the maximum number of degrees of freedom is 12 for single shell models and 7 for two-component
models. Using Pearson's chi-square distribution table 
(https://www.statology.org/chi-square-distribution-table), we find that the critical
chi-square values for 5\% significance level and degrees of freedom 7 and 
12 correspond to 14.1 and 21.0 respectively. The critical values decrease 
with decrease in the degrees of freedom. For the 
majority of our stars, the calculated chi-square statistic lies well above the critical value for both
single shell and two-component models. Whenever possible, the model with the least $\chi^{2}$ value has
been chosen as the best-fit for a particular source. However, the chi-square statistic cannot be unambiguously used to decide between our models. 
Given the large number of free parameters, p vis-\'a-vis the
number of observed data points, a much larger dataset is required for the use
of the reduced chi-square method to unambiguously decide the goodness of fit.

\subsection{Photometric magnitudes, extinction and stellar model atmospheres}

The photometric magnitudes were corrected for the combined effects of Galactic and 
LMC reddening using a mean $E(B-V)$ = 0.08 \citep{keller2006}. From the 
ultraviolet to the near-infrared, we used the LMC's average extinction curve and 
{$R_\mathrm{V}$ = 3.41} as derived by \citet{gordon2003}. In the mid-infrared, 
we applied the extinction law by \citet{gao2013}. The DUSTY code requires the spectral shape of
the source illuminating the dusty shell. \citet{kamath2015} have derived the
effective temperatures ($T_{\mathrm eff}$), gravities (log\,$g$) and metallicities $[Fe/H]$ of the 
objects used in the study. Hence, for the purpose of modeling, we used the Kurucz stellar atmosphere models 
closest to these parameters from http://kurucz.harvard.edu/grids.html.  

\subsection{Estimating the circumstellar mass}

The DUSTY code outputs the SED, normalized to the bolometric flux, F$_{\rm bol}$. We determined F$_{\rm bol}$~by 
scaling the model SED to match the de-reddened SED of our sources. 
The luminosity and dust mass ($M_{\mathrm d}$) in the  circumstellar component was computed for each model.  
The distance (d) to the LMC was adopted to be 50 kpc. 
We estimated the luminosity for each model as: 
\begin{equation}
L =  4\pi d^{2} F_{\mathrm bol}.  
\label{lum_eq}
\end{equation}

For objects obeying a r$^{-2}$ density distribution, the dust mass in the circumstellar component is given by:
\begin{equation}
M_{d} = 4\pi R^{2}_{in} Y(\tau_{100}/\kappa_{100}).
\label{dustmass_eq}
\end{equation}
From this, we inferred  the total mass (gas$+$dust), $M_{\mathrm gd}$ = M$_{\rm d}\delta$ \citep{sarkar2006}
where, $\tau_{100}$
is the shell optical depth at 100 $\micron$, $\kappa_{100}$ is the dust mass absorption coefficient, $\kappa$, at
100 $\micron$ and $\delta$ is the gas-to-dust ratio. We assume $\kappa_{100} = 34$\,cm$^{2}$g$^{-1}$, as in \citet{sarkar2006}. 
Although this equation is valid for any wavelength, we selected the latter to lie in the far-infrared range, where the  
SED is dominated by optically-thin thermal emission and therefore the optical depth is better constrained by the model than 
that at short wavelengths.
We note that $\kappa$ is poorly constrained in general. It may be different for warm ($\gsim$ 300 K) and 
cold dust ($\lsim$ 300 K), e.g. \citep{demyk2018}. We have therefore chosen to use the same value of $\kappa_{100}$ in Eq.\,\ref{dustmass_eq} for the cool shell and warm disk.

The inner radius of the dust shell, $R_{\mathrm in}$ scales as $L^{1/2}$ where $L$
is the luminosity \citep{ivezic2012}. DUSTY computes $R_{\mathrm in}$ for 10$^{4}L_{\odot}$; we use our estimated luminosity
(Eq.\,\ref{lum_eq}) for each source, to infer $R_{\mathrm in}$ for the dust shell/disk in it. $M_{\mathrm d}$ also scales in proportion to $Y$ (discussed below). 

\subsubsection{Thickness of the circumstellar component}
\label{thickness}
The estimated dust mass scales in proportion to $Y$ (Eq.\,\ref{dustmass_eq}) and may therefore be taken as a
a lower limit to the actual mass in the shell. To understand this, we look at the post-RGB objects, J051920.18-722522.1 (shell-source) and 
J045755.05-681649.2 (disk-source). 
Increasing the thickness, $Y$ of the cold outer shell by a factor of 10 in these objects, increased the total shell mass by the same factor (Table\,\ref{mod-tbl}) but did not cause any noticeable change in the fit to the observed SEDs (Figs. 4.2 and 7.4). While $R_{\mathrm in}$ is constrained by $T_{\mathrm d}$ which is an input parameter, we cannot constrain $R_{\mathrm out}$ in our SED modeling. 
This is because the dust temperature decreases with radius -- thus the outer regions of the shell emit dominantly at wavelengths longward of the maximum wavelength for which we have data, $\sim$24\,$\micron$. In this paper, we adopt the minimum $Y$ value needed to fit the 24\,$\micron$~photometry.

\subsection{Gas-to-dust ratio}

The derived mass-loss rate depends on the assumed gas-to-dust ratio. Considering that RGB stars are much less luminous
than their AGB counterparts, the gas-to-dust ratio may deviate from the typical value of 200 for the post-AGBs. 
The dust composition around the post-RGBs may be different from that around the post-AGBs. The gas-to-dust ratio may also depend on the metallicity
of the galaxy, e.g. \citet{loon1999}, \citet{nanni2019}. \citet{loon1999}  use a value of 500 for the LMC. 
\citet{roman2014} found gas-to-dust ratios of
380$^{+250}_{-130}$ in the LMC.
Determining the gas-to-dust ratio as a function of fundamental stellar parameters 
(e.g., luminosity and metallicity) and evolutionary phase is still a distant goal \citep{sahai2009}. In
this paper, we adopt $\delta$ = 200. A higher value of $\delta$ would imply 
a proportionate increase in the derived masses of the ejecta (gas+dust).

\subsection{Expansion velocity and mass loss rate}

We have no reliable way of estimating the expansion velocity of the cool shells in post-RGB objects.  For (the only example of) a post-RGB object where this information is available -- the Boomerang Nebula -- the cold outer shell is expanding at 165 \kms. Typical expansion velocities of the shells around post-AGB and PNe are lower, $\sim10-30$ \kms \citep{nyman1992}, although collimated outflows in pre-planetary nebulae can have expansion velocities comparable to that in the Boomerang's cool shell e.g., \citep{sanchez2012, bujarrabal2001}
We have therefore assumed an expansion velocity of 50 \kms for the cool shells for the post-RGB sources in order to estimate their ages. The thickness of the model shell provides an ejection time-scale, and together with the model shell mass ($M_{\mathrm gd}$), an estimate for the mass loss rate ($dM/dt$). Our estimate of $dM/dt$ is sensitive to both the extent of the dust shell and the assumed expansion velocity (it is possible that the dust shell is much larger, as in the Boomerang Nebula.)

For the warm inner disk in these objects, the expansion velocity is likely to be much lower. For example, in the Boomerang nebula, the FWHM of the emission from the central waist/disk region is 4.3 \kms \citep{sahai2017}, which may suggest an expansion velocity of 2.2 \kms. \citet{jura1999} suggest FWHM $<$ 5 \kms from molecular gas observations around evolved red giant stars. We have therefore assumed an expansion velocity of 2 \kms~to derive 
the ages and mass-loss rates for the warm inner disks in post-RGB. 

In the post-AGB phase, the cold outer shell is most likely a remnant of the previous AGB phase. Since typical AGB expansion velocity is 15 \kms \citep{olofsson1993}, we adopt this value for the cool post-AGB shells. For the inner disk, we use an expansion velocity of 2 \kms, similar to the post-RGBs. 
The SED of some post-AGB sources such as J050632.10-714229.8  (\S\,\ref{J050632}) indicate the presence of `nested shells'. Following the original AGB phase, a post-AGB wind may develop. Radiation pressure from the hot central post-AGB star on the less dense post-AGB wind may accelerate the dust to velocities of upto 150 \kms \citep{szczerba1993}. Since, our post-AGB sources are relatively cool ($T_{\mathrm eff}$ $<$ 10000 K), we assume a smaller value of 50 \kms for the expansion velocity of the warm inner post-AGB shells.

\section {Models of post-RGB objects}\label{prgbmod}
The detailed modeling of the post-RGBs objects listed in Table\,\ref{srcs} is discussed below. Our procedure for determining the best-fit model is explained in detail for J043919.30-685733.4, together with a discussion of additional models demonstrating the effect of varying different input parameters. 
The physical parameters and derived model parameters corresponding to each of the post-RGB models 
are given in Table\,\ref{mod-tbl}. We check 
for the robustness of the fits by varying the different input model parameters. 
All model fits are shown in their respective Figure Sets.
The adopted best-fit models are flagged with $\dagger$ symbol. Corresponding to each figure set,
a sample figure displaying the best-fit model is shown. 
Important physical parameters of the adopted best-fits are summarized in Table\,\ref{model_param}.

\subsection{J043919.30-685733.4} 
\label{J043919.30}

This object is classified as a post-RGB shell source by \citet{kamath2015}, with $T_{\mathrm eff}$ = 6313K, log\,$g$ = 1.5 
and $[Fe/H]$ = $-$2.0. We used a Kurucz stellar atmosphere model with $T_{\mathrm eff}$ = 6250 K, log\,$g$ = 1.5 and $[Fe/H]$ = $-$ 2.0.
In Fig. 2.1 we  
show the observed and model SEDs obtained by varying $Y$ = 8, 100 and 500. We 
assumed the standard MRN grain size distribution. With $T_{\mathrm d}$(in) = 550 K and $\tau$ = 1.5, 
we find a reasonable fit to the SED, except in the mid-IR region ($\sim$ 5 $-$ 12 $\micron$).
We find a lower limit on $Y\sim100$, from the MIPS 24 $\micron$ flux. 

Taking $Y$ = 100 as the lower limit to the shell thickness, we investigated whether variation in the   
value of $T_{\mathrm d}$  can improve the fit in the mid-IR region. 
Models with $T_{\mathrm d}$ (in) = 145, 175 and 500 K, Fig. 2.2 were computed using a standard MRN grain size distribution, and $\tau$ = 0.8. 
We find that the discrepancy
in the near-IR persists, except for the model with $T_{\mathrm d}$ (in) = 500 K.  However, in this
case, the model shows a large silicate feature which is not observed -- although using larger size grains in
the shell may suppress this emission feature, the 
far-IR flux too falls much more rapidly than that indicated by the MIPS 24 $\micron$ data. Since the shell is optically thin 
at all wavelengths longward of $\sim 1\micron$, the model SED is not sensitive to the shell geometry. 
We conclude that a single-shell cannot fit the SED adequately, and resolving these discrepancies requires a two-component model.

Following the two-component approach outlined in Sec.\,\ref{modsed}, we found that a good fit to the SED could be obtained with a disk fraction $\sim$ 0.35, together with a cold outer shell that dominates the emission in $\lambda\gtrsim10\micron$ region. This cold shell has $T_{\mathrm d}$ (in) = 130 K, $Y$ = 20, $\tau$ = 0.95 and standard MRN grain size distribution (blue curve in Fig. 2.3 and 2.5).
$Y$ = 20 is a lower limit for the shell thickness.  For the warm inner disk with $Y$ = 1.4 and an r$^{-2}$ density, alternate model fits were obtained 
assuming standard MRN grain size distribution, model \#\,1: $T_{\mathrm d}$ (in) = 1000 K, yellow curve, Fig. 2.3 and modified 
MRN grain size distribution: $a_{\mathrm min}$ = 1 $\micron$ and $a_{\mathrm max}$ = 25 $\micron$, model \#\,2: $T_{\mathrm d}$ (in) = 800 K, yellow curve, Fig. 2.5) in the disk.

The sensitivity of the SED to different $a_{\mathrm min}$ and $a_{\mathrm max}$ grain sizes in the 
inner disk is shown in Fig. 2.7.  A smaller $a_{\mathrm min}$ 
(by a factor of 10) would cause a deficit in the near-IR range.
Similarly, decreasing a$_{\max}$ (by a factor of 5) causes a deficit between 5 and 10 $\micron$. Increasing $a_{\mathrm max}$ however, does not appear to effect the SED.  

For both models \#\,1 and \#\,2 (Figs. 2.3 and 2.5), there is a mismatch with the IRAC4\,(8 $\micron$) flux. 
We convolved the model flux with the IRAC4 response curve and estimated the magnitude at 8 $\micron$ for 
models \#\,1 (14.96 mag) and \#\,2 (14.76 mag). In comparison, the observed magnitude at 8 $\micron$ (corrected for extinction) is significantly lower (13.84).
The IRAC4 bandpass covers a strong PAH emission feature which is not included in the DUSTY code. If we were to include the 
excitation of PAHs, it is possible that a better fit would be obtained in this region. 

We show the sensitivity of the SEDs to varying the $Y$ parameter for models
\#\,1 and \#\,2 in Figs. 2.4 and 2.6. Increasing the thickness causes a mismatch in the mid-IR and with WISE band 3 (W3/12 $\micron$) flux.

In the two-component models (models \#\,1 and \#\,2), it was assumed that the shell is illuminated 
by the radiation from the central star and captures $\sim$ 0.65 of the bolometric flux of the star. The disk 
was treated as a separate entity illuminated by the central star 
and capturing ($\sim$ 0.35) of the bolometric flux from the star. The output was the sum 
of the modeled fluxes in each instance. Finally, we constructed the correctly illuminated models
Table\,\ref{mod-tbl}, model \#\,1,c \& \#\,2,c and Figs. 2.8 \& 2.9
following the prescription in Sec.\,\ref{modsed}.

The fit in the far-IR is identical for the two models; we compared the model and observed
SEDs in the near and mid-IR region to choose our best-fit model. The model SED is closer to
the observed J,H,K fluxes in the case of model \#\,1,c, and we adopt this as the best-fit for J043919.30-685733.4. This choice is re-affirmed by a
comparison of reduced chi-square in the two cases, model \#\,1,c (18.1) and
model \#\,2,c (30.4). In the light of the above discussion, we omitted the IRAC4\,(8 $\micron$) flux and the W4$^{\prime}$ flux while calculating the reduced chi-square. The latter is an upper limit. Our estimated 
luminosity, $L$ =  116 $L_{\odot}$ from model \#\,1,c  is in close agreement with the observed luminosity of the object, $L$ =  106 $L_{\odot}$ (KWVW15).  
The total mass accumulated in the shell is 5.2 $\times$ 10$^{-3}$ $M_{\odot}$ over a relatively short period of 
$\sim$ 200 $-$ 300 years at a mass loss rate of 7.1$\times 10^{-6} - 1.1\times 10^{-5}M_{\odot}\mathrm{yr}^{-1}$ (Table\,\ref{mod-tbl}). The inner disk is very young ($\sim$ 2 years)
with a mass of $\sim 2.19 \times 10^{-8} M_{\odot}$ and a mass loss rate of 1.18$\times10^{-8}$ $M_{\odot}\mathrm{yr}^{-1}$.

\subsection{J051347.57-704450.5}
\label{J051347.57}

This object is classified as a post-RGB shell source by KWVW15, with $T_{\mathrm eff}$ = 4500 K, log\,$g$ = 1.0 and
$[Fe/H]$ = $-$0.5. We used a Kurucz stellar atmosphere model with these parameters for  
modeling the SED of the object. Our adopted best-fit model corresponds to 
$T_{\mathrm d}$ (in) = 250 K, modified MRN grain size distribution: $a_{\mathrm min}$ = 0.1 $\micron$,
$a_{\mathrm max}$ = 0.25 $\micron$, $\tau$ = 0.40 and $Y$ = 3.0 (Fig. 3.1; $\chi^{2}$ = 69.2). In Figs. 3.2 $-$ 3.5, we show
the effects of varying $T_{\mathrm d}$(in), $\tau$, grain sizes and $Y$.
We find that the GSC 2.2 R-band flux is much below the modeled SED. 
Three R-band magnitudes are available for the object from Vizier, R = 16.74 $\pm$ 0.44
(GSC 2.2 Catalogue/STScI 2001), R1 = 15.76 and R2 = 16.29 (USNO-B 1.0 Catalog); \citet{monet2003}. We plotted the corresponding extinction corrected fluxes in the SED. The USNO-B 1.0 R-band fluxes are closer in agreement with the modeled SED. Mismatch between the modeled SED and the I-band flux is seen in several of our objects and is discussed in \S\,\ref{RIdata}. Since a single shell model provided satisfactory fits to the optical and IR fluxes, we did not try a two-component model for this star. 

The luminosity of the object is estimated to be 776 $L_{\odot}$. 
In comparison, KWVW15 estimated a luminosity of 712 $L_{\odot}$ by integrating the flux under the observed SED and a photospheric luminosity of 
840 $L_{\odot}$. From our best-fit model, the total mass accumulated in the shell is 4.66 $\times10^{-5} M_{\odot}$
over a period of 18 years at a mass loss rate of 2.59 $\times10^{-6} M_{\odot}\mathrm{yr}^{-1}$.

\subsection{J051920.18-722522.1}
\label{J051920.18}

This object is classified as a post-RGB shell source by KWVW15, with $T_{\mathrm eff}$ = 4500 K, log\,$g$ = 1.5 and 
$[Fe/H]$ = $-$1.5. We used a Kurucz stellar atmosphere model with these parameters. 
A fit to the U,B,V and far-IR fluxes could be obtained using $T_{\mathrm d}$ (in) = 200 K, MRN grain size distribution, $\tau$ = 0.75 and $Y$ = 20 (model \#\,1,s; $\chi^{2}$ = 89.1).
Figs. 4.3 $-$ 4.6 show the effects of varying $T_{\mathrm d}$ (in), $\tau$, grain sizes and $Y$ in model \#\,1,s. For all cases, the R and I-band fluxes remain well below the modeled SED (\S\,\ref{RIdata}).
To account for this deficit, we assumed a two-component
model (model \#\,2) with a warm ($T_{\mathrm d}$ (in) = 500 K), inner disk, in addition to a cold outer shell at 110 K. This disk
was approximated as a wedge-shaped fraction ($\sim$ 0.4) of a sphere with $\tau$ = 0.4, $Y$ = 2.0 and modified
MRN grain size distribution: $a_{\mathrm min}$ = 0.5 $\micron$ and $a_{\mathrm max}$ = 20 $\micron$. The cold outer shell is
characterised by $\tau$ = 1 $\micron$, $Y$ = 20 and MRN grain size distribution.
We constructed the correctly illuminated spectrum (model \#\,2,c; $\chi^{2}$ = 17.5) for the shell in the two solid angles (Fig. 4.1). Increasing the thickness of the cold outer shell by a factor of 10, does
not cause any appreciable change in the fit to the SED (Fig. 4.2, \S\,\ref{thickness}).
  
Based on visual inspection and a comparison of reduced chi-square values, we adopted model \#\,2,c as the best-fit. From this, we estimated luminosity,
$L$ =  582 $L_{\odot}$. In comparison, KWVW15 estimated a luminosity of 496 $L_{\odot}$ by integrating the flux under the observed SED. The total mass in the shell is estimated to be 3.41 $\times10^{-2} M_{\odot}$ accumulated over $\sim$ 645 $-$ 792 years at a 
mass loss rate of $\sim 2.1\times10^{-5} - 2.6\times10^{-5} M_{\odot}\mathrm{yr}^{-1}$. The inner disk is much younger (15.9 years) with a 
mass of 3.01$\times10^{-5} M_{\odot}$ and a mass loss rate of 1.89$\times10^{-6} M_{\odot}\mathrm{yr}^{-1}$. 
 
\subsection{J053930.60-702248.5}
\label{J053930.60}

This object is classified as a post-RGB shell source by KWVW15, with $T_{\mathrm eff}$ = 4315K, log\,$g$ = 0.5 and
$[Fe/H]$ = $-$0.9. We used a Kurucz stellar atmosphere model with $T_{\mathrm eff}$ = 4250 K , log\,$g$ = 0.5 , $[Fe/H]$ = $-$1.0
for modeling the SED of this object. A fit to the 
optical and IR fluxes could be obtained using $T_{\mathrm d}$ (in) = 300 K, MRN grain size distribution, $\tau$ = 0.7 and $Y$ = 10 (Fig. 5.1; $\chi^{2}$ = 30.4). Figs. 5.2 $-$ 5.5, show the effects of varying $T_{\mathrm d}$(in), $\tau$, grain sizes and $Y$. We find that the reduced chi-square is least for a modified MRN grain size distribution, $a_{\mathrm min}$ = 0.005 $\micron$ and $a_{\mathrm max}$ = 0.50 $\micron$ 
(Table\,\ref{mod-tbl}, Fig. 5.4). Visual inspection of the model plot in this case reveals that this is due to better agreement with ALLWISE W4$^{\prime}$ data. However, the latter model shows a mismatch with the 
IRAC4\,(8$\micron$) flux. Hence, we adopt the model with standard MRN grain size distribution as the best-fit.
For the same reason, we also reject the model with $\tau$ = 0.8 (Fig. 5.3) even though the model
has a lower chi-square value than our adopted best-fit. Since a single shell model provided satisfactory fits to the optical and IR fluxes, we did not try a two-component model for this star.

Our estimated luminosity (295 $L_{\odot}$) is in good agreement with the observed (248 $L_{\odot}$) and photospheric luminosity (294 $L_{\odot}$) values of KWVW15. From our best
fit model, the total mass accumulated in the shell is 5.8 $\times10^{-5} M_{\odot}$ over a period of 27.2 years at a mass loss rate of 2.13 $\times10^{-6} M_{\odot}\mathrm{yr}^{-1}$.

\subsection{J045555.15-712112.3}
\label{J045555.15}

This object is classified as a post-RGB disk source by \citet{kamath2015}, with $T_{\mathrm eff}$ = 10000 K, log\,$g$ = 2.0 and $[Fe/H]$ = $-$0.5. We used a Kurucz stellar atmosphere model with these parameters. 
Fig. 6.1 shows a one-component model fit to the observed SED (model \#\,1,s)
using Sil-Ow grains, $T_{\mathrm d}$ (in) = 550 K, a$_{\min}$ = 0.005 $\micron$, a$_{\max}$ = 1.0 $\micron$, $\tau$ = 2.3 and $Y$ = 20. The effects of varying $T_{\mathrm d}$(in), $\tau$, grain sizes and $Y$ are shown in Figs. 6.5 $-$ 6.8.
Our model fit (\#\,1,s) is deficit in the near-IR region and additionally, there is a mismatch with the IRAC2\,(4.5 $\micron$), W2\,(4.6 $\micron$) and IRAC3\,(5.8 $\micron$) fluxes. 

For a better fit in the near-IR region, we assumed a two-component model (model \#\,2) with a warm, $T_{\mathrm d}$ (in) = 1200 K, thin inner disk, in addition to the cold outer shell at 550 K. The disk composed of Sil-Ow grains was approximated as a wedge-shaped fraction ($\sim$ 0.1) of a sphere with $\tau$ = 0.5, $Y$ = 10 and grain size distribution: $a_{\mathrm min}$ = 0.005 $\micron$ and $a_{\mathrm max}$ = 0.1 $\micron$.  We constructed the correctly illuminated spectrum for the shell in the two solid angles (\#\,2,c, Fig. 6.2). 
We tried an alternate two-component model (model \#\,3) using graphite grains (grf-DL) for the warm inner disk ($T_{\mathrm eff}$ = 800 K) and a combination of Sil-Ow (0.8) and grf-DL (0.2) for the cold outer shell ($T_{\mathrm eff}$ = 500 K). The inner disk was approximated to be a wedge-shaped fraction (0.1) of a sphere with $\tau$ = 0.7, $Y$ = 5 and MRN grain size distribution: $a_{\mathrm min}$ = 0.005 $\micron$ and $a_{\mathrm max}$ = 0.25 $\micron$. The cold shell
is modeled using $\tau$ = 1.8, $Y$ = 2 and MRN grain size distribution. Fig. 6.3 shows the correctly illuminated two-component model (\#\,3,c). Besides the continuing 
mismatch at IRAC2\,(4.5 $\micron$), W2\,(4.6 $\micron$) and IRAC3\,(5.8 $\micron$), the fit remained
unsatisfactory in the near-IR.

Further, we investigated the possibility that a stellar companion dominates the emission at and around 4.6 $\micron$ in models \#\,2 and \#\,3, with suitable adjustments in the disk and the shell parameters, by fitting a blackbody curve to the flux at 4.6 $\micron$. We find that a blackbody with 
$T_{\mathrm eff}$ = 800 K, and $L$ =  39 $L_{\odot}$ would account for the flux at 4.6 $\micron$. Since a main-sequence companion star with such a low $T_{\mathrm eff}$ would have a very low luminosity (i.e., much lower than 39 $L_{\odot}$), we conclude models including a stellar companion cannot explain the discrepancy at 4.6 $\micron$.

The relatively low flux at 4.6 $\micron$ may result from the presence of a deep absorption feature at 4.6 $\micron$. Such a feature has been observed in YSOs \citep{lacy1984}, and is believed to be due to solid phase CO and cyano group molecules. J045555.15-712112.3  was classified originally as a possible YSO by \citet{gruendl2009}. KWVW15 re-classified it (and many other objects also originally classified as YSOs), using an automated spectral typing pipeline (STP) that focusses on the 
Balmer-line region for the determination of $T_{\mathrm eff}$, for objects with $T_{\mathrm eff}>8000$K \citep{kamath2014}.
However, an examination of the optical spectrum (downloaded from VizieR) reveals a lack of reliable discernible spectral features in this region. It is therefore plausible that for this case, the STP fit did not provide a good estimate of $T_{\mathrm eff}$ and hence log\,$g$. A higher log\,$g$ 
value would support the classification of J045555.15-712112.3 as a possible YSO.

In the scenario that the object is indeed a post-RGB source as classified by KWVW15 and not a YSO, we explored models which may give a better fit at 4.6 $\micron$. 
The two-component model \#\,4 is shown in Fig. 6.4. The warm inner disk, $T_{\mathrm d}$ (in) = 1000 K, composed of Sil-Ow grains was approximated to be a wedge-shaped fraction (0.4) of a sphere with $\tau$ = 1.0, $Y$ = 10 and MRN grain size distribution. The cold outer shell, $T_{\mathrm d}$ (in) = 300 K has Sil-Ow grains, $\tau$ = 2.5 and $Y$ = 2 and MRN grain size distribution. The model provides a good fit at 4.6 $\micron$ but is discrepant with the observed values in the near-IR and
at IRAC3\,(5.8 $\micron$) and IRAC4\,(8$\micron$). The discrepancy at IRAC bands 3 and 4 may be due to
PAH emissions and has been discussed further in Sec.\,\ref{pahs}. Since the inner disk is optically thick, we cannot apply the ``correctly illuminated model" formalism to this object.

We calculated the reduced chi-square for each of the models, 
model \#\,1,s/$\chi^{2}$ = 619.8, model \#\,2,c/$\chi^{2}$ = 1125.5
and model \#\,3,c/$\chi^{2}$ = 1318.3, model \#\,4/$\chi^{2}$ = 116.6. 
Based on the above discussion, if the object is indeed a post-RGB source,
we adopt model \#\,4 as the best-fit to the observed SED. 

From this model, we estimated a luminosity of 465$L_{\odot}$ for J045555.15-712112.3. KWVW15 estimated a luminosity of 454 $L_{\odot}$ by integrating the flux under the observed SED and a photospheric luminosity of 
191 $L_{\odot}$. They attributed this difference in luminosity estimates to either a non-spherically symmetric dust envelope or substantial flux contribution from another object coincident with J045555.15-712112.3 in the sky. 
The total mass accumulated 
in the shell is 2.84$\times10^{-4} M_{\odot}$ over $\sim$ 7.9 years at a mass loss rate of 
3.59$\times10^{-5} M_{\odot}\mathrm{yr}^{-1}$.
The mass accumulated in the disk is 3.08$\times10^{-6} M_{\odot}$ over 131 years at a mass loss rate of 2.35$\times10^{-8} M_{\odot}\mathrm{yr}^{-1}$. 

\subsection{J045755.05-681649.2}

This object is classified as a post-RGB disk source by KWVW15, with $T_{\mathrm eff}$ = 4927K, log\,$g$ = 1.5 and
$[Fe/H]$ = $-$0.3. We used a Kurucz stellar atmosphere model with  $T_{\mathrm eff}$ = 5000 K, log\,$g$ = 1.5 and $[Fe/H]$ = $-$0.3. We first modeled the SED of the object, assuming a single
dust shell surrounding the central star (model \#\,1,s, Fig. 7.1).
A fit to the optical and far-IR fluxes could be obtained using $T_{\mathrm d}$ (in) = 500 K, 
MRN grain size distribution (a$_{\min}$ = 0.005 $\micron$, a$_{\max}$ = 0.25 $\micron$),
$\tau$ = 1.0 and $Y$ = 20 (model \#\,1,s). The I-band flux is well below the modeled SED.
Additionally, there is a mismatch at W2\,(4.6 $\micron$) 
and IRAC4\,(8 $\micron$). These discrepancies have been discussed further in (\S\,\ref{discuss}). The mismatch at IRAC4 may be attributed to PAH emission, similar to 
J043919.30-685733.4.  Figs. 7.5 $-$ 7.7
show the observed and modeled SED of the object along with effects of varying 
$T_{\mathrm d}$ (in), $\tau$, and $Y$.

To obtain a better fit in the mid-IR region, we tried a two-component model (model \#\,2), i.e. a warm inner disk, approximated as an
axially symmetric wedge-shaped fraction (0.4) of a sphere and a cold outer shell. Assuming the fraction that is not
covered by the disk, reasonable fit to the far-IR flux distribution ($\lambda\gtrsim10\micron$) could be obtained for
for $T_{\mathrm d}$ (in) = 400 K, $Y$ = 30, $\tau$ = 0.9 and modified MRN grain size distribution: $a_{\mathrm min}$ = 0.1 $\micron$,
$a_{\mathrm max}$ = 1.0 $\micron$. For the warm inner disk, $T_{\mathrm d}$ (in) = 1300 K, $Y$ = 2, $\tau$ = 0.5 and modified MRN grain
size distribution: $a_{\mathrm min}$ = 0.005 $\micron$, $a_{\mathrm max}$ = 2.0 $\micron$ was used.
We constructed the correctly illuminated model for the shell in the two solid angles
(\#\,2,c). The output modeled flux distribution is shown in Fig. 7.2.

Since, both the one and two-component models could not provide a fit at IRAC4\,(8 $\micron$), we omitted
the flux at this point for the reduced chi-square calculation. Further, the ALLWISE W4$^{\prime}$ flux is an 
upper limit and is also omitted from the chi-square calculation. The reduced chi-square is least for
the single shell model with $T_{\mathrm d}$ (in) = 550K, $\chi^{2}$ = 20.7 (Table\,\ref{mod-tbl}). This
value is comparable with $\chi^{2}$ = 24.7 for the two-component model \#\,2,c. Since the two fits appear
similar, we overplotted
the two models (Fig. 7.3). On visual inspection, we find that the latter provides a better fit in 
the region of the mid-IR IRAC bands, 3$\micron - 6\micron$ with a lower value for the residual sum (98.8) than
in the former case (186.4). Hence, we adopt the two-component model \#\,2,c as the best-fit for
J045755.05-681649.2. Increasing the thickness of the outer shell by a factor of 10 in model \#\,2,c does
not cause any appreciable change in the fit to the SED (Fig. 7.4, \S\,\ref{thickness}).

From our best-fit model, the luminosity is estimated to be $L$ =  217 $L_{\odot}$. In comparison, KWVW15 estimated a luminosity of 190 $L_{\odot}$ by integrating the flux under the observed SED and a photospheric luminosity of 
215 $L_{\odot}$. The total mass accumulated in the shell 
is $\sim$ 5.73 $\times$ 10$^{-5} M_{\odot}$ over a short period of $\sim$ 30 $-$ 40 years at a mass 
loss rate of $\sim$ 7 $\times 10^{-7} - 9 \times 10^{-7} M_{\odot}\mathrm{yr}^{-1}$.
The inner disk is much younger, $\sim$ 1.9 yr. with a mass of $\sim 9.64 \times 10^{-9} M_{\odot}$. 
and a mass loss rate of $\sim 5 \times 10^{-9} M_{\odot}\mathrm{yr}^{-1}$.

\subsection{J050257.89-665306.3}
This object is classified as a post-RGB disk source by KWVW15, with $T_{\mathrm eff}$ = 4586K, log\,$g$ = 0.5 and
$[Fe/H]$ = $-$0.5. We used a Kurucz stellar atmosphere model with Teff = 4500 K, log\,$g$ = 0.5 and $[Fe/H]$ = $-$0.5. A single shell model could not provide a fit to the near and mid-IR fluxes (model \#\,1,s; Fig. 8.1,  $\chi^{2}$ = 141.77). Hence, we tried two-component models with a warm inner disk and a cold outer shell. A fit to the optical
and IR flux could be obtained under two scenarios: a two-component model in which the disk and shell are composed of Sil-Ow grains (model \#\,2) and another two-component model in which the disk is composed of amC-Hn grains and shell is composed of Sil-Ow grains (model \#\,3). The discrepancies at I-band and IRAC4\,(8 $\micron$) have
been discussed in \S\,\ref{discuss}. In calculating the reduced chi-square, we have omitted the flux at
8$\micron$.

In model \#\,2, the inner disk ($T_{\mathrm d}$ (in) = 1200 K) is composed of Sil-Ow grains with modified 
MRN grain size distribution: $a_{\mathrm min}$ = 0.3 $\micron$ and $a_{\mathrm max}$ = 5.0 $\micron$, 
$\tau$ = 0.5 and $Y$ =  3.0. The outer shell ($T_{\mathrm d}$ (in) = 240 K) is composed of
Sil-Ow grains with modified MRN grain size distribution: $a_{\mathrm min}$ = 0.005 $\micron$ and 
$a_{\mathrm max}$ = 1.0 $\micron$, $\tau$ = 1.2 and $Y$ = 10.0. The disk was approximated as 
an axially symmetric wedge-shaped fraction (0.4) of a sphere. The correctly illuminated model (\#\,2,c, $\chi^{2}$ = 48.4) is presented in Fig. 8.2. 

In model \#\,3, the inner disk ($T_{\mathrm d}$ (in) = 1000 K) is composed of amC-Hn grains with modified
MRN grain size distribution: $a_{\mathrm min}$ = 0.005 $\micron$ and $a_{\mathrm max}$ = 0.5 $\micron$. 
$\tau$ = 0.6 and $Y$ =  2.0. The outer shell ($T_{\mathrm d}$ (in) = 250 K) is composed of
Sil-Ow grains with MRN grain size distribution: $a_{\mathrm min}$ = 0.005 $\micron$ and $a_{\mathrm max}$ = 0.25 $\micron$, $\tau$ = 1.7 and $Y$ =  10.0. The disk was approximated as an
axially symmetric wedge-shaped fraction (0.3) of a sphere. The correctly illuminated model (\#\,3,c, $\chi^{2}$ = 43.9) is presented in Fig. 8.3.

The reduced chi-square is comparable for models \#\,2,c and \#\,3,c. On visual
inspection we find that model \#\,2,c shows better agreement with the optical
(U,B) fluxes (Fig. 8.4). Hence we adopt the two-component
model \#\,2,c as the best fit for J050257.89-665306.3.

From our correctly illuminated model \#\,2,c (Table\,\ref{mod-tbl}), 
the luminosity is estimated to be $L$ =  303 $L_{\odot}$.
In comparison, KWVW15 estimated a luminosity of 267 $L_{\odot}$ by integrating the flux under the observed SED and a photospheric luminosity of 
177 $L_{\odot}$.
The total mass accumulated in the shell is 2.24$\times10^{-4} M_{\odot}$ over a period of $\sim$ 30 $-$ 38 years at a mass loss rate of $\sim$ 3$\times10^{-6} M_{\odot}$ yr$^{-1}$. The inner disk is much younger ($\sim$ 4.3yr) with a mass of $\sim$ 5.77$\times10^{-8} M_{\odot}$ and a mass loss rate of $\sim$ 1.3$\times10^{-8} M_{\odot}$ yr$^{-1}$. 

\subsection{J055102.44-685639.1}
\label{J055102.44}
This object is classified as a post-RGB disk source by KWVW15, with $T_{\mathrm eff}$ = 7625 K, log\,$g$ = 1.0 and
$[Fe/H]$ = $-$0.5. We used a Kurucz stellar atmosphere model with $T_{\mathrm eff}$ = 7500 K, log\,$g$ = 1.0 and $[Fe/H]$ = $-$0.5.
The near-IR flux distribution of this object suggests the presence of a warm inner disk. A single shell
model (model \#\,1,s, $\chi^{2}$ = 600.2) could not provide a fit to the optical and near-IR fluxes. We therefore tried two component models with a warm inner disk and cold outer shell made up of Sil-Ow grains. 
In these models, \#\,2a and \#\,2b, 
the inner disk was approximated as a fraction (0.4 and 0.3 respectively) of a sphere. 
However, we could not obtain a simultaneous fit to the optical and near-IR fluxes. The model fits \#\,1,s, \#\,2a and \#\,2b are shown in Fig. 9.1.

The near-IR flux of the object indicates a dust temperature in excess of 1500 K. This would exceed the sublimation temperature of silicate grains. We therefore tried amorphous carbon grains (amC-Hn) for the inner disk (model \#\,3,  $\chi^{2}$ = 220.0). In the DUSTY input, the dust sublimation temperature was changed to 2500 K.  Amorphous carbon has a higher sublimation temperature than silicate grains. While amorphous carbon has been observed in the dusty disk around C-rich post-AGB stars, e.g. HR4049 \citep{acke2013}, we do not expect to see it in the circumstellar environment of a post-RGB star because such dust is believed to form when the C/O ratio is $>$1 in the star's atmosphere, following formation of C via 3-$\alpha$ nucleosynthesis and (the third) dredge-up -- events that occur at the centers of AGB stars. In the case of J055102.44-685639.1, we find that the disk is optically thick ($\tau$ = 1.0) and amC-Hn grains provide a reasonable fit to the SED in the near and mid-infrared (Fig. 9.2). 
The cold outer shell has a combination of warm silicates and  silicon carbide. The carbon-rich circumstellar chemistry may be explained if the post-RGB star is a CH giant in a binary system that formed when the post-RGB progenitor accreted carbon-rich matter from a more massive AGB companion (now a WD) before undergoing Common Envelope (CE) ejection. Such a scenario may also account for the presence of PAH molecules in the circumstellar environment of some post-RGB stars (see \S\,\ref{pahs}).

Our choice of a two-component model (\#\,3) as the best fit is re-affirmed by a lower value of the reduced chi-square in this case, $\chi^{2}$ = 147.3. The disk was approximated as an axially symmetric wedge-shaped fraction (0.3) of a sphere with amC-Hn grains, T${\rm d}$ = 2000 K, $Y$ = 7.0, $\tau$ = 1.0 and modified MRN grain
size distribution: $a_{\mathrm min}$ = 0.005 $\micron$, $a_{\mathrm max}$ = 0.05$\micron$. The outer shell (Sil-Ow/0.4 and SiC-Pg/0.6) has $T{\mathrm d}$ = 350 K, $Y$ = 3.0, $\tau$ = 12.0 and modified MRN grain size distribution: $a_{\mathrm min}$ = 0.005$\micron$, $a_{\mathrm max}$ = 0.07 $\micron$. At the large optical depth of the outer shell, the SiC feature goes into absorption (Fig. 9.2).
Further, since the inner disk is optically thick, we cannot apply the ``correctly illuminated model" formalism to this object.  
 
 From the best-fit model \#\,3 (Table\,\ref{mod-tbl}), we estimated a luminosity of 621$L_{\odot}$. In comparison,
 KWVW15 estimated 452$L_{\odot}$ and 815$L_{\odot}$ respectively for the observed and photospheric luminosity of the object. The total mass accumulated in the shell 
is 3.05 $\times$ 10$^{-3}$ $M_{\odot}$ over a period of $\sim$ 12.5 years at a mass 
loss rate of $\sim$ 2.43$\times10^{-4} M_{\odot}\mathrm{yr}^{-1}$.
The inner disk is  younger, $\sim$ 8.8 yr. with a mass of $\sim$ 1.99$\times10^{-8} M_{\odot}$ and a mass loss rate of $\sim 2.27\times10^{-9} M_{\odot}\mathrm{yr}^{-1}$. 

\section {Models of post-AGB objects}\label{pagbmod}

The detailed modeling of the post-AGB objects listed in Table\,\ref{srcs} is discussed below. 
The physical parameters and derived model parameters corresponding to the post-AGB models 
are given in Table\,\ref{mod-tbl-pagb}. We check 
for the robustness of the fits by varying the different input model parameters. 
All model fits are shown in their respective Figure Sets.
The adopted best-fit models are flagged with $\dagger$ symbol. Corresponding to each figure set,
a sample figure displaying the best-fit model is shown. 
Important physical parameters of the adopted best-fits are summarized in Table\,\ref{model_param}.

\subsection{J050632.10-714229.8}
\label{J050632}
This object is classified as a post-AGB shell source by KWVW15, with $T_{\mathrm eff}$ = 7614 K,
log\,$g$ = 0.5 and $[Fe/H]$ = $-$0.4. However, based on high resolution spectra, \cite{aarle2013} obtained a
much lower $T_{\mathrm eff}$ (= 6750 K), log\,$g$ = 0.5 and a lower $[Fe/H]$ (= $-$1.0). We prefer the 
\citet{aarle2013} values based on their high resolution data. Hence, we used a Kurucz stellar atmosphere model 
with $T_{\mathrm eff}$ = 7000 K, log\,$g$ = 0.5 and $[Fe/H]$ = $-$1.0. Fig. 10.1 shows a one-component model fit to observed SED using Sil-Ow grains, standard MRN grain size distribution, $T_{\mathrm d}$ (in) = 200 K, $\tau$ = 0.6 and $Y$ = 20 (model \#\,1,s). 
This model is deficient in the mid-IR region, particularly at 
IRAC3\,(5.8 $\micron$) and IRAC4\,(8 $\micron$).

Since J050632.10-714229.8 is a post-AGB star, we modeled the SED using the `nested shells' approach outlined in Sec.\,\ref{modsed}. Fig. 10.2 
shows the fit obtained in the mid-IR using
Sil-Ow (grain type \#\,1) in the inner shell, $T_{\mathrm d}$ (in) = 400 K, standard MRN grain size distribution, $\tau$ = 0.4 and $Y$ = 2. It also shows that changing the grains from warm (Sil-Ow) to cold silicates (Sil-Oc) doesn't alter the fit significantly in the 
mid-IR, so we use Sil-Ow grains for the inner shell.  
The circumstellar environment of a post-AGB star may have a mix of carbon and silicate dust grains. 
Fig. 10.3 shows the fits obtained using different grain combinations. We find that 
Sil-Ow (\#\,1) at 400 K, a combination of amC-Hn/0.1 and Sil-Ow/0.9 (\#\,2) at 350 K and
a combination of grf-DL/0.3 and Sil-Ow/0.7 (\#\,3) at 350 K  
(Fig. 10.4) all provide good fits to the 
observed SED in the mid-IR. However, neither combination
of grains could provide a satisfactory fit to the flux from 4.6 $\micron$ $-$ 8 $\micron$. 
These discrepancies are further discussed in \S\,\ref{discuss}. 
Using each of the above best-fit warm inner-shell models, we obtained best-fit models for the cold outer shell
(nested models \#\,1; $\chi^{2}$ = 181.6; \#\,2; $\chi^{2}$ = 164.2 and \#\,3; $\chi^{2}$ = 154.7; Table\,\ref{mod-tbl-pagb}). Fig. 10.5 shows the complete correctly illuminated model SED of 
J050632.10-714229.8 for each grain combination. 

The reduced chi-square is least for model \#\,3. Hence, we adopt this as the best-fit model.
Our estimated luminosity (5434$L_{\odot}$) is in good agreement with KWVW15's observed
luminosity for the object (4910$L_{\odot}$). They estimated a much higher photospheric luminosity 
of 7606$L_{\odot}$. The latter may be in error since it is based on their estimate 
of $T_{\mathrm eff}$ and metallicity from low resolution spectra which is significantly different from that
based on high resolution spectra of \cite{aarle2013}. From model \#\,3, the total mass
accumulated in the outer shell, the shell age and mass loss rate respectively are 
5.91$\times10^{-2} M_{\odot}$, 1020 yrs.,  5.81$\times10^{-5} M_{\odot}\mathrm{yr}^{-1}$. 

\subsection{J051848.84-700247.0}
\label{J051848.84}
This object is classified as a post-AGB shell source by KWVW15, with $T_{\mathrm eff}$ = 6015 K, log\,$g$ = 0.0 and $[Fe/H]$ = $-$1.0 based on their STP.  However, Table 3 in 
KWVW15 suggests that using high resolution spectra \citet{aarle2013} estimate log\,$g$ = 0.5. We could
not find the object in the paper by \citet{aarle2013}. Hence, we used a Kurucz stellar atmosphere model with $T_{\mathrm eff}$ = 6000 K, log\,$g$ = 0.0 and $[Fe/H]$ = $-$1.0.
Based on VizieR, for the R-band magnitude we used 14.80 $\pm$ 0.44 (GSC 2.2 Catalogue/STScI 2001)\footnote{We did not use KWVW15's 
R-band magnitude (14.31) because we could not find a corresponding reference.} We modeled the SED of the
object (Fig. 11.1, $\chi^{2}$ = 50.4) using $T_{\mathrm d}$ (in) = 350 K, standard MRN grain size distribution,
$\tau$ = 3.2 and $Y$ = 20. Figs. 11.2 $-$ 11.4 show the 
effects of varying $T_{\mathrm d}$, $\tau$ and $Y$ . Since a single shell model provided 
satisfactory fits to the
optical and IR fluxes, we did not try a two-component model for this object. 

We estimate a luminosity of 6210$L_{\odot}$. KWVW15 obtained vastly different values for
the observed and photospheric luminosity of the object, 4477$L_{\odot}$ and 14112$L_{\odot}$ respectively.
Their observed luminosity is obtained by integrating the flux under the observed SED. In the event of circumstellar
dust and reddening, the former may be significantly less than the actual luminosity. Their derived 
photospheric luminosity depends on the extinction correction applied to the observed V-magnitude. 
KWVW15's formalism attempts to account for both interstellar and circumstellar reddening. 
If the V-magnitude is over-corrected, they would derive a higher luminosity for the object.

From our best-fit model, the total mass accumulated in the shell is 1.76$\times10^{-2} M_{\odot}$
over 1100 years at a mass-loss rate of 1.6$\times10^{-5} M_{\odot}\mathrm{yr}^{-1}$.

\subsection{J051906.86-694153.9}
\label{J051906.86}
This object is classified as a post-AGB shell source by KWVW15,
with $T_{\mathrm eff}$ = 5613K, log\,$g$ = 0.0 and $[Fe/H]$ = $-$1.3. The WISE and ALLWISE W3 and W4 data are discrepant for this source, with ALLWISE reporting upper limits significantly lower than the WISE magnitudes. Given the good agreement between the Spitzer MIPS\,24\,\micron~and the WISE W4 fluxes, we have chosen to ignore the ALLWISE W3 and W4 data for this source. 
We used a Kurucz stellar atmosphere model with $T_{\mathrm eff}$ = 5500 K, log\,$g$ = 0.0 and $[Fe/H]$ = $-$1.5.
A single shell model could not simultaneously fit the observed flux distribution from the optical to the far-IR.  
Similar to the case of the post-AGB star, J050632.10-714229.8, we assumed a pair of nested shells. 
A fit to the optical, mid- and far-IR data was obtained 
using amC-Hn grain type, standard MRN grain size distribution, $T_{\mathrm d}$ (in) = 2000 K, $\tau$ = 0.35 and $Y$ = 2.0 (Fig. 12.1).  The radiation emerging from this warm shell 
illuminates a cold outer shell of
SiC-Pg grains with modified MRN grain size distribution: $a_{\mathrm min}$ = 2.3 $\micron$, $a_{\mathrm max}$ = 3.0 $\micron$, $T_{\mathrm d}$ (in) = 160 K, $\tau$ = 0.07 and $Y$ =  2.0. 
Fig. 12.2 ($\chi^{2}$ = 21.3) shows the correctly illuminated nested shell model. In calculating the
reduced chi-square for this source, we used the WISE W3 and W4 fluxes. 

We estimate a luminosity of 2018$L_{\odot}$. This is in good agreement with the observed luminosity estimated
by KWVW15 (2052$L_{\odot}$). However, as in the case of the post-AGB star, J051848.84-700247.0, KWVW15 report 
vastly different values for the observed and photospheric luminosities of the object, 2052$L_{\odot}$ and 4246$L_{\odot}$ respectively. From our best-fit model, the total mass accumulated in the shell is 1.78$\times10^{-4}\,M_{\odot}$  over a period of 51.4 years at a mass loss rate of 3.46$\times10^{-6} M_{\odot}\mathrm{yr}^{-1}$.

\subsection{J053250.69-713925.8}

This object is classified as a post-AGB shell source by KWVW15, with $T_{\mathrm eff}$ = 6073K,
log\,$g$ = 1.0 and $[Fe/H]$ = $-1.1$. Based on high resolution spectra, \cite{aarle2013} obtained 
$T_{\mathrm eff}$ = 5500 K, log\,$g$ = 0.0 and $[Fe/H]$ = $-$1.0. We prefer the 
\cite{aarle2013} values based on their high resolution data. Hence, we used a Kurucz stellar atmosphere model 
with $T_{\mathrm eff}$ = 5500 K, log\,$g$ = 0.0 and $[Fe/H]$ = $-$1.0. A single shell model gave
a reasonable fit to the near, mid and far-IR data (Fig. 13.1, $\chi^{2}$ = 214.4). The fit 
was obtained for $T_{\mathrm d}$ (in) = 260 K, using a mix of silicate (Sil-Ow/0.2) and
graphite (grf-DL/0.8) grains,
modified MRN grain size distribution: $a_{\mathrm min}$ = 0.005 $\micron$, $a_{\mathrm max}$ = 0.50 $\micron$,
$\tau$ = 0.9 and $Y$ = 2. The presence of graphite dust indicates that it is a carbon-rich post-AGB star. Hence, 
the mismatch at IRAC3\,(5.8 $\micron$) and IRAC4\,(8 $\micron$) may be attributed to PAH emissions.

We estimate a luminosity of 4657.0$L_{\odot}$. This is in good agreement with the observed luminosity
estimated by KWVW15 (4223$L_{\odot}$). However, similar to the case of the post-AGBs, J051848.84-700247.0 (\S\,\ref{J051848.84})
and J051906.86-694153.9 (\S\,\ref{J051906.86}), KWVW15 estimated a much higher photospheric luminosity, 11056$L_{\odot}$.
From our best-fit model, the total mass accumulated in the shell is 6.45$\times10^{-3}M_{\odot}$ over 
a period of 144 years at a mass loss rate of 4.48$\times10^{-5} M_{\odot}\mathrm{yr}^{-1}$. 

\subsection{J045623.21-692749.0}
\label{J045623}
This object is classified as a post-AGB disk source by KWVW15, with $T_{\mathrm eff}$ = 4500 K, log\,$g$ = 0.0 and
$[Fe/H]$ = $-$1.0. We used a Kurucz stellar atmosphere model with these parameters. A single shell model (Fig. 14.3)
could not provide a fit to the far-IR data.
An overall fit from the optical to the far-IR
could be obtained under two scenarios: (1) a nested shells model (\#\,1) as in the case of the post-AGB shell sources J050632.10-714229.8 and J051906.86-694153.9 and (2) a two-component model with a warm inner disk and a cold outer shell (model \#\,2). 

In case of the former (\#\,1), the inner shell is composed of Sil-Ow grains
with modified MRN grain size distribution: $a_{\mathrm min}$ = 0.005 $\micron$ and $a_{\mathrm max}$ = 5.0 $\micron$, 
$T_{\mathrm d}$ (in) = 1100 K, $\tau$ = 0.6 and $Y$ = 3.0. The outer shell is composed of Sil-Ow grains with
standard MRN grain size distribution, $T_{\mathrm d}$ (in) = 150 K,
$\tau$ = 0.10 and $Y$ = 2.0. Fig. 14.1 ($\chi^{2}$ = 100.2) shows the correctly illuminated nested shells model. 

In the second scenario (\#\,2), the warm inner disk ($T_{\mathrm d}$ (in) = 1100 K) is composed of amC-Hn grains and the cold outer shell ($T_{\mathrm d}$ (in) = 400 K) is made up of Sil-Ow grains. The amC-Hn disk 
was approximated as a wedge-shaped fraction (0.5) of a sphere with $\tau$ = 0.5, $Y$ = 2 and modified MRN grain size distribution: $a_{\mathrm min}$ = 0.005 $\micron$ and $a_{\mathrm max}$ = 2.0 $\micron$. The Sil-Ow shell has modified MRN grain size distribution: $a_{\mathrm min}$ = 0.40 $\micron$ and $a_{\mathrm max}$ = 1.0 $\micron$, $\tau$ = 0.22 and $Y$ =  15. The correctly illuminated model SED is shown in Fig. 14.2 ($\chi^{2}$ = 195.2).

Based on the reduced chi-square values, we adopt the nested shells model (\#\,1) as the best-fit. The estimated luminosity based on this model is 6598$L_{\odot}$. KWVW15 gave an observed luminosity estimate of 6864$L_{\odot}$ and and a photospheric luminosity of 7131$L_{\odot}$.
The total mass accumulated in the outer shell is estimated to be 8.25$\times10^{-4} M_{\odot}$ in 225 years at 
a mass-loss rate of 3.7$\times10^{-6} M_{\odot}\mathrm{yr}^{-1}$.

\subsection{J051418.09-691234.9}
This object has been classified as a post-AGB disk source (KWVW15), a population II Cepheid variable \citep{soszynski2008} and a RV Tauri star \citep{aarle2011}. KWVW15 estimated $T_{\mathrm eff}$ = 6112 K, log\,$g$ = 0.5 and
$[Fe/H]$ = $-$1.6. We adopted Kurucz stellar atmosphere model with $T_{\mathrm eff}$ = 6000 K, log\,$g$ = 0.5 and $[Fe/H]$ = $-$1.5.
For the R magnitude we used 13.78\footnote{We did not use KWVW15's R-band magnitude (13.96) because we could not find a corresponding reference.} from the NOMAD catalog \citep{zacharias2005}.

A reasonable fit
to the near, mid and far-IR data (Fig. 15.1, $\chi^{2}$ = 98.0) is obtained assuming an inner disk ($T_{\mathrm d}$ (in) = 1100 K) composed of amorphous carbon (amC-Hn) with modified MRN grain size distribution: $a_{\mathrm min}$ = 0.005 $\micron$, $a_{\mathrm max}$ = 2.0 $\micron$, $\tau$ = 5.0 and $Y$ = 15.0. The disk is assumed to be a wedge-shaped fraction (0.25) of a sphere. The outer shell ($T_{\mathrm d}$ (in) = 600 K) is composed of silicate (Sil-Ow/0.4) and graphite (grf-DL/0.6) grains with standard MRN grain size distribution, $\tau$ = 0.4 and $Y$ = 30.0. 
Since the inner disk is optically thick, similar to the post-RGB star, J055102.44-685639.1
(\S\,\ref{J055102.44}) we cannot apply the ``correctly illuminated model" formalism to this object. 

The variable nature of the star may explain the 
difference between the 
WISE (ALLWISE) and IRAC data at their bands 1 and 2 ($\sim$3.5 $\micron$ and 4.5 $\micron$) as also the mismatch between model and observed I-band data. ALLWISE photometry at bands 1(3.4 $\micron$) and 2(4.6 $\micron$) is generally an improvement over the WISE data at these wavelengths (\S\,\ref{wiseall}).  
From our best-fit model, the luminosity of the object is 7763$L_{\odot}$. KWVW15 obtained 6667$L_{\odot}$ for the observed luminosity and 4703$L_{\odot}$ for the photospheric luminosity. 
The total mass accumulated in the shell is estimated to be 1.21$\times10^{-3}$$M_{\odot}$ over 800 years at a mass-loss rate of 1.51$\times10^{-6}$$M_{\odot}$ yr$^{-1}$. The total mass in the disk is 4.67$\times10^{-5}$$M_{\odot}$ over 220 years at a
mass-loss rate of 2.12$\times10^{-7}$$M_{\odot}\mathrm{yr}^{-1}$.

\subsection{J055122.52-695351.4}
This object has been classified as a post-AGB disk source (KWVW15), a 
Population II Cepheid variable \citep{soszynski2008} and a RV Tauri star \citep{aarle2011}. 
KWVW15 estimated $T_{\mathrm eff}$ = 6237 K, log\,$g$ = 1.5 and
$[Fe/H]$ = $-$2.5 for the star.  We used a Kurucz stellar atmosphere model with $T_{\mathrm eff}$ = 6250 K, log\,$g$ = 1.5 and $[Fe/H]$ = $-$2.5. The difference between the 
WISE (ALLWISE) and IRAC fluxes at bands 1 and 2 ($\sim$3.5 $\micron$ and 4.5 $\micron$) may be
attributed to the variable nature of the object. 

A reasonable fit to the near, mid and far-IR data could be obtained assuming a single shell model (Fig. 16.1, $\chi^{2}$ = 41.8) with a mix of graphite (grf-DL/0.7) and amorphous carbon (amC-Hn/0.3) grains, modified MRN grain size distribution: $a_{\mathrm min}$ = 0.05 $\micron$, $a_{\mathrm max}$ = 0.3 $\micron$, $T_{\mathrm d}$ (in) = 450 K, $\tau$ = 0.78 and $Y$ = 6.

We obtained a luminosity of 4116$L_{\odot}$ for the object in close agreement with the observed (3780$L_{\odot}$) and photospheric (3657$L_{\odot}$) luminosity estimated by KWVW15.
The total mass accumulated in the shell is estimated to be 
1.18$\times10^{-3} M_{\odot}$ in 243 years at a mass-loss rate of 4.86$\times10^{-6} M_{\odot}\mathrm{yr}^{-1}$.

\subsection{J052519.48-705410.0}
\label{J052519.48}
This object has been classified as a post-AGB disk source (KWVW15), a 
Population II Cepheid variable \citep{soszynski2008} and a RV Tauri star \citep{aarle2011}. KWVW15 estimated $T_{\mathrm eff}$ = 8117, log\,$g$ = 1.0 and 
$[Fe/H]$ = $-$0.5 . We used a Kurucz stellar atmosphere model with $T_{\mathrm eff}$ = 8000 K , log\,$g$ = 1.0 and $[Fe/H]$ = $-$0.5. There is a significant difference between the 
WISE (ALLWISE) and IRAC fluxes at bands 1 and 2 ($\sim$3.5 $\micron$ and 4.6 $\micron$) which may (at least partly) be
due to the photometric variability. We have chosen to fit the IRAC data, and have excluded the WISE(ALLWISE) fluxes at 3.6 $\micron$, 4.6 $\micron$ and
12 $\micron$ for estimating the reduced chi-square.

Single shell models \#\,1 and \#\,2 (Fig. 17.1) could not
provide simultaneous fit to the near, mid and far-IR data. Using Sil-Ow grains a
fit could be obtained to the near and far-IR fluxes (model \#\,1) while single shell grf-DL model (model \#\,2) 
provided a fit to the near-IR and IRAC (3.6 $\micron$, 4.5 $\micron$ and 5.8 $\micron$) fluxes.

Hence we tried two-component model fits to the observed SED. Fig. 17.2 shows the best-fit 
two-component model for J052519.48-705410.0 (model \#\,3, $\chi^{2}$ = 35.9).
The warm inner disk ($T_{\mathrm d}$ (in) = 800 K) is approximated as a wedge-shaped fraction (0.35) of a sphere and
is made up of Sil-Ow grains with standard MRN grain size distribution, $\tau$ = 1.0 and $Y$ = 20. The outer
shell ($T_{\mathrm d}$ (in) = 600K)  is made up of grf-DL grains with modified MRN grain size distribution: $a_{\mathrm min}$ = 0.005 $\micron$, $a_{\mathrm max}$ = 0.5 $\micron$, $\tau$ = 0.8 and $Y$ = 2.0. Since the inner disk is
optically thick, we cannot apply the ``correctly illuminated model" formalism to this object. 

From our best-fit model \#\,3, we estimated a luminosity of 3804$L_{\odot}$. KWVW15 obtained
3219$L_{\odot}$ and 4943$L_{\odot}$ respectively for the observed and photospheric luminosities. 
The total mass in the disk and shell was estimated to be 3.64$\times10^{-5} M_{\odot}$ and 
1.25$\times10^{-4} M_{\odot}$ respectively.
The age of the disk is estimated to be
451 years while the shell is $\sim$ 20 years old. The mass loss rates for the disk and shell respectively are
8.08$\times10^{-8} M_{\odot}\mathrm{yr}^{-1}$ and 6.12$\times10^{-6} M_{\odot}\mathrm{yr}^{-1}$.

\section{Discussion}
\label{discuss}

\subsection{Common envelope evolution and ejecta mass}

When the primary in a binary system overflows its Roche lobe and the 
resulting mass transfer proceeds too rapidly to be accreted by
the compact companion, a CE system results. 
CEE \citep{paczynski1976, ivanova2013} may give rise to close binary systems.
Two scenarios may exist --- a rapid plunge in of the companion or a slow spiral-in phase. 
In the former, if enough orbital energy is deposited to the CE via dynamical friction, the whole envelope is ejected on a
dynamical timescale \citep{paczynski1976}. The second case provides an alternate
route for the envelope ejection \citep{ivanova2002, podsiadlowski2010}
over 'several dynamical time scales' \citep{ivanova2013, clayton2017}. 
Recently, a third scenario has been proposed by \citet{glanz2018} wherein  dust driven winds similar to those observed in AGB stars may
lead to the ejection of the CE.

Following the ejection of the CE, dust formation may occur in the 
expanding gas \citep{lu2013, iaconi2020} which may explain the
presence of circumstellar dust shells in these systems. Some fraction of the ejected mass may also fall back and interact with the binary leading to the formation of circumbinary disks \citep{kashi2011}.

\citet{lu2013} discuss dust formation in the CE ejecta of binary systems wherein the giant is 
on the RGB (they call it the first giant branch or FGB) and the companion is a 1\,\ms~degenerate star. Since RGB stars are expected to be oxygen-rich
following the first dredge-up, they use only olivine, pyroxene, and quartz-type silicates and 
iron grains for discussion. They find that silicate and iron-rich dust in the CE ejecta 
can form over a very wide range of radial distances ($10^{14}-10^{18}$ cm) from the RGB star (Fig. 5, \citet{lu2013}) which
easily encompasses our $R_{\mathrm in}$ values for the cold dust shells from our best-fit models (Table\,\ref{mod-tbl}). In L\"u et al.'s models, 
the mass of dust produced in the CE ejecta of giants with masses 1$-$7\,$M_{\odot}$ ranges from about 10$^{-9}$$M_{\odot}$ to 10$^{-2}$$M_{\odot}$. The lower end of this mass range corresponds to CEE occurring at the base of the RGB and the model parameter $\gamma = 0.2$ (see eqn. 2 of  \citet{lu2013} for a definition of $\gamma$). Higher values of $\gamma$ 
result in CE ejecta with relatively higher densities, and thus more efficient dust formation.  Thus, the upper end of the ejecta mass range corresponds to CEE occurring anywhere between the base and the tip of the RGB with $\gamma\ge0.3$, or at the tip of the RGB with $\gamma\ge0.2$ (Fig. 2, \citet{lu2013}).

For our sample of post-RGB objects, the values of total mass, $M_{\mathrm gd}$ in the outer shell range from about 4$\times10^{-5}$\,$M_{\odot}$ to  3$\times10^{-2}$\,$M_{\odot}$
(Table\,\ref{model_param}) -- inferred from our model dust masses, $M_{\mathrm d}$ that lie in the range 2$\times10^{-7}$\,$M_{\odot}$--$1.5\times10^{-4}$\,$M_{\odot}$. 
We compare our dust mass estimates with those of \citet{lu2013} in Fig.\,\ref{dustmass_lu}.
Since our derived  masses are likely lower limits to the total dust mass, its most likely that either $\gamma\ge0.3$ and/or CEE has occurred near or at the tip of the RGB for our post-RGB sources. \citet{lu2013}were pessimistic about the difficulty to observe
the dust produced in CE ejecta, since ``the distance of dust formation
in the CE ejecta is between $\sim10^{14}$ and $\sim10^{18}$ cm and is relatively
far away from the FGB star"; however our study clearly shows that dust is relatively easily detected in our sample of post-RGB sources between few$\times10^{14}$ and $\sim10^{18}$ cm. Dust formation in CE ejecta wherein the primary is a RGB star has also been studied by \citet{iaconi2020}
using a 3D SPH hydrodynamic simulation. Their range of dust masses obtained were comparable to that of \citet{lu2013}.

\subsection{``Disk" and ``Shell" source classification}
KWVW15 classified the post-RGB and post-AGB sources into two groups -- ``disk" and ``shell" \citep{vanwinckel2003}, based on their SEDs. 
 The ``disk" sources are inferred to have a stable compact circumbinary disk and are confirmed binaries (with orbital periods, P$_{orb}\sim100-2000$\,d), whereas the ``shell" sources have not resulted (so far) in any clear detected binary orbit \citep{hrivnak2011}, suggesting that these are either 
single-stars or have companions with relatively very wide orbits.

However, our modeling shows that the KWVW15 ``shell"/``disk" classification is not robust. All ``disk" sources in our sample require the presence of ``shells". In addition, we find the presence of a disk in some ``shell" sources (J043919.30-685733.4 and J051920.18-722522.1), and we exclude the presence of a disk in some ``disk" sources (J045623.21-692749.0 and J045623.21-692749.0).

We find that the disk fractions are surprisingly large (typically 0.3--0.4), implying disks with large opening angles ($\sim41\arcdeg\pm6\arcdeg$ ) and hence
geometrically thick structures.
The large opening angles appear to be roughly consistent with the gas density of the ejected envelope as seen in numerical simulations of CEE, relatively soon after CEE occurs e.g., see Fig.\,1 of \citet{garcia-segura18}. It appears unlikely that the ``disk" structures in our models represent the stable compact circumbinary disks envisaged by \citet{vanwinckel2003}, although it is possible that the latter reside within the former. The inner rims of circumbinary disks may be puffed up, as suggested by the modeling of near and mid-IR interferometric imaging of the disk structure in the post-AGB star IRAS 08544-4431 \citep{deroo2007}, and may mimic the physical effect of absorbing a substantial part of the direct stellar radiation within a relatively large solid angle.

\subsubsection{Source Luminosities}
We summarize our derived model stellar luminosities, together with those provided by KWVW15 in Tables \ref{lum} (for post-RGB sources) and \ref{lum} (for post-AGB sources). We find that our values are in reasonable agreement with KWVW15's observed values. For a few objects where KWVW15's observed and photospheric values are very discrepant, our model values provide a more accurate estimate of the stellar luminosities.
    
\subsubsection{Ages of the dusty disks and shells}
A comparison of the ages of the disk and shells can be potentially useful for constraining details of the physical mechanism responsible for the transition of RGB stars directly (i.e., without ascending the AGB) towards the PN phase, which has been argued to be CEE. For example, in the Boomerang Nebula, the age of the central disk structure is estimated to be 1925 yr, less than the age of the shell of $\gtrsim3840$\,yr, which implies that the disk was formed after the (inferred) companion entered the CE, resulting in the ejection of the primary red giant star's envelope and the subsequent merging of the companion with the primary's core. In general, we find that the disk age is generally lower than the shell age for the objects in our study with the exception of the post-RGB object, J045555.15-712112.3 and the post-AGB object, J052519.48-705410.0.
Our estimate for the age depends on (i) the assumed expansion velocities for the disk and the shell respectively (V$_{\rm exp}$) and (ii) on the thickness of the shell/disk ($Y$). 

Our derived $Y$
value is likely a lower limit to the estimated shell thickness. We illustrate this point with the post-RGB objects, J045755.05-681649.2 ($Y$ = 30) and J051920.18-722522.1 ($Y$ = 20).  In Figs. 4.2 and 7.4, we plot the correctly illuminated two-component models for the objects with the thickness of the outer shell increased by a factor of 10, i.e. $Y$ = 300 and 200 respectively. No noticeable change is observed in the fits. The derived model parameters corresponding to these fits are given in \ref{mod-tbl}. 
In the absence of data beyond 24 $\micron$, it is quite plausible that the shell is even more extended (i.e. the actual value of Y is much larger than in our model)
For example, in the post-AGB object, IRAS 22036+5306, where the modeled SED included data up to $\sim$200\,$\micron$~the outer radius of the shell is a factor $\sim40$ larger than the inner radius \citep{sahai2006}. A thicker outer shell
in J045555.15-712112.3 would translate into a larger value for the shell age. 

The post-AGB star, J052519.48-705410.0 has been classified as a Cepheid \citep{soszynski2008} and a RV Tauri variable \citep{aarle2011}. The variable nature of this object is also evidenced by the difference between the IRAC and WISE (ALLWISE) fluxes (\S\,\ref{J052519.48}). The disk being older
than the shell is consistent with the RV Tauri classification of the object \citep{waters1993}. 
In RV Tauri systems, the hot dust is believed to be related to the presence of a companion and may be the remnant of a phase of high mass loss in the past or rapid mass transfer during Roche lobe overflow.
The outer shell is usually related to more recent mass-loss history \citep{vanwinckel1999, waters1993}.

\subsubsection{RV Tauri post-AGB stars}
\label{rvtau}
The post-AGB disk stars, J051418.09-691234.9, J052519.48-705410.0, J055122.52-695351.4 are 
known RV Tauri variables \citep{aarle2011}. Galactic RV Tauri variables with infrared excess are
almost always associated with circumstellar disks, e.g. \citep{lloyd1995, vanwinckel1999, gielen2007}
The RV Tauri stars in the LMC may also be expected to show similar morphology \citep{manick2018}. However, our model SEDs show that although J052519.48-705410.0 and J055122.52-695351.4 harbour a warm inner disk in addition to the outer shell, the observed SED of the RV Tauri star, J055122.52-695351.4 did not require a circumstellar disk, as it could be fitted by a single shell model (Fig. 16.1).

\subsection{Systematic discrepancies in SED Modeling}
In this section, we discuss some systematic discrepancies between the model and observed SEDs and their implications. Some of these may be due to the presence of dust emission features -- JWST/MRS spectroscopy will be needed to test this hypothesis. Other discrepancies may be related to systematic uncertainties in the photometry.

\subsubsection{PAH emission}
\label{pahs}
We find systematic discrepancies between the model and observed SEDs in the 5.8--8 $\micron$ region for some of our post-RGB sources -- J043919.30-685733.4, J045555.15-712112, J045755.05-681649.2 and J050257.89-665306.3.These discrepancies may be explained by the presence of 
PAHs in the dust disks/shells of our sources.

We discuss two possibilities that may account for the presence of PAH molecules in oxygen-rich post-RGB stars. Both of
these have been discussed by \citet{jura2006} for the case of HD233517, an evolved oxygen-rich red giant showing 
PAH features in the circumstellar environment. In the first scenario, since the dusty post-RGB stars in our sample are most likely in close binary systems, the companion of the post-RGB star which may have been more massive than the latter on the MS, may have evolved to become a C-rich AGB star and transferred C-rich material to the post-RGB star progenitor, making it carbon-rich\footnote{this mechanism is believed to be responsible for dwarf carbon stars and CH stars}; then, the post-RGB progenitor, when it evolved to become a red giant underwent a CE interaction with the companion (then a WD), ejecting C-rich material that made PAHs. Another possible scenario is that when the more massive star of the binary pair became a red giant, it engulfed the secondary, and the mass transfer and resulting ejection of circumstellar matter via CEE that followed harbored the required physical conditions for the formation of PAHs. For example, Fischer-Tropsch
catalysis on the surface of iron grains \citep{willacy2004} converted much of the carbon initially contained in CO into hydrocarbons, and subsequently, analogous to the atmospheric chemistry in Jupiter \citep{wong2003}, these
simple hydrocarbons were converted into PAHs.

PAH emission is believed to be the result of excitation of the PAH molecules and/or photodissociation of the CO molecule in dense torii. The latter is believed to account for the presence of PAH emission features in O-rich circumstellar environments of post-AGB stars, PPNe  and PNe \citep{beintema1996, waters1998, matsuura2004, cerrigone2009, guzman2011}.  It is widely accepted that UV flux is required for both these mechanisms. However, UV flux cannot account for the excitation of PAH molecules in the circumstellar environment of post-RGB stars, because of their low T$_{eff}$. \citet{li2002} argue that UV flux is not imperative for the excitation of PAH molecules -- depending on the sizes of the PAH molecules, these may be excited by photons covering a wide range of wavelengths. In particular, it is possible for PAH molecules to be excited by photons longwards of UV wavelengths -- in fact, PAH emission may be observed even if the effective temperature of the exciting star is as low as 3000 K.

\subsubsection{Silicate emission feature}
The $\sim$10\,$\micron$~amorphous silicate emission feature is visible in the model spectra of several of our objects. 
We illustrate the effects of choice of silicate grains
(Sil-Ow, Sil-Oc), dust temperature at the inner boundary of the dust shells ($T_{\mathrm d}$) and maximum grain size on the
silicate emission feature, using the post-AGB (shell) source, J050632.10-714229.8 as an example. Changing the grains from ``warm" (Sil-Ow) to ``cold" (Sil-Oc) silicates doesn't do much to alter the fit in the 
mid-IR (Fig. 10.2). The silicate emission feature is also not very sensitive to changing $T_{\mathrm d}$ (Fig. 10.7). 
Incorporating larger Sil-Ow grains in the inner shell has the effect of broadening the wings of the 
silicate emission feature while reducing its intensity. Fig. 10.8 shows the fits 
for different values of $a_{\mathrm max}$. 

\subsubsection{R and I band data}
\label{RIdata}
For several of our post-RGB and post-AGB sources, the model SEDs do not provide good matches to the R- and I-band data.
The R magnitudes, taken from the GSC\,2.2 catalog (as in KWVW15), often have large error bars, e.g. the post-RGB sources, J051347.57-704450.5 and J051920.18-722522.1. The modeled SEDs do not agree with the extinction corrected Rmag even after taking into account the large error bars. For the post-RGB source J051347.57-704450.5, in addition to the GSC 2.2 data, we plotted the  extinction corrected USNO-B 1.0 R1 and R2 fluxes (Fig. Set 3). We find that the latter are in better agreement with the modeled SED.

KWVW15 compiled I magnitudes from the MCPS Catalog for the LMC \citep{zaritsky2004}. In 
Table\,\ref{tbl-denisetc} we compare the I-band
magnitudes from \citet{zaritsky2004} with data from the DENIS catalog \citep{cioni2000} and
USNO-B 1.0 \citep{monet2003}. The \citet{zaritsky2004} data is in good agreement with the DENIS data with the exception of the post-RGB source, J050257.89-665306.3 ($\Delta$Imag $\sim$ 1 mag). The corrected DENIS Imag for this object is in good agreement with our modeled SED, Fig. 8.2. In most cases, the USNO-B 1.0 magnitudes are brighter than the \citet{zaritsky2004} data and would show better agreement with our modeled SEDs.

\subsubsection{WISE and ALLWISE photometry}
\label{wiseall}
The WISE magnitudes from the All-Sky Data Release \citet{cutri2012} and the ALLWISE magnitudes \citep{cutri2013} are listed in Table \ref{srcs}.
A mismatch between modeled and observed SEDs at WISE band 2 (W2/4.6 $\micron$) was observed for the post-RGB sources, J043919.30-685733.4, J045555.15-712112.3 and J045755.05-681649.2. 
Hence, we additionally plotted the ALLWISE 
magnitudes of the sources from Vizier.
Because of processing improvements in compiling the photometric data and the addition of a second coverage of the sky in the short wavelength bands, ALLWISE achieves better sensitivity in bands 1 and 2. The latter data shows good agreement with our modeled SEDs  for J043919.30-685733.4 (Figs. 2.8, 2.9), J045755.05-681649.2 (Fig. 7.2) and J045555.15-712112.3 (Fig. 6.4). 
Similarly, we notice a mismatch between modeled and
observed SEDs for the post-AGB ``disk'' sources in the WISE band 1 and 2 fluxes. 
When the mismatch between WISE and ALLWISE data significantly 
exceeds the error in the observed photometry, it may be due to source variability. Three of the 
post-AGB disk sources, J051418.09-691234.9, J052519.48-705410.0, J055122.52-695351.4 are known
RV Tauri variables (\S\,\ref{rvtau}).

\section{Conclusions}\label{conclude}

We have modeled the SEDs of a select sample of post-RGB and post-AGB objects (8 in each class) in the LMC,  drawn from a large list of such sources classified as possessing ``disks" or ``shells" based on the shape of their SEDs. A circumstellar disk suggests binarity. Close binary systems undergoing CEE can result in the
the premature evolution of an RGB star into a PNe. We have used the 1-D radiative transfer code, DUSTY, approximating disk structures as wedge-shaped fractions of a sphere (disk-fractions). We have derived the total dust mass (and using an assumed gas-to-dust ratio, the total mass of gas) in the disks and  shells, and set constraints on the dust grain composition and sizes. Our main conclusions are listed below:
\begin{enumerate}
    \item The masses of ejecta in the disks of post-RGBs (disks/inner shells of post-AGBs) lie in the range $\sim9\times10^{-9}-3\times10^{-5}$ \ms~($\sim9\times10^{-9}-7\times10^{-4}$ \ms) and
    in the outer shells from $\sim4\times10^{-5}-3\times10^{-2}$ \ms~($\sim1\times10^{-4}-6\times10^{-2}$ \ms). On average, the shell masses in 
    post-RGBs are less than those in post-AGBs, with the caveat that a substantial amount of mass in both types of objects may lie in an extended cold shell, that is detectable only at wavelength longwards of the MIPS\,24 $\micron$ band. 
    \item The disk fractions are surprisingly large (typically 0.3--0.4), implying that the disks are geometrically thick structures with a substantial opening angle ($\sim41\arcdeg\pm6\arcdeg$). The large opening angles appear to be roughly consistent with the latitudinal  variation of gas density in the ejected envelope as seen in numerical simulations of CEE, relatively soon after CEE occurs.
    \item We find evidence that for some post-RGB sources the ejected matter may be carbon-rich, even though it is expected to be oxygen-rich. For J055102.44-685639.1, the inner disk is composed of amorphous carbon grains and the outer shell has a mix of silicate and silicon carbide grains. In addition, for J043919.30-685733.4, J045555.15-712112, J045755.05-681649.2 and J050257.89-665306.3, PAH emission may account for the discrepancy between model and observed SEDs. The presence of PAHs provides independent support for the hypothesis of binary interaction leading to the formation of post-RGB objects. 
    \item We find that the published classification of these objects as ``shell" or ``disk" sources is not robust. Our modeling shows
    the presence of a disk is (a) required in some ``shell" sources (the post-RGB sources: J043919.30-685733.4
    and J051920.18-722522.1), and (b) not required in some ``disk" sources (the post-AGB sources: J045623.21-692749.0 and J055122.52-695351.4).
    \item Comparison of our model dust mass values with the predictions of models of CEE on the RGB that produce dust suggest that CEE occurred near or at the tip of the RGB for the post-RGB sources in our sample. 
\end{enumerate}

\begin{acknowledgements}

We thank our anonymous referees for their meticulous reviews of our paper, which have helped us improve it. G.S. would like to acknowledge financial support from the Department of Science and Technology (DST), Government of India, through a grant numbered SR/WOS-A/PM-93/2017. R.S.'s contribution to the research described here was carried out at the Jet Propulsion Laboratory, California Institute of Technology, under a contract with NASA, and funded in part by NASA via ADAP awards, and multiple HST GO awards from the Space Telescope Science Institute.
\end{acknowledgements}

\clearpage 
\movetabledown=6cm
\begin{rotatetable}
\begin{deluxetable*}{{cccccccccccccccccccccccc}}
\setlength{\tabcolsep}{1pt}
\tabletypesize{\scriptsize}
\tablenum{1}
\tablecaption{Photometric Data \label{srcs}}
\tablehead{
\colhead{Object} & \colhead{Type} & \colhead{U} & \colhead{B} & \colhead{V} & \colhead{R} & \colhead{I} & \colhead{J} & \colhead{H} & \colhead{K} & \colhead{[3.6]} & \colhead{[4.5]} & \colhead{[5.8]} & \colhead{[8.0]} & \colhead{[24]$^{\rm a}$} & \colhead{[24]$^{\rm b}$ }& \colhead {W1} & \colhead{W2} & \colhead{W3} & \colhead{W4} & \colhead{W1$^{\prime}$} & \colhead{W2$^{\prime}$} & \colhead{W3$^{\prime}$} & \colhead{W4$^{\prime}$}} 
\startdata
\cutinhead{post-RGB stars}
J043919.30-685733.4 & Shell & 20.093 & 19.475 & 18.805 & 18.100 & 17.900 & 17.206 & 16.597 & 16.333 & 16.068 & 15.741 &...& 13.856 & 9.263 & 9.307 & 16.416 & 16.259 & 13.07$^{\dagger}$ & 9.380$^{\dagger}$ & 16.194 & 15.894 & 12.938 & 8.988$^{\dagger}$\\
J051347.57-704450.5 & Shell & 19.909 & 18.510 & 17.077 & 16.740 & 15.683 & 14.661 & 13.933 & 13.719 & 13.661 & 13.728 & 13.625 & 13.563 &  9.257 &  9.310 &  ... & ... & ... & ... & ... & ... & ... & ... \\
    &   & ... & ... & ... & 15.76  & ... & ... & ... & ... & ... &
        & ... & ... & ... & ... &  ... & ... & ... & ... & ... & ... & ... & ...       \\
     &               & ... & ... & ... & 16.29 & ... & ... & ... & ... & ... &
        & ... & ... & ... & ... &  ... & ... & ... & ... & ... & ... & ... & ...  \\
J051920.18-722522.1 & Shell & 19.408 & 18.650 & 17.426 & 17.020 & 16.152 & 15.154 & 14.540 & 14.396 & 13.977 & 13.785 & 13.159 & 12.057 & 8.050 & 8.053 & 14.084 & 13.807 & 10.882 & 8.462 & 14.069 & 13.789 & 10.752 & 8.297\\
J053930.60-702248.5 & Shell & ... & 19.934 & 18.243 & 18.600 & 16.915 & 15.810 & 14.981 & 14.756 &  14.611 & 14.686 & 14.405 & 14.160 & 9.568 & 9.345 & 14.907 & 15.130 & 11.350 & 9.302 & 14.804 & 15.206 & 11.274 & 9.015\\
J045555.15-712112.3 & disk & 18.216 & 18.455& 18.165 & 17.610 & ... & 16.790 & 16.350 & 16.011 & 14.232 & 14.190 & 11.518 & 9.974 & 7.323 & 7.197 & 14.034 & 13.774 & 8.696 & 6.833 & 14.035 & 13.814 & 8.762 & 6.896 \\
J045755.05-681649.2 & disk & 20.118 & 19.563 & 18.321 & ... & 17.205 & 16.369 & 15.651 & 15.470 & 15.059 & 14.922 & 14.557 & 12.480 & 9.626 & 9.732 & 15.233 & 15.460 & 11.489 & 9.825$^{\dagger}$ & 15.144     & 15.094 & 11.253 & 9.190$^{\dagger}$\\
\enddata
\tablecomments{\\
1) [3.6], [4.5], [5.8], [8.0] photometric values correspond to $Spitzer$ IRAC band
1--4 observations.\\
2) [24]$^{\rm a}$, [24]$^{\rm b}$: $Spitzer$ MIPS band 1 observations corresponding to epoch 1 and 2 respectively \citep{meixner2006}, retrieved
from https://irsa.ipac.caltech.edu.\\
3) WISE magnitudes: W1\,(3.4\,$\micron$), W2\,(4.6\,$\micron$), W3\,(12\,$\micron$), W4\,(22\,$\micron$)\\ 
4) ALLWISE magnitudes: W1$^{\prime}$\,(3.4\,$\micron$), W2$^{\prime}$\,(4.6\,$\micron$), W3$^{\prime}$\,(12\,$\micron$), W4$^{\prime}$\,(22\,$\micron$)\\ 
5) $^{\dagger}$: Magnitudes are upper limits}
\end{deluxetable*}
\end{rotatetable}

\startlongtable
\begin{deluxetable*}{{ccccccccccccc}}
\setlength{\tabcolsep}{2pt}
\tabletypesize{\scriptsize}
\tablenum{2}
\tablecaption{Models of post-RGB objects \label{mod-tbl}}
\tablehead{
\colhead{$T_{\mathrm d}$(in)\tablenotemark{a} }& \colhead{$T_{\mathrm d}$(out)\tablenotemark{b} }& \colhead{$Y$\tablenotemark{c}} & \colhead{$R_{\mathrm in}$\tablenotemark{d}} & \colhead{$\tau$\tablenotemark{e}} & \colhead{$a_{\mathrm min}$} & \colhead{$a_{\mathrm max}$} & \colhead{$F_{\rm bol}$} & \colhead{$L$} & \colhead{$M_{\mathrm gd}$\tablenotemark{f} }& \colhead{Age\tablenotemark{g} }& \colhead{$dM/dt$\tablenotemark{h}} & \colhead{$\chi^{2}$}\\
\colhead{(K)} & \colhead{(K)} &  & \colhead{(10$^{14}$cm) }&  & \colhead{($\micron$)} & \colhead{($\micron$)} & \colhead{(\fbolunit)} & \colhead{(\ls)} & \colhead{(\ms)} & \colhead{(yr)} & \colhead{(\my)} & }
\startdata
\cutinhead{{\bf J043919.30-685733.4} (shell source)}
\multicolumn{12}{l}{{\bf Single shell (one-component) models with Sil-Ow grains}} & \\
\hline
550 & 161  & 8   & 1.82 & 1.5 & 0.005 & 0.25 & 1.8$\times10^{-15}$ & 140.0 & 1.45$\times10^{-5}$ & 8.09 &  1.79$\times10^{-6}$ & 87.6\\
550 & 65.6 & 100 & 1.79 & 1.5 & 0.005 & 0.25 & 1.8$\times10^{-15}$ & 140.0 & 1.74$\times10^{-4}$ & 112  & 1.55$\times10^{-6}$ & 61.4\\
550 & 38.4 & 500 & 1.79 & 1.5 & 0.005 & 0.25 & 1.8$\times10^{-15}$ & 140.0 & 8.72$\times10^{-4}$ & 565  & 1.54$\times10^{-6}$ & 68.6\\
145 & 27.2 & 100 & 26.3 & 0.8 & 0.005 & 0.25 & 1.5$\times10^{-15}$ & 116.0 & 2.01$\times10^{-2}$ & 1650 & 1.22$\times10^{-5}$ & 48.3\\
175 & 32   & 100 & 16   & 0.8 & 0.005 & 0.25 & 1.5$\times10^{-15}$ & 116.0 & 7.51$\times10^{-3}$ & 1010 & 7.45$\times10^{-6}$ & 47.8\\
500 & 66.7 & 100 & 1.79 & 0.8 & 0.005 & 0.25 & 1.5$\times10^{-15}$ & 116.0 & 9.32$\times10^{-5}$ & 112  & 8.30$\times10^{-7}$ & 43.6\\
\hline
\multicolumn{12}{l}{{\bf Two-component model, \#\,1:} cold outer shell: Sil-Ow and warm inner disk: Sil-Ow (disk fraction = 0.35)} & 18.8\\
\enddata
\tablecomments{\\
The inner disk is approximated as an axially symmetric wedge-shaped fraction of a sphere. This fraction is referred to as {\bf ``disk fraction"} The acronymns for the dust grain compositions refer to 
`warm' and `cold' silicates (Sil-Ow and Sil-Oc) from \citet{ossenkopf1992},
graphite grains (grf-DL) from \citet{draine1984}, amorphous carbon (amC-Hn) from \citet{hanner1988} and silicon carbide (SiC-Pg) from \citet{pegourie1988}.\\
{\bf shell:} Parameters for the outer shell in two-component models\\
{\bf disk:} Parameters for the inner disk in two-component models\\
{\bf shell(low-lat):} Correctly illuminated models, part of shell in the solid angle not covered by the disk \\
{\bf shell(high-lat):} Correctly illuminated models, part of shell in the solid angle covered by the disk \\
{\bf suffix `s':} Single shell model\\
{\bf suffix `c':} Correctly illuminated two-component model\\
{$^{\dagger}$:} Adopted (best-fit) model
\tablenotetext{a}{The (input) dust temperature at shell inner radius}
\tablenotetext{b}{The (output) dust temperature at shell outer radius}
\tablenotetext{c}{The (input) ratio of the dust shell's outer to inner radius}
\tablenotetext{d}{The (inferred) inner radius of the dust shell}
\tablenotetext{e}{The (input) dust shell's optical depth at $0.55\micron$}
\tablenotetext{f}{The (inferred) circumstellar (gas+dust) mass using $\delta$ = 200}
\tablenotetext{g}{The (inferred) dust shell's age}
\tablenotetext{h}{The (inferred) mass-loss rate characterizing the dust shell}}
\end{deluxetable*}

\clearpage
\startlongtable
\begin{deluxetable*}{{ccccccccccccc}}
\setlength{\tabcolsep}{2pt}
\tabletypesize{\scriptsize}
\tablenum{3}
\tablecaption{Models of post-AGB objects \label{mod-tbl-pagb}}
\tablehead{
\colhead{$T_{\mathrm d}$(in)\tablenotemark{a} }& \colhead{$T_{\mathrm d}$(out)\tablenotemark{b} }& \colhead{$Y$\tablenotemark{c}} & \colhead{$R_{\mathrm in}$\tablenotemark{d}} & \colhead{$\tau$\tablenotemark{e}} & \colhead{$a_{\mathrm min}$} & \colhead{$a_{\mathrm max}$} & \colhead{$F_{\rm bol}$} & \colhead{$L$} & \colhead{$M_{\mathrm gd}$\tablenotemark{f} }& \colhead{Age\tablenotemark{g} }& \colhead{$dM/dt$\tablenotemark{h}} & \colhead{$\chi^{2}$}\\
\colhead{(K)} & \colhead{(K)} &  & \colhead{(10$^{14}$cm) }&  & \colhead{($\micron$)} & \colhead{($\micron$)} & \colhead{(\fbolunit)} & \colhead{(\ls)} & \colhead{(\ms)} & \colhead{(yr)} & \colhead{(\my)} & }
\startdata
\cutinhead{{\bf J050632.10-714229.8} (shell source)} 
\multicolumn{13}{l}{\bf{Single shell (one-component) model with Sil-Ow grains, \#\,1,s}} \\ 
\hline
200 & 62.4 & 20 & 81.1 & 0.60 & 0.005 & 0.25 & 7$\times10^{-14}$ & 5434 & 2.87$\times10^{-2}$ & 3260 & 8.83$\times10^{-6}$ & 231.1\\
\hline
\multicolumn{13}{l}{Varying the dust temperature at the shell's inner radius} \\ 
\cline{1-5}
180 & 57.1 & 20 & 106.0 & 0.60 & 0.005 & 0.25 & 7$\times10^{-14}$ & 5434 & 4.93$\times10^{-2}$ & 4260 & 1.16$\times10^{-5}$ & 248.7\\
220 & 67.4 & 20 & 64.7 & 0.60 & 0.005 & 0.25 & 7$\times10^{-14}$ & 5434 & 1.83$\times10^{-2}$ & 2600 & 7.05$\times10^{-6}$ & 253.9\\
\hline
\multicolumn{13}{l}{\bf{Nested Shell Models: The warm inner shell}} \\
\cline{1-6}
\multicolumn{13}{l}{grain type \#\,1: Sil-Ow; varying the temperature at the inner radius of the warm shell} \\
\cline{1-7}
450 & 286 & 2 & 15.4 & 0.4 & 0.005 & 0.25 & 7$\times10^{-14}$ & 5434 & 6.92$\times10^{-5}$ & 9.77 & 7.08$\times10^{-6}$ & 1690.1\\
\enddata
\tablecomments{\\
The inner disk is approximated as an axially symmetric wedge-shaped fraction of a sphere. This fraction is referred to as ``disk fraction". The acronymns for the dust grain compositions refer to 
`warm' and `cold' silicates (Sil-Ow and Sil-Oc) from \citet{ossenkopf1992},
graphite grains (grf-DL) from \citet{draine1984}, amorphous carbon (amC-Hn) from \citet{hanner1988} and silicon carbide (SiC-Pg) from \citet{pegourie1988}.\\
{\bf shell:} Parameters for the outer shell in two-component models\\
{\bf disk:} Parameters for the inner disk in two-component models\\
{\bf shell(low-lat):} Correctly illuminated models, part of shell in the solid angle not covered by the disk \\
{\bf shell(high-lat):} Correctly illuminated models, part of shell in the solid angle covered by the disk \\
{\bf suffix `s':} Single shell model\\
{\bf suffix `c':} Correctly illuminated two-component model\\
{\bf $^{\dagger}$: }Adopted (best-fit) model
\tablenotetext{a}{The (input) dust temperature at shell inner radius}
\tablenotetext{b}{The (output) dust temperature at shell outer radius}
\tablenotetext{c}{The (input) ratio of the dust shell's outer to inner radius}
\tablenotetext{d}{The (inferred) inner radius of the dust shell}
\tablenotetext{e}{The (input) dust shell's optical depth at $0.55\micron$}
\tablenotetext{f}{The (inferred) circumstellar (gas+dust) mass using $\delta$ = 200}
\tablenotetext{g}{The (inferred) dust shell's age}
\tablenotetext{h}{The (inferred) mass-loss rate characterizing the dust shell}}
\end{deluxetable*}

\movetabledown=7cm
\movetableright=-1cm
\begin{rotatetable}
\begin{deluxetable*}{@{\extracolsep{5pt}}cc|ccccccc|ccccccc|ccc}
\setlength{\tabcolsep}{1pt}
\tabletypesize{\scriptsize}
\tablecolumns{19}
\tablenum{4}
\vskip 1cm
\tablecaption{Important parameters derived from the best-fit models\label{model_param}}
\tablehead{
\colhead{} &\colhead{} & \multicolumn{7}{c}{Inner disk/shell} & \multicolumn{7}{c}{Outer shell} & \colhead{}  &\colhead {} &\colhead{} \\
\cline{3-9} \cline{10-16}\\
\colhead{Object} & \colhead{Disk-Fraction}& \colhead{Dust Composition} & \colhead{$T_{\mathrm d}$(in)\tablenotemark{a}} &\colhead{$\tau$\tablenotemark{b}} & \colhead{$a_{\mathrm min}$\tablenotemark{c}} & \colhead{$a_{\mathrm max}$\tablenotemark{d}} &\colhead{ $Y$\tablenotemark{e}} & \colhead{$M_{\mathrm gd}$\tablenotemark{f}} & \colhead{Dust Composition} & \colhead{$T_{\mathrm d}$(in)\tablenotemark{a}} & \colhead{$\tau$\tablenotemark{b} }& \colhead{$a_{\mathrm min}$\tablenotemark{c}} & \colhead{$a_{\mathrm max}$\tablenotemark{d}} & \colhead{$Y$\tablenotemark{e}} & \colhead{$M_{\mathrm gd}$\tablenotemark{f}} & \colhead{$L$\tablenotemark{g}} & \colhead{$\frac{M_{\mathrm gd}(disk)}{M_{\mathrm gd}(shell)}$}  & \colhead{$\chi^{2}$\tablenotemark{h}}\\
\colhead{}& \colhead{} & \colhead{} & \colhead{(K)} &\colhead{} & \colhead{($\micron$)} & \colhead{($\micron$)} &\colhead{} & \colhead{(\ms)} & \colhead{} & \colhead{(K)} & \colhead{} & \colhead{($\micron$)} & \colhead{($\micron$)} &\colhead{} & \colhead{(\ms)} & \colhead{(\ls)} &\colhead{} &\colhead{}}
\startdata
\cutinhead{post-RGB stars}
\multicolumn{13}{l}{{\bf shell sources}} \\
\hline 
J043919.30-685733.4 & 0.35 & Sil-Ow/1.0 & 1000 & 0.5 & 0.005 & 0.25 & 1.4 & 2.19$\times10^{-8}$ & Sil-Ow/1.0 & 130 & 0.65 & 0.005 & 0.25 & 20 & 5.2$\times10^{-3}$ & 116 & 4.21$\times10^{-6}$ & 18.1\\
J051347.57-704450.5 & ... & ... & ... & ... & ... & ... & ... & ... & Sil-Ow/1.0 &250 & 0.40 & 0.1 & 0.25& 3.0 & 4.66$\times10^{-5}$ & 776 & ... & 69.2\\
J051920.18-722522.1 & 0.4 & Sil-Ow/1.0  & 500 & 0.4 & 0.5 & 20 & 2.0 & 3.01 $\times\,10^{-5}$ & Sil-Ow/1.0 &110 & 0.65 & 0.005 & 0.25 & 20 & 3.41$\times10^{-2}$ & 582 & 8.83$\times10^{-4}$ & 17.5\\
J053930.60-702248.5 & ... & ... & ... & ... & ... & ... & ... & ... & Sil-Ow/1.0 &300 & 0.70 & 0.005 & 0.25 & 10 & 5.81$\times10^{-5}$ & 295 & ... & 30.4\\
\hline
\multicolumn{19}{l}{{\bf disk sources}}\\
\hline
J045555.15-712112.3 & 0.4 & Sil-Ow/1.0 & 1000 & 1.0 & 0.005 & 0.25 & 10.0 & 3.08$\times10^{-6}$ & Sil-Ow/1.0 & 300 & 2.5 & 0.005 & 0.25 & 2.0 & 2.84$\times10^{-4}$ & 465 & 1.08$\times10^{-2}$ & 116.6\\
J045755.05-681649.2 & 0.4 & Sil-Ow/1.0 &1300 & 0.5 & 0.005 & 2.0 & 2.0 & 9.64$\times10^{-9}$ &  Sil-Ow/1.0 &400 & 0.6 & 0.1 & 1.0 & 30.0 & 5.73$\times10^{-5}$ & 217 & 1.68$\times10^{-4}$ & 24.7\\
\enddata
\tablecomments{\\
The acronymns for the dust grain compositions refer to 
`warm' silicates (Sil-Ow) from \citet{ossenkopf1992},
graphite grains (grf-DL) from \citet{draine1984}, amorphous carbon (amC-Hn) from \citet{hanner1988} and silicon carbide (SiC-Pg) from \citet{pegourie1988}.
\tablenotetext{a}{The (input) dust temperature at shell (disk) inner radius}
\tablenotetext{b}{The dust shell's (disk's) optical depth at 0.55\,$\micron$}
\tablenotetext{c}{The minimum dust grain size}
\tablenotetext{d}{The maximum dust grain size}
\tablenotetext{e}{The thickness of the disk/shell}
\tablenotetext{f}{The (inferred) circumstellar (gas+dust) mass using $\delta$ = 200}
\tablenotetext{g}{The (inferred) luminosity}
\tablenotetext{h}{The reduced chi-square value}}
\end{deluxetable*}
\end{rotatetable}

\begin{deluxetable*}{cccc}
\tablenum{5}
\tablecaption{I-band photometry\label{tbl-denisetc}}
\tablehead{
\colhead{Object} & \colhead{Zaritsky (Imag)} & \colhead{DENIS (Imag)} & \colhead{USNO-B 1.0 (Imag)}}
\startdata
\cutinhead{post-RGB stars}
J043919.30-685733.4 & 17.900 & 17.658 & ... \\
J051347.57-704450.5 & 15.683 & 15.699 & ... \\
J051920.18-722522.1 & 16.152 & 16.169 & 15.87 \\
J053930.60-702248.5 & 16.915 & 16.917 & ... \\
J045555.15-712112.3 & ... & ... & ... \\
J045755.05-681649.2 & 17.205 & 17.352 & 17.73 \\
J050257.89-665306.3 & 17.563 & 16.536 & 16.34 \\
J055102.44-685639.1 & 17.821 & 17.866 & 17.16 \\
\cutinhead{post-AGB}
J050632.10-714229.8 & 13.780 & 13.886 & 13.53 \\
J051848.84-700247.0 & 13.852 & 13.845 & ...  \\
J051906.86-694153.9 & 14.473 & ...  & ... \\
J053250.69-713925.8 & 13.803 & 13.877        & 13.60 \\
J045623.21-692749.0 & 13.123 & 13.154        & 12.85 \\
J051418.09-691234.9 & 13.853 & 13.856        & 13.38 \\
J055122.52-695351.4 & 14.009 & 14.077        & 13.64 \\
J052519.48-705410.0 & 14.276 & 14.575        & 14.50 \
\enddata
\end{deluxetable*}

\begin{deluxetable*}{@{\extracolsep{5pt}}cccc}
\tablenum{6}
\tablecaption{Luminosities of the sources\label{lum}}
\tablehead{
\colhead{Object} & \colhead{Best-fit Model} & \multicolumn{2}{c}{KWVW15}  \\
\cline{2-2} \cline{3-4}\\
 \colhead{} & \colhead{} & \colhead{Observed} & \colhead{Photospheric} \\
 \colhead{}  & \colhead{$L_{\odot}$} & \colhead{$L_{\odot}$} & \colhead{$L_{\odot}$}}
\startdata
\cutinhead{post-RGB stars}
J043919.30-685733.4 & 116 & 106 & 159 \\
J051347.57-704450.5 & 776 & 712 & 840 \\
J051920.18-722522.1 & 582 & 496 & 507 \\
J053930.60-702248.5 & 295 & 248 & 294 \\
J045555.15-712112.3 & 465 & 454 & 191 \\
J045755.05-681649.2 & 217 & 190 & 215 \\
J050257.89-665306.3 & 303 & 267 & 177 \\
J055102.44-685639.1 & 621 & 452 & 815 \\
\cutinhead{post-AGB stars}
J050632.10-714229.8 & 5434 & 4910 & 7606\\
J051848.84-700247.0 & 6210 & 4477 & 14112\\
J051906.86-694153.9 & 2018 & 2052 & 4246 \\
J053250.69-713925.8 & 4657 & 4223 & 11056\\
J045623.21-692749.0 & 6598 & 6864 & 7131\\
 J051418.09-691234.9 & 7763 & 6667 & 4703\\
J055122.52-695351.4 & 4116 & 3780 & 3657 \\
J052519.48-705410.0 & 3804 & 3219 & 4943\\
\enddata
\end{deluxetable*}

\clearpage
\renewcommand{\thefigure}{1}
\begin{figure}[h]
	\centering
	\input{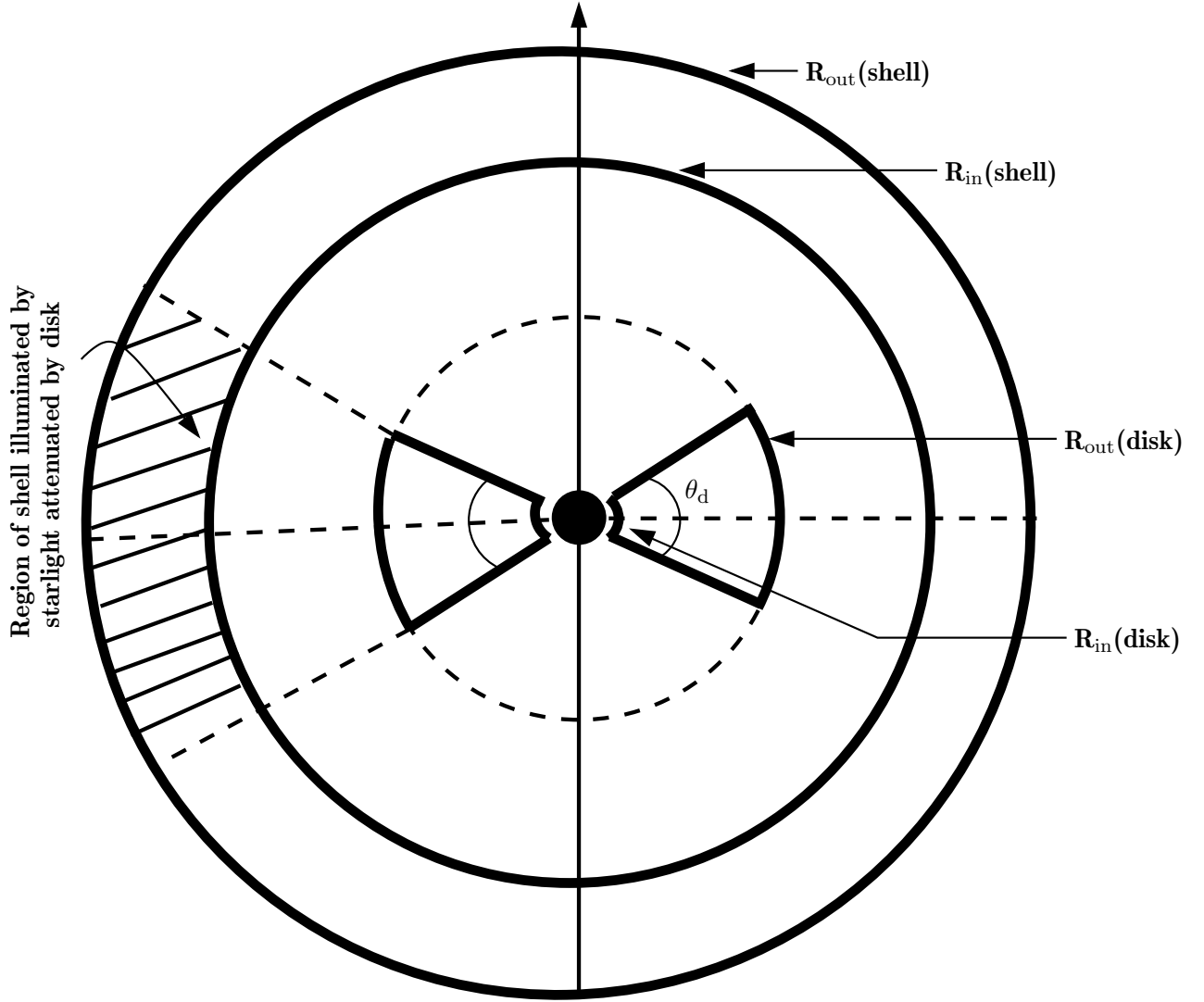}
	\caption{Illustration showing post-RGB dust geometry with an inner disk and an outer shell.  $R_{\mathrm in}$ and $R_{\mathrm out}$
	are the inner and outer radius of the disk (shell). A disk with an opening
	angle $\theta_{\rm d}$ intercepts Sin($\theta_{\rm d}$/2) (referred to as disk-fraction in this paper) of starlight.}
	\label{cartoon}
\end{figure}

\clearpage
\figsetstart
\figsetnum{2}
\figsettitle{ J043919.30-685733.4 (post-RGB shell source)}

\figsetgrpstart
\figsetgrpnum{2.1}
\figsetgrptitle{One-component fits}
\figsetplot{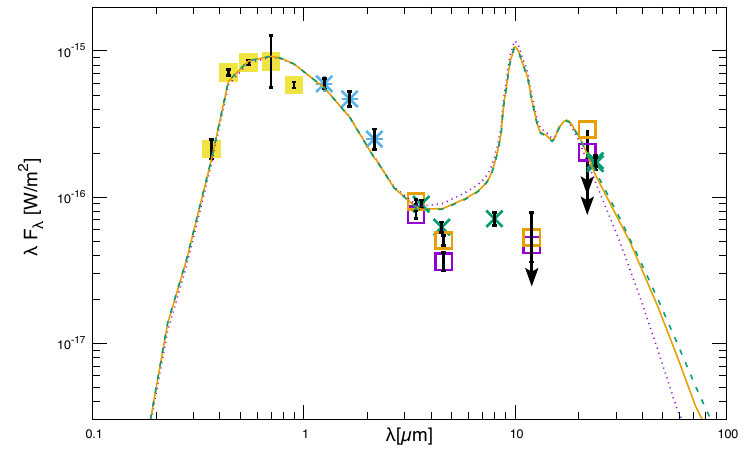}
\figsetgrpnote{One-component model fits to the observed SED (black solid curve) 
are obtained using Sil-Ow grain-type, standard MRN grain size distribution, $T_{\mathrm d}$ (in) = 550 K, 
$\tau$ = 1.5 and varying shell thickness, $Y$ = 8 (purple dotted curve, $\chi^{2}$ = 87.6), 100 (orange solid curve, $\chi^{2}$ = 61.4) and 500 (green dashed curve, $\chi^{2}$ = 68.6).}
\figsetgrpend

\figsetgrpstart
\figsetgrpnum{2.2}
\figsetgrptitle{One-component fits}
\figsetplot{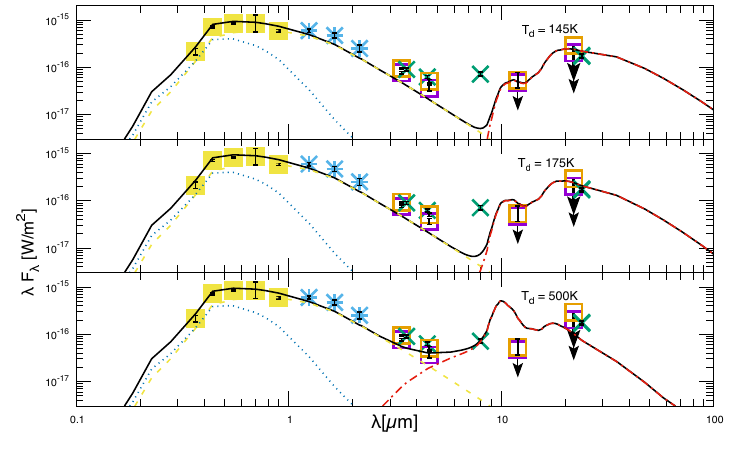}
\figsetgrpnote{One-component model fits to the observed SED
(black solid curve) are obtained for 
varying dust temperatures at the inner shell boundary, $T_{\mathrm d}$ (in) = 145 K ($\chi^{2}$ = 48.3), 175 K ($\chi^{2}$ = 47.8) and 500 K ($\chi^{2}$ = 43.6).
The shell has Sil-Ow grain-type with standard MRN grain size distribution, optical depth, $\tau$ = 0.8 and $Y$ =  100. The attenuated 
stellar flux (yellow dashed curve), scattered light flux (blue dotted curve) and thermal emission (red dash-dotted curve) are also plotted in each case. }
\figsetgrpend

\figsetgrpstart
\figsetgrpnum{2.3}
\figsetgrptitle{Two-component fit}
\figsetplot{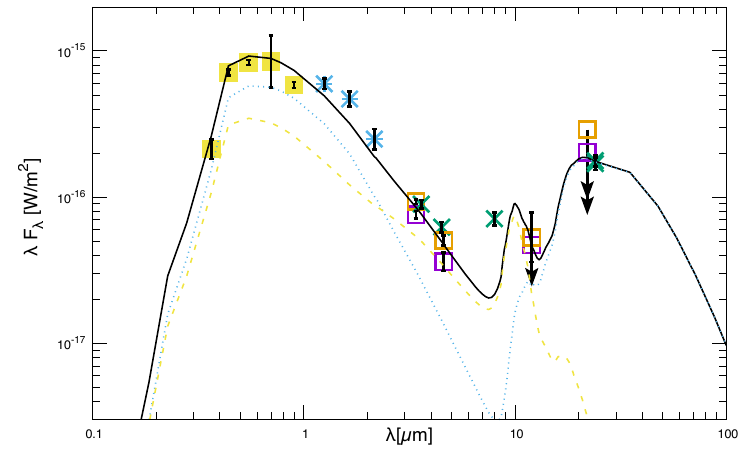}
\figsetgrpnote{Two-component fit, model \#\,1, $\chi^{2}$ = 18.8: the Sil-Ow cold outer shell (blue dotted curve) and
the Sil-Ow warm inner disk (yellow dashed curve) 
components have standard MRN grain size distribution. Using $T_{\mathrm d}$ (in) = 130 K, $Y$ = 20
and $\tau$ = 0.95, a fit was obtained to the far-infrared 
fluxes (i.e., $\gtrsim10$\,$\micron$). For the warm inner disk component, we used 
$T_{\mathrm d}$ (in) = 1000 K, $Y$ = 1.4, $\tau$ = 0.5.  The sum of the two components is shown in black solid curve.}
\figsetgrpend

\figsetgrpstart
\figsetgrpnum{2.4}
\figsetgrptitle{Two-component fit}
\figsetplot{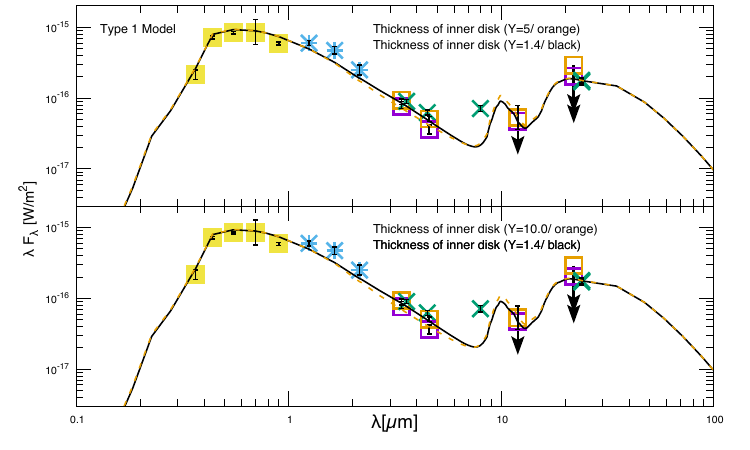}
\figsetgrpnote{Two-component fits (orange dashed curve) to the observed SED for varying thickness 
of the inner disk, $Y$ = 5 ($\chi^{2}$ = 34.4), 10 ($\chi^{2}$ = 39.4) along with  model \#\,1 (black solid curve/$Y$ = 1.4, $\chi^{2}$ = 18.8).}
\figsetgrpend

\figsetgrpstart
\figsetgrpnum{2.5}
\figsetgrptitle{Two-component fit}
\figsetplot{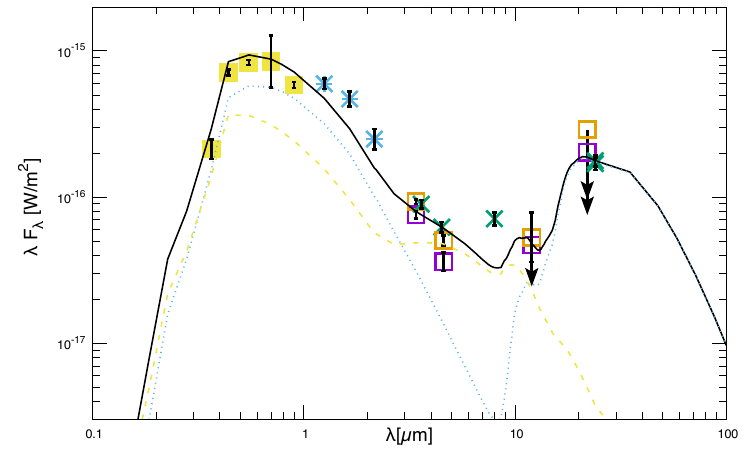}
\figsetgrpnote{Two-component fit, model \#\,2, $\chi^{2}$ = 25.8: the Sil-Ow cold outer shell (blue dotted curve) has standard 
MRN grain size distribution and the Sil-Ow warm inner disk component (yellow dashed curve) has large grains, using 
modified MRN grain size distribution: $a_{\mathrm min}$ = 1 $\micron$ and $a_{\mathrm max}$ = 25 $\micron$. 
Using $T_{\mathrm d}$ (in) = 130 K, $Y$ = 20 and $\tau$ = 0.95, a fit was obtained to the far-infrared 
fluxes (i.e. $\gtrsim10$\,$\micron$). For the warm inner disk component, we used $T_{\mathrm d}$ (in) = 800 K, 
$Y$ = 1.4, $\tau$ = 0.25. The sum of the two components is shown in black solid curve (model \#\,2).}
\figsetgrpend

\figsetgrpstart
\figsetgrpnum{2.6}
\figsetgrptitle{Two-component fit}
\figsetplot{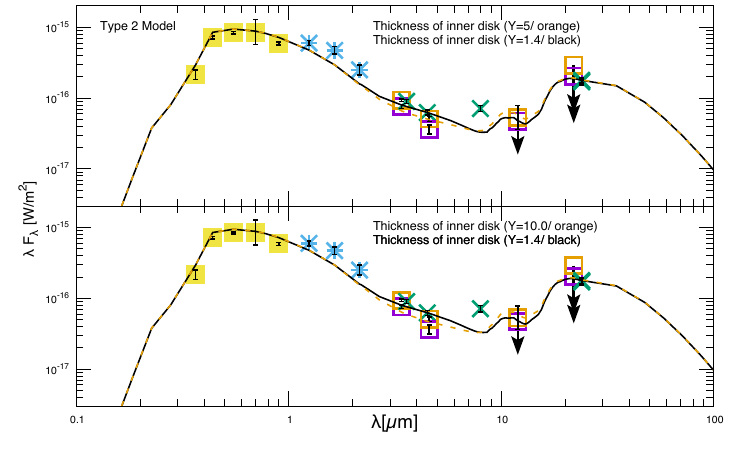}
\figsetgrpnote{Two-component model fits (orange dashed curve) to the observed SED for varying thickness of the 
inner disk, $Y$ = 5 ($\chi^{2}$ = 40.9), 10 ($\chi^{2}$ = 45.8) along with model \#\,2 
(black solid curve/$Y$ = 1.4, $\chi^{2}$ = 25.8).}
\figsetgrpend

\figsetgrpstart
\figsetgrpnum{2.7}
\figsetgrptitle{Two-component fit}
\figsetplot{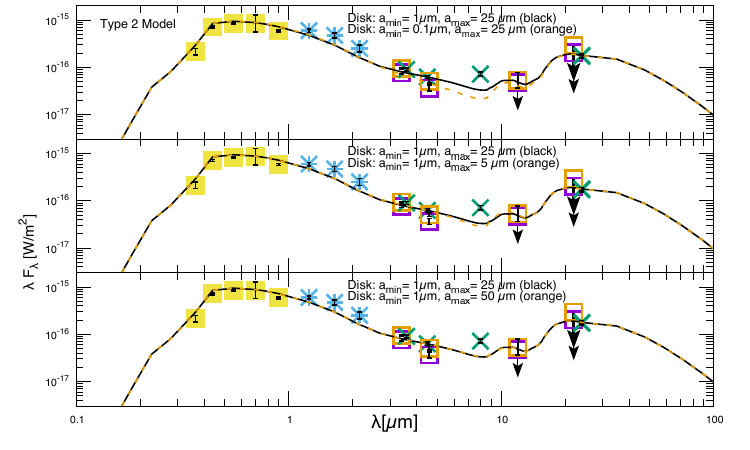}
\figsetgrpnote{In the two-component model (\#\,2; black solid curve), a modified MRN grain size 
distribution: $a_{\mathrm min}$ = 1 $\micron$ and $a_{\mathrm max}$ = 25 $\micron$ in the warm inner disk 
is used to obtain a fit to the near-IR fluxes.  The effects of varying $a_{\mathrm min}$ 
and $a_{\mathrm max}$ on the model SED are shown (orange dashed curves).}
\figsetgrpend

\figsetgrpstart
\figsetgrpnum{2.8}
\figsetgrptitle{Two-component fit}
\figsetplot{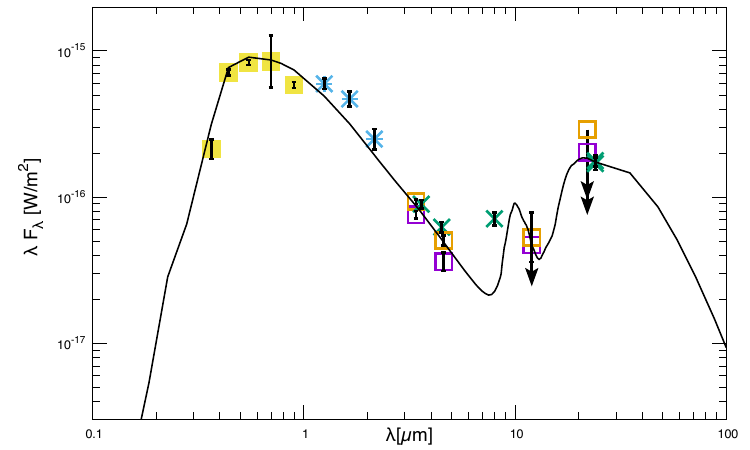}
\figsetgrpnote{Two-component model fit ($\chi^{2}$ = 18.1) to the observed SED, in which the shell was assumed to 
be divided into two parts. The first part (0.65 solid angle) received 
the incident starlight as well as the scattered and thermal emission from the inner disk. The second 
part (0.35 solid angle) received the attenuated, scattered and thermal emission from the inner disk. 
The DUSTY code was run separately for each part of the shell. The outputs were proportionately added to 
obtain the correct model SED corresponding to model \#\,1 (model \#\,1,c).}
\figsetgrpend

\figsetgrpstart
\figsetgrpnum{2.9}
\figsetgrptitle{Two-component model fit}
\figsetplot{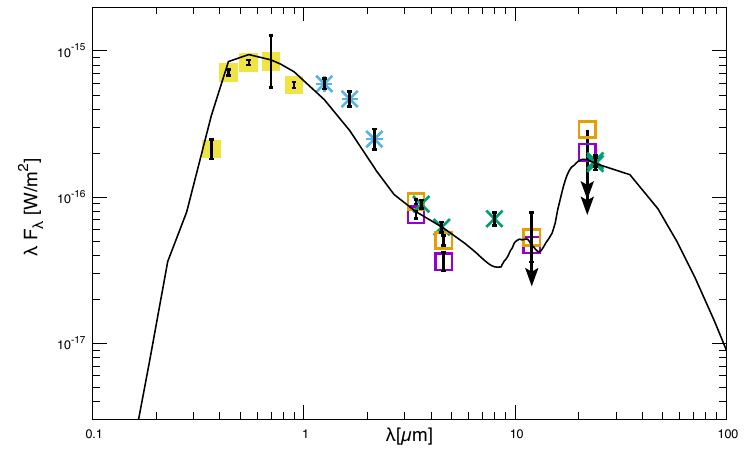}
\figsetgrpnote{Two-component model fit ($\chi^{2}$ = 30.4) to the observed SED, in which the shell was assumed 
to be divided into two parts. The first part (0.65 solid angle) received the 
incident starlight as well as the scattered and thermal emission from the inner disk. The second part 
(0.35 solid angle) received the attenuated, scattered and thermal emission from the inner disk. The DUSTY code 
was run separately for each part of the shell. The outputs were proportionately added to obtain the correct 
model SED corresponding to model \#\,2 (model \#\,2,c). }
\figsetgrpend

\figsetend

\renewcommand{\thefigure}{2}
\begin{figure}
\plotone{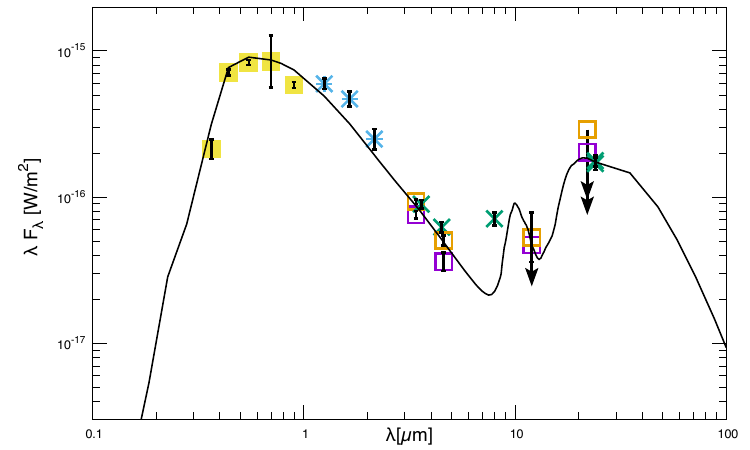}
\caption{The figure shows the adopted two-component best-fit, model \#\,1,c ($\chi^{2}$ = 18.1) for the post-RGB (shell) source, J043919.30-685733.4. 
The observed fluxes are de-reddened for Galactic and LMC reddening. 
U,B,V,R,I (yellow), 2MASS J,H,K (cyan) data are plotted along with WISE (purple) and ALLWISE (orange) photometry 
and data from the SAGE-LMC Survey (green) which covers the IRAC and MIPS bands. The error bars
and upper limits (arrows) are indicated in black.  All model fits obtained for the source including the adopted best-fit are available in the corresponding Figure Set.}
\end{figure}

\figsetstart
\figsetnum{3}
\figsettitle{J051347.57-704450.5 (post-RGB shell source)}

\figsetgrpstart
\figsetgrpnum{3.1}
\figsetgrptitle{One-component fit}
\figsetplot{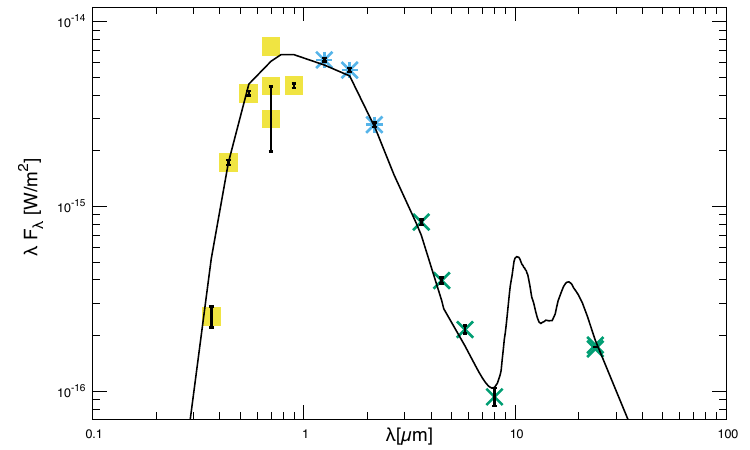}
\figsetgrpnote{Adopted one-component best-fit model ($\chi^{2}$ = 69.2). Fit 
is obtained using $T_{\mathrm d}$ (in) = 250 K, Sil-Ow grain type with modified MRN grain size distribution: $a_{\mathrm min}$ = 0.1 $\micron$ and $a_{\mathrm max}$ =  0.25 $\micron$, $\tau$ = 0.40 
and $Y$ = 3.}
\figsetgrpend

\figsetgrpstart
\figsetgrpnum{3.2}
\figsetgrptitle{One-component fits}
\figsetplot{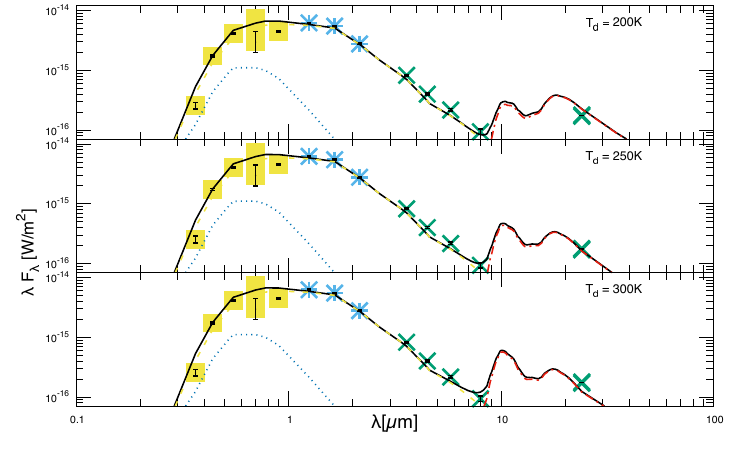}
\figsetgrpnote{One-component model fits to the observed SED for varying
dust temperatures at the inner shell boundary, $T_{\mathrm d}$ (in) = 200 K ($\chi^{2}$ = 74.7), 250 K ($\chi^{2}$ = 71.7) and 300 K ($\chi^{2}$ = 74.2). The attenuated stellar flux (yellow dashed curve), scattered light flux (blue dotted curve) and thermal emission (red dash-dotted curve) are also plotted in each case. We
adopted the fit with $T_{\mathrm d}$ (in) = 250 K.}
\figsetgrpend

\figsetgrpstart
\figsetgrpnum{3.3}
\figsetgrptitle{One-component fits}
\figsetplot{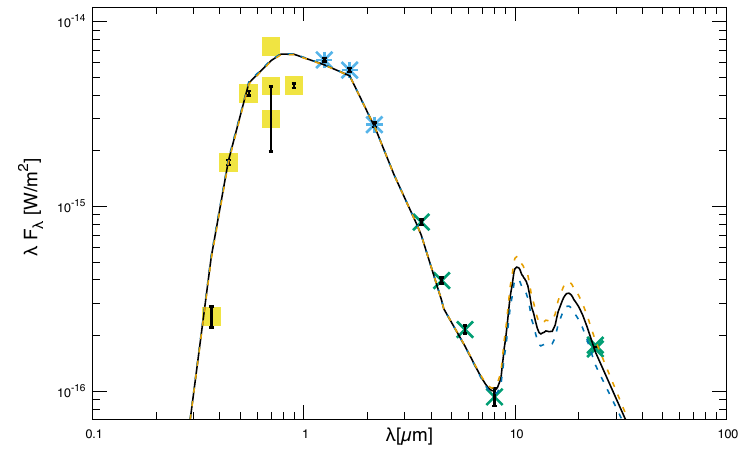}
\figsetgrpnote{One-component model fits to the observed SED for varying shell optical depth
at 0.55 $\micron$: $\tau$ = 0.30/blue dashed curve ($\chi^{2}$ = 75.6) and $\tau$ = 0.40/orange dashed curve ($\chi^{2}$ = 69.2) along with the best-fit model, $\tau$ = 0.35/black solid curve ($\chi^{2}$ = 71.7).}
\figsetgrpend

\figsetgrpstart
\figsetgrpnum{3.4}
\figsetgrptitle{One-component fits}
\figsetplot{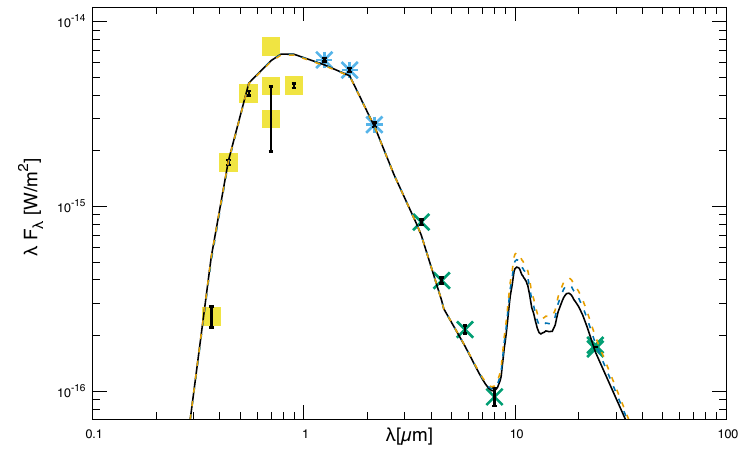}
\figsetgrpnote{One-component model fits to the observed SED 
for varying minimum and maximum grain sizes, blue dashed curve ($\chi^{2}$ = 69.5): $a_{\mathrm min}$ = 0.05 $\micron$, $a_{\mathrm max}$ = 0.25 $\micron$; orange dashed curve ($\chi^{2}$ = 69.3):
$a_{\mathrm min}$ = 0.1 $\micron$, $a_{\mathrm max}$ = 0.35 $\micron$ along with the best-fit 
model, black solid curve ($\chi^{2}$ = 71.7): $a_{\mathrm min}$ = 0.1 $\micron$, $a_{\mathrm max}$ = 0.25 $\micron$.}
\figsetgrpend

\figsetgrpstart
\figsetgrpnum{3.5}
\figsetgrptitle{One-component fits}
\figsetplot{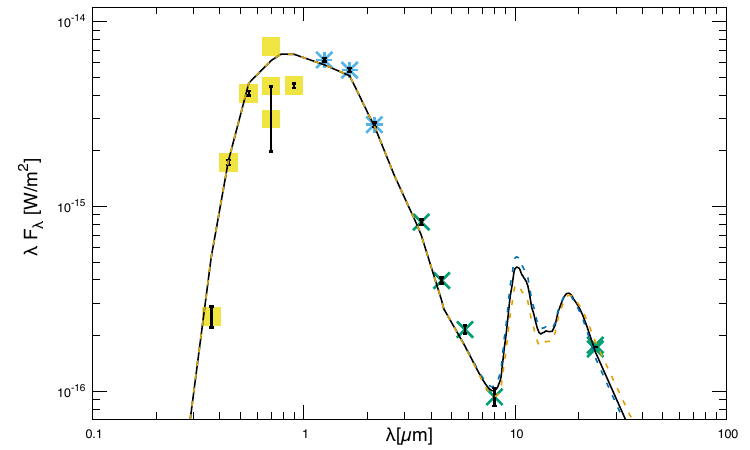}
\figsetgrpnote{One-component model fits to the observed SED for varying shell thickness, $Y$ = 2/blue dashed curve ($\chi^{2}$ = 72.2)and $Y$ = 10/orange dashed curve ($\chi^{2}$ = 71.7) along with the best-fit model, $Y$ = 3/black solid curve ($\chi^{2}$ = 71.7).}
\figsetgrpend

\figsetend

\renewcommand{\thefigure}{3}
\begin{figure}
\plotone{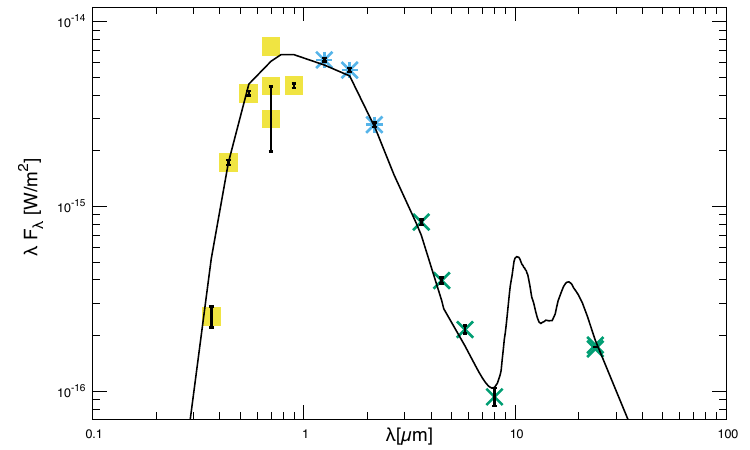}
\caption{The figure shows the adopted one-component best-fit model ($\chi^{2}$ = 69.2) for the post-RGB (shell) source, J051347.57-704450.5. The observed fluxes are de-reddened for Galactic and LMC 
reddening. U,B,V,R,I (yellow), 2MASS J,H,K (cyan) data are plotted along with 
data from the SAGE-LMC Survey (green) which covers the
IRAC and MIPS bands. The error bars are indicated in black. All model fits obtained for the source including the adopted best-fit are available in the corresponding Figure Set.}
\end{figure}

\figsetstart
\figsetnum{4}
\figsettitle{J051920.18-722522.1 (post-RGB shell source)}

\figsetgrpstart
\figsetgrpnum{4.1}
\figsetgrptitle{Two-component fit}
\figsetplot{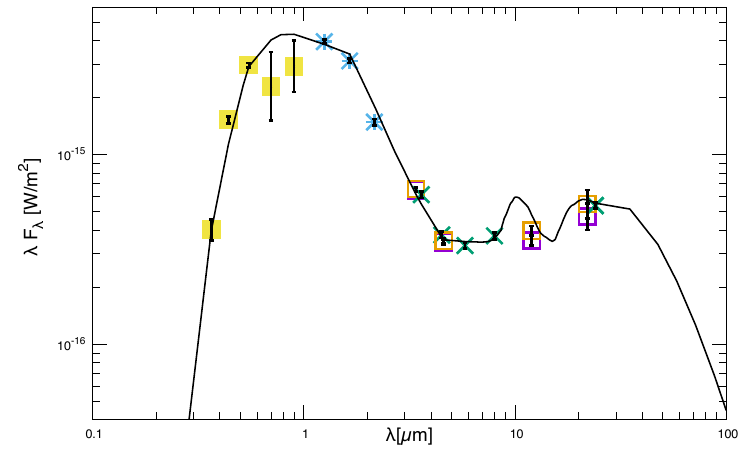}
\figsetgrpnote{Two-component model fit (disk fraction = 0.4; $\chi^{2}$ = 17.5) to the observed SED. The fit (model \#\,2,c) corresponds to the correctly illuminated model \#\,2. The inner disk ($T_{\mathrm d}$ (in) = 500 K) is composed  of Sil-Ow grains with modified
MRN grain size distribution: $a_{\mathrm min}$ = 0.5 $\micron$, $a_{\mathrm max}$ = 20 $\micron$, $\tau$ = 0.4 and $Y$ = 2.0.  The cold outer shell ($T_{\mathrm d}$ (in) = 110 K) is composed of Sil-Ow grains with standard MRN grain size distribution, $\tau$ = 0.65 and $Y$ = 20.}
\figsetgrpend

\figsetgrpstart
\figsetgrpnum{4.2}
\figsetgrptitle{Two-component fit}
\figsetplot{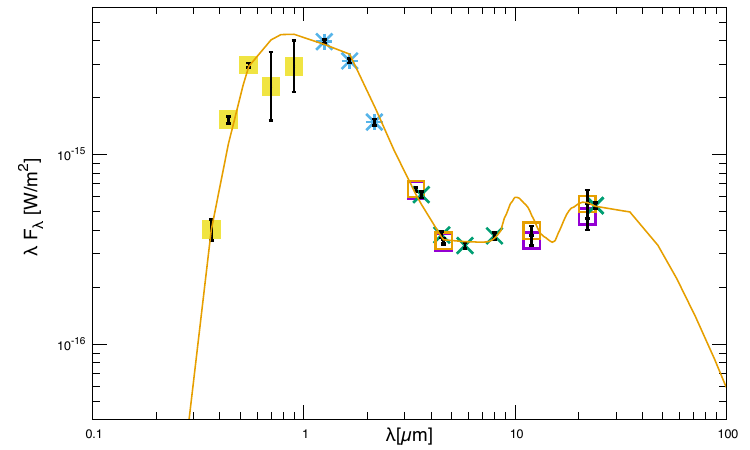}
\figsetgrpnote{Two-component model
fit to the observed SED. The fit (orange solid curve, $\chi^{2}$ = 16.9) is obtained by increasing the thickness of the outer shell by a factor of 10: $Y$ = 200 in model \#\,2,c (\S\,\ref{thickness}).}
\figsetgrpend

\figsetgrpstart
\figsetgrpnum{4.3}
\figsetgrptitle{One-component fits}
\figsetplot{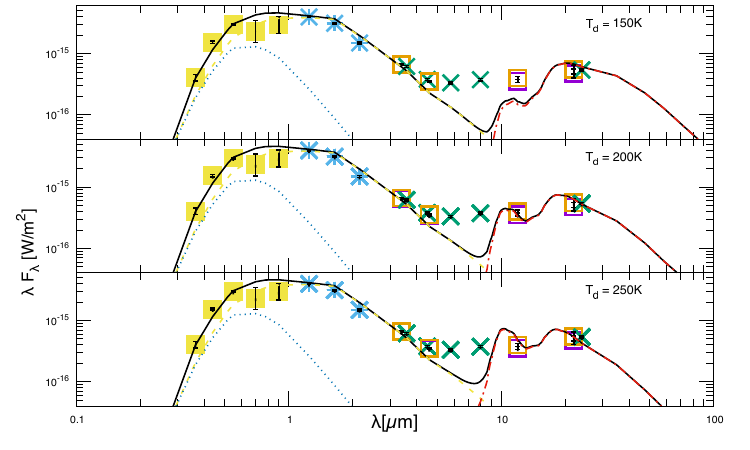}
\figsetgrpnote{One-component model fits to the observed SED for varying 
dust temperatures at the inner shell boundary, $T_{\mathrm d}$ (in) = 150 K ($\chi^{2}$ = 103.2), 200 K (model \#\,1,s, $\chi^{2}$ = 89.1) and 250 K ($\chi^{2}$ = 88.1).
The attenuated stellar flux (yellow dashed curve),
scattered light flux (blue dotted curve) and thermal emission (red dash-dotted curve) are also plotted in each case.}
\figsetgrpend

\figsetgrpstart
\figsetgrpnum{4.4}
\figsetgrptitle{One-component fits}
\figsetplot{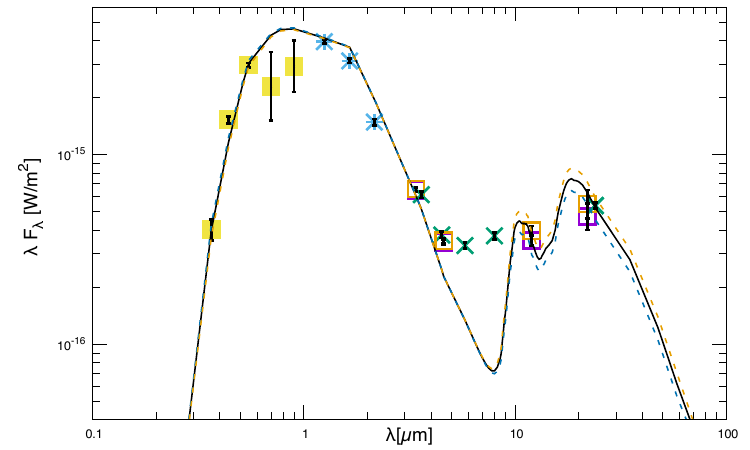}
\figsetgrpnote{One-component model fits to the observed SED for varying shell optical depth
at 0.55 $\micron$: $\tau$ = 0.65/blue dashed curve ($\chi^{2}$ = 93.5)and $\tau$ = 0.85/orange dashed curve ($\chi^{2}$ = 90.6) along with model \#\,1,s,  $\tau$ = 0.75/black solid curve ($\chi^{2}$ = 89.1).}
\figsetgrpend

\figsetgrpstart
\figsetgrpnum{4.5}
\figsetgrptitle{One-component fits}
\figsetplot{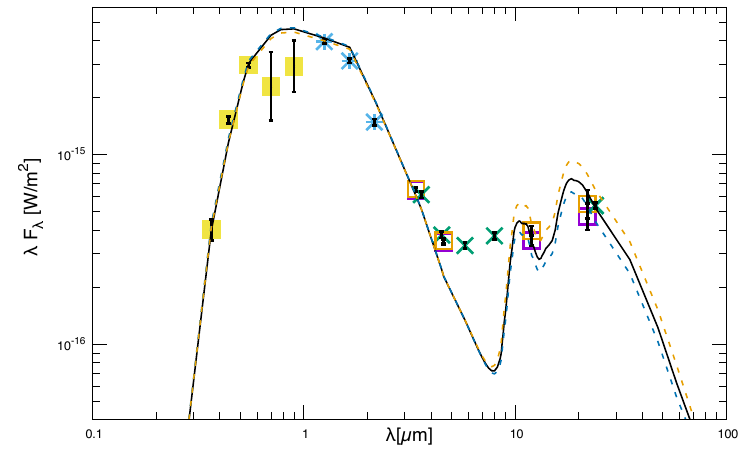}
\figsetgrpnote{One-component model fits to the observed SED for varying 
minimum and maximum grain sizes, blue dashed curve ($\chi^{2}$ = 94.6): $a_{\mathrm min}$ = 0.05 $\micron$, $a_{\mathrm max}$ = 0.25 $\micron$; orange dashed curve ($\chi^{2}$ = 91.9): $a_{\mathrm min}$ = 0.005 $\micron$, $a_{\mathrm max}$ = 0.50 $\micron$ along with
model \#\,1,s/black solid curve ($\chi^{2}$ = 89.1): $a_{\mathrm min}$ = 0.005 $\micron$, $a_{\mathrm max}$ = 0.25 $\micron$.}
\figsetgrpend

\figsetgrpstart
\figsetgrpnum{4.6}
\figsetgrptitle{One-component fits}
\figsetplot{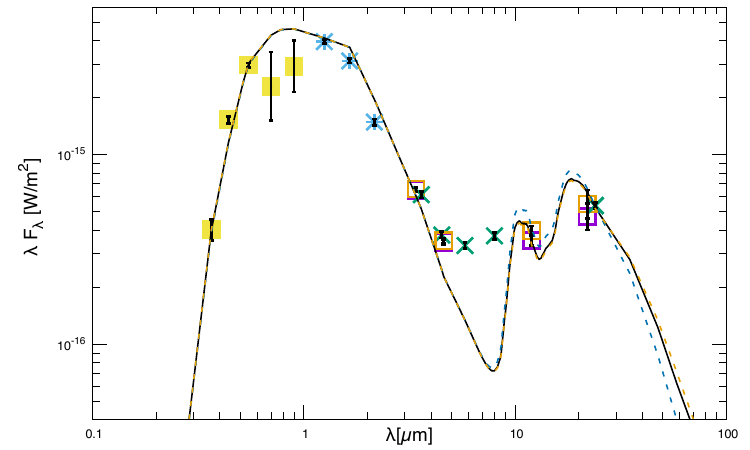}
\figsetgrpnote{One-component model fits to the observed SED for varying shell thickness, $Y$ = 5/blue dashed curve ($\chi^{2}$ = 88.2)and $Y$ = 40/orange dashed curve ($\chi^{2}$ = 89.5) along with model \#\,1,s, $Y$ = 20/black solid curve ($\chi^{2}$ = 89.1).}
\figsetgrpend

\figsetgrpstart
\figsetgrpnum{4.7}
\figsetgrptitle{Two-component fits}
\figsetplot{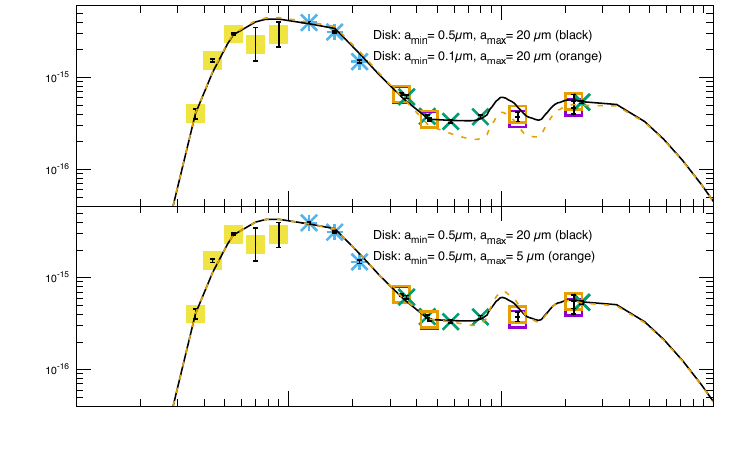}
\figsetgrpnote{Two-component model fits to the observed SED 
for varying minimum and maximum grain sizes in the inner disk: a$_{\min}$ = 0.1 $\micron$, $a_{\mathrm max}$ = 20 $\micron$ ($\chi^{2}$ = 53.2); a$_{\min}$ = 0.5 $\micron$, $a_{\mathrm max}$ = 5 $\micron$ ($\chi^{2}$ = 21.1) along with model \#\,2, black solid curve ($\chi^{2}$ = 17.3): a$_{\min}$ = 0.5 $\micron$, $a_{\mathrm max}$ = 20 $\micron$.}
\figsetgrpend

\figsetend

\renewcommand{\thefigure}{4}
\begin{figure}
\plotone{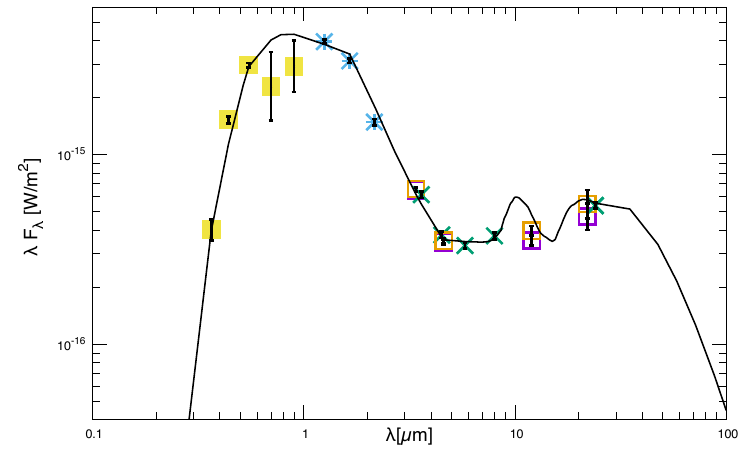}
\caption{The figure shows the adopted two-component best-fit model (disk fraction = 0.4; $\chi^{2}$ = 17.5) for the post-RGB (shell) source, J051920.18-722522.1. The fit (model \#\,2,c) corresponds to the correctly illuminated model \#\,2.
The observed fluxes are de-reddened for Galactic and LMC 
reddening. U,B,V,R,I (yellow), 2MASS J,H,K (cyan) data are plotted along with WISE (purple) and ALLWISE
photometry (orange) and 
data from the SAGE-LMC Survey (green) which covers the
IRAC and MIPS bands. The error bars are indicated in black. All model fits obtained for the source including the adopted best-fit are available in the corresponding Figure Set.}
\end{figure}

\figsetstart
\figsetnum{5}
\figsettitle{J053930.60-702248.5 (post-RGB shell source)}

\figsetgrpstart
\figsetgrpnum{5.1}
\figsetgrptitle{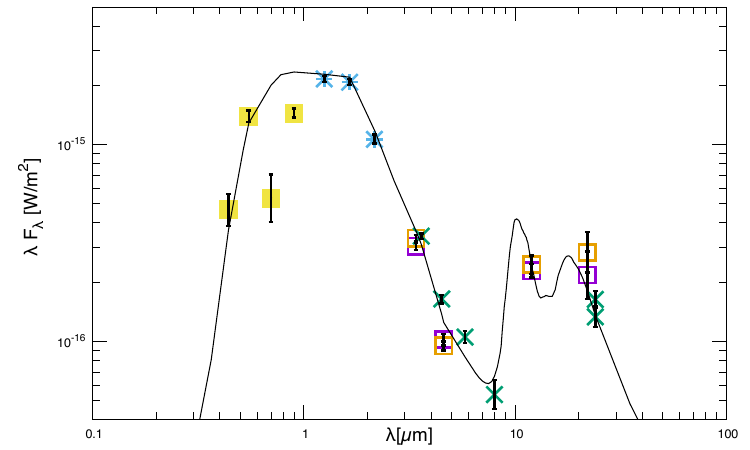}
\figsetplot{One-component fit}
\figsetgrpnote{One-component model fit ($\chi^{2}$ = 30.4) to the observed SED. The fit was 
obtained using $T_{\mathrm d}$ (in) = 300 K, Sil-Ow grain type, standard MRN grain size distribution, $\tau$ = 0.7 and $Y$ = 10.}
\figsetgrpend

\figsetgrpstart
\figsetgrpnum{5.2}
\figsetgrptitle{One-component fits}
\figsetplot{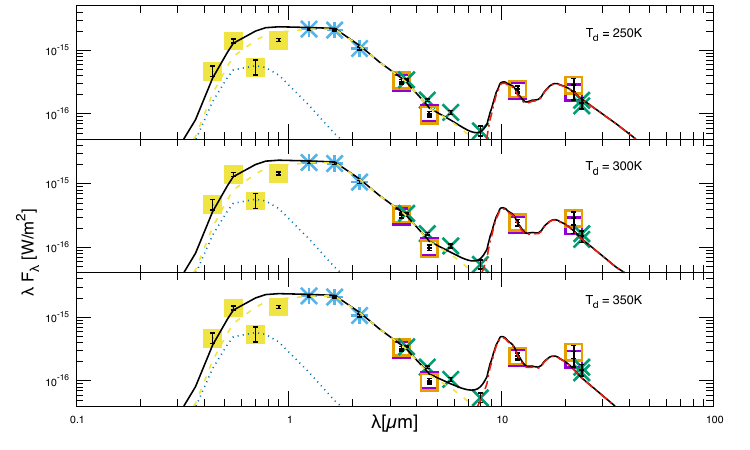}
\figsetgrpnote{One-component model fits to the observed SED for varying
dust temperatures at the inner shell boundary, $T_{\mathrm d}$ (in) = 250 K ($\chi^{2}$ = 30.8), 300 K ($\chi^{2}$ = 30.4) and 350 K ($\chi^{2}$ = 32.0). The attenuated stellar flux (yellow dashed curve), scattered light flux (blue dotted curve) and thermal emission (red dash-dotted curve) are also plotted in each case.
We adopted the fit with $T_{\mathrm d}$ (in) = 300 K.}
\figsetgrpend

\figsetgrpstart
\figsetgrpnum{5.3}
\figsetgrptitle{One-component fits}
\figsetplot{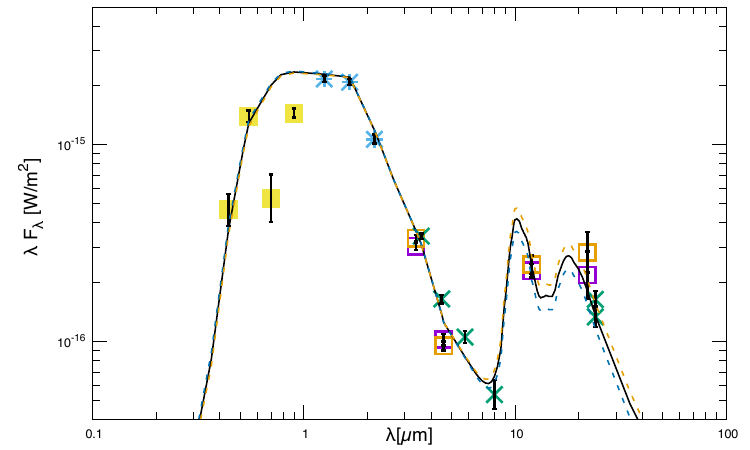}
\figsetgrpnote{One-component model fits to the observed SED for varying shell optical depth
at 0.55 $\micron$: $\tau$ = 0.60/blue dashed curve ($\chi^{2}$ = 32.7) and $\tau$ = 0.80/orange dashed curve ($\chi^{2}$ = 29.2) along with the  
best-fit model, $\tau$ = 0.70/black solid curve ($\chi^{2}$ = 30.4).}
\figsetgrpend

\figsetgrpstart
\figsetgrpnum{5.4}
\figsetgrptitle{One-component fits}
\figsetplot{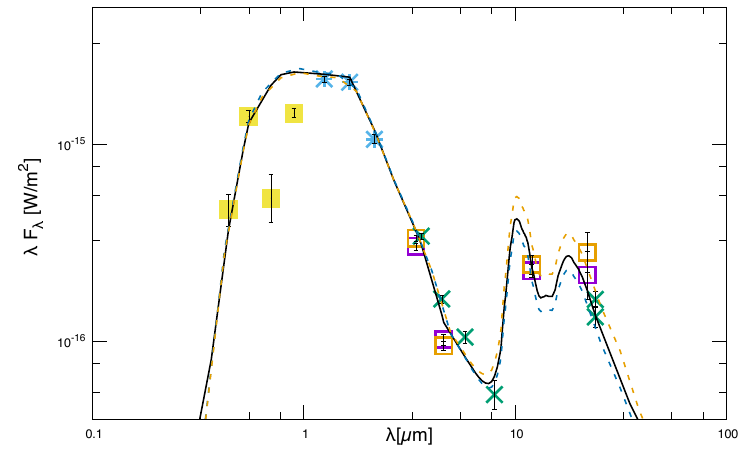}
\figsetgrpnote{One-component model fits to the observed SED for varying 
minimum and maximum grain sizes, blue dashed curve ($\chi^{2}$ = 32.9): $a_{\mathrm min}$ = 0.05 $\micron$, $a_{\mathrm max}$ = 0.25 $\micron$; orange dashed curve ($\chi^{2}$ = 27.8): $a_{\mathrm min}$ = 0.005 $\micron$, $a_{\mathrm max}$ = 0.50 $\micron$ along with
model \#\,1,s/black solid curve ($\chi^{2}$ = 30.4): $a_{\mathrm min}$ = 0.005 $\micron$, $a_{\mathrm max}$ = 0.25 $\micron$.}
\figsetgrpend

\figsetgrpstart
\figsetgrpnum{5.5}
\figsetgrptitle{One-component fits}
\figsetplot{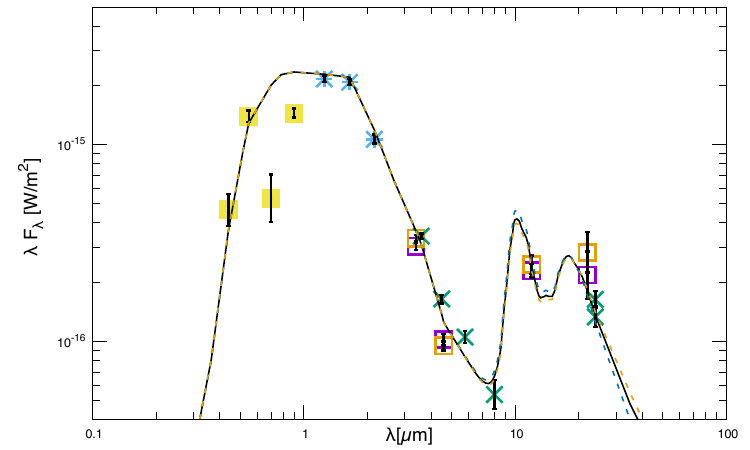}
\figsetgrpnote{One-component model fits to the observed SED for varying shell thickness, $Y$ = 5/blue dashed curve ($\chi^{2}$ = 30.7) and 
$Y$ = 20/orange dashed curve $\chi^{2}$ = 30.4) along with the best-fit model, $Y$ = 10/black solid curve ($\chi^{2}$ = 30.4).}
\figsetgrpend

\figsetend

\renewcommand{\thefigure}{5}
\begin{figure}
\plotone{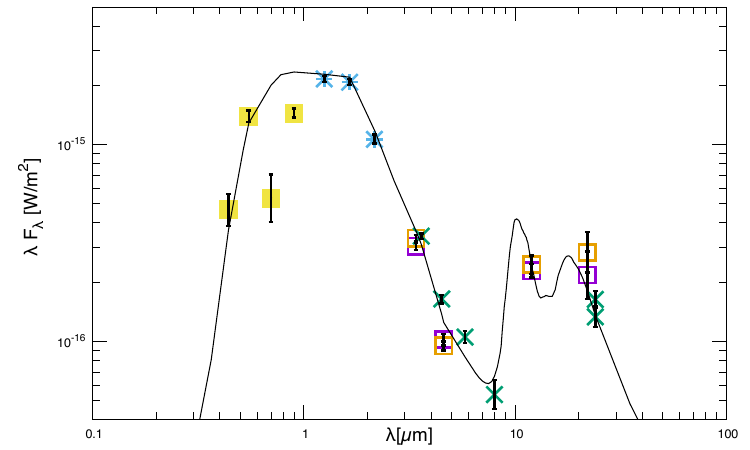}
\caption{The figure shows the adopted one-component model fit ($\chi^{2}$ = 30.4) to the observed SED of the post-RGB (shell) source, J053930.60-702248.5. The observed fluxes are de-reddened for Galactic and LMC 
reddening. B,V,R,I (yellow), 2MASS J,H,K (cyan) data are plotted along with WISE (purple) and ALLWISE (orange)
photometry and data from the SAGE-LMC Survey (green) which covers the
IRAC and MIPS bands. The error bars are indicated in black. All model fits obtained for the source including the adopted best-fit are available in the corresponding Figure Set.}
\end{figure}

\figsetstart
\figsetnum{6}
\figsettitle{J045555.15-712112.3 (post-RGB disk source)}

\figsetgrpstart
\figsetgrpnum{6.1}
\figsetgrptitle{One-component fit}
\figsetplot{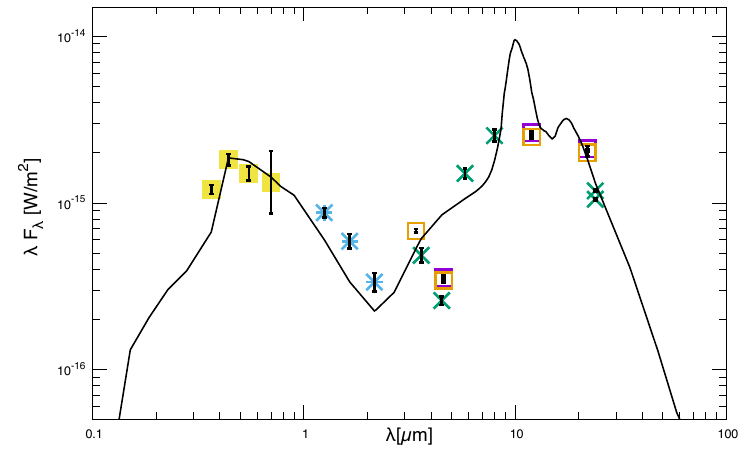}
\figsetgrpnote{One-component model fit ($\chi^{2}$ = 619.8) to the observed SED. The fit 
corresponds to $T_{\mathrm d}$ (in) = 550 K, Sil-Ow grain type with modified
MRN grain size distribution: $a_{\mathrm min}$ = 0.005 $\micron$, $a_{\mathrm max}$ = 1 $\micron$, $\tau$ = 2.3 and $Y$ =  20 (model \#\,1,s).}
\figsetgrpend

\figsetgrpstart
\figsetgrpnum{6.2}
\figsetgrptitle{Two-component fit}
\figsetplot{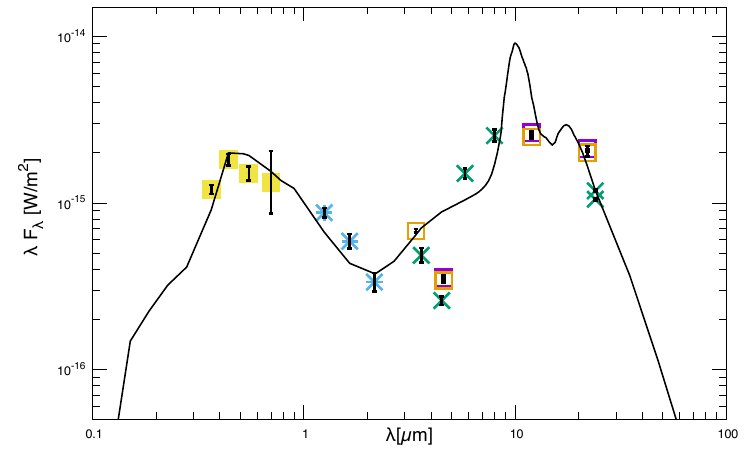}
\figsetgrpnote{Two-component model fit (disk fraction = 0.1, $\chi^{2}$ = 1125.5) to the observed SED. The fit (model \#\,2,c)
corresponds to the correctly illuminated model \#\,2. The inner disk/$T_{\mathrm d}$ (in) = 1200 K is composed of Sil-Ow grains with modified MRN grain size distribution: 
$a_{\mathrm min}$ = 0.005 $\micron$, 
$a_{\mathrm max}$ = 0.1 $\micron$, $\tau$ = 0.5 and $Y$ = 10. The cold outer shell/$T_{\mathrm d}$ (in) = 550 K has Sil-Ow grains with modified MRN grain size distribution: $a_{\mathrm min}$ = 0.005 $\micron$, $a_{\mathrm max}$ = 1.0 $\micron$, $\tau$ = 2.1 and $Y$ = 20.}
\figsetgrpend

\figsetgrpstart
\figsetgrpnum{6.3}
\figsetgrptitle{Two-component fit}
\figsetplot{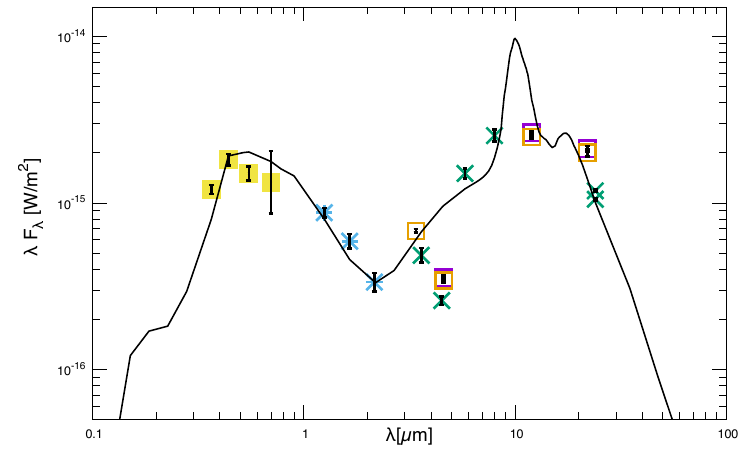}
\figsetgrpnote{Two-component model fit (disk fraction = 0.1, $\chi^{2}$ = 1318.3) to the observed SED. 
The fit (model \#\,3,c) corresponds to the correctly illuminated model \#\,3.
The inner disk/$T_{\mathrm d}$ (in) = 800 K is composed of grf-DL grains with standard MRN grain size distribution, 
$\tau$ = 0.7 and $Y$ = 5. The cold outer shell/$T_{\mathrm d}$ (in) = 500 K, 
has mixed grain chemistry, Sil-Ow/0.8 and grf-DL/0.2, standard MRN grain size distribution, $\tau$ = 1.8 and $Y$ = 2. }
\figsetgrpend

\figsetgrpstart
\figsetgrpnum{6.4}
\figsetgrptitle{Two-component fit}
\figsetplot{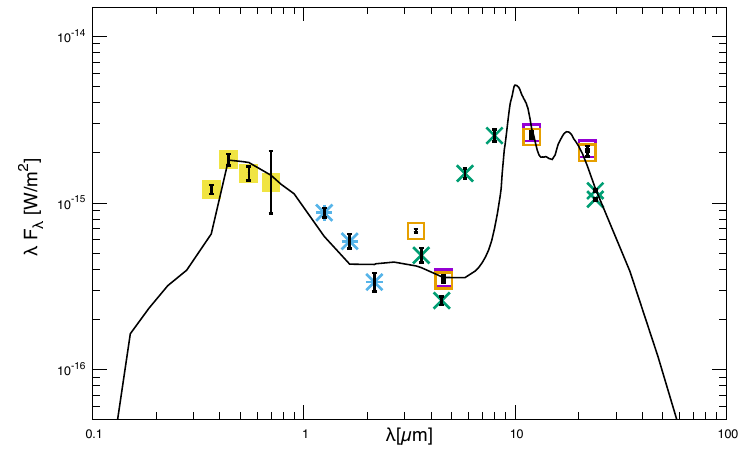}
\figsetgrpnote{Two-component model fit (model \#\,4, disk fraction = 0.4, $\chi^{2}$ = 116.6) to the observed SED.
The inner disk/$T_{\mathrm d}$ (in) = 1000 K is composed of Sil-Ow grains with standard MRN grain size distribution, 
$\tau$ = 1.0 and $Y$ = 10. The cold outer shell/$T_{\mathrm d}$ (in) = 300 K, 
has Sil-Ow grains, standard MRN grain size distribution, $\tau$ = 2.5 and $Y$ = 2.0.}
\figsetgrpend

\figsetgrpstart
\figsetgrpnum{6.5}
\figsetgrptitle{One-component fits}
\figsetplot{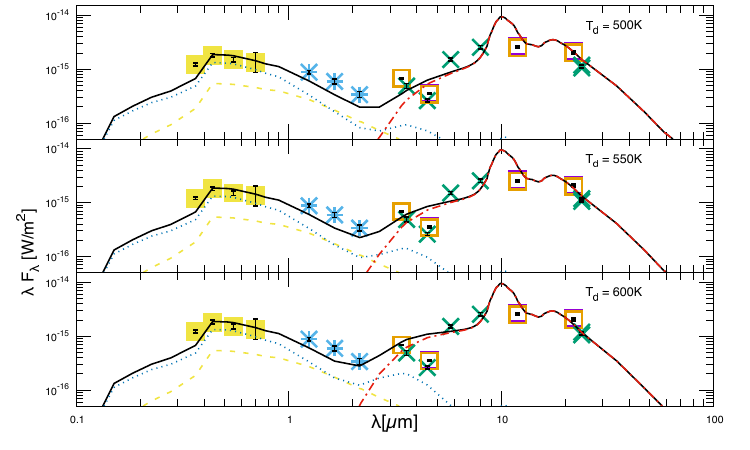}
\figsetgrpnote{One-component model fits to the observed SED for varying
dust temperatures at the inner shell boundary, $T_{\mathrm d}$ (in) = 500 K ($\chi^{2}$ = 423.8), 550 K (model \#\,1,s, $\chi^{2}$ = 619.8) and 600 K ($\chi^{2}$ = 1033.9).
The attenuated stellar flux (yellow dashed curve),
scattered light flux (blue dotted curve) and thermal emission (red dash-dotted curve) are also plotted in each case.}
\figsetgrpend

\figsetgrpstart
\figsetgrpnum{6.6}
\figsetgrptitle{One-component fits}
\figsetplot{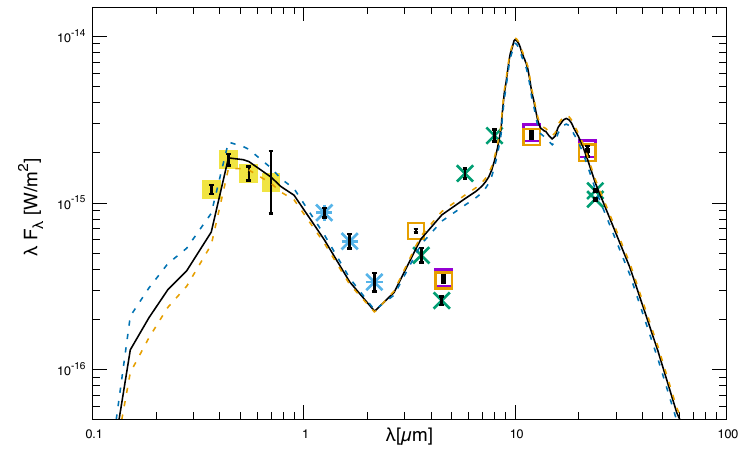}
\figsetgrpnote{One-component model fits to the observed SED for varying shell optical depth
at 0.55 $\micron$: $\tau$ = 2.0 ($\chi^{2}$ = 463.8)/blue dashed curve and $\tau$ = 2.5 ($\chi^{2}$ = 727.4)/orange dashed curve along with  model \#\,1,s, $\tau$ = 2.3 ($\chi^{2}$ = 619.8)/black solid curve.}
\figsetgrpend

\figsetgrpstart
\figsetgrpnum{6.7}
\figsetgrptitle{One-component fits}
\figsetplot{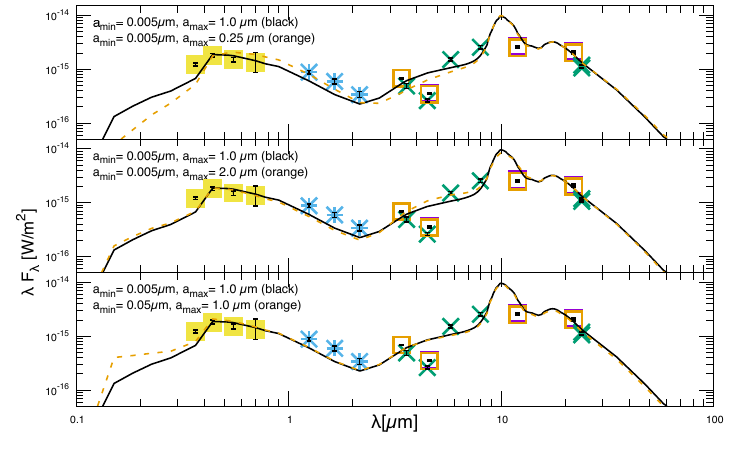}
\figsetgrpnote{One-component model fits (orange) to the observed SED for varying minimum and maximum grain sizes: $a_{\mathrm min}$ = 0.005 $\micron$, $a_{\mathrm max}$ = 0.25 $\micron$  ($\chi^{2}$ = 341.0); 
$a_{\mathrm min}$ = 0.005 $\micron$, $a_{\mathrm max}$ = 2.0 $\micron$ ($\chi^{2}$ = 1132.3);
$a_{\mathrm min}$ = 0.05 $\micron$, $a_{\mathrm max}$ = 1.0 $\micron$ ($\chi^{2}$ = 554.4) along with
model \#\,1,s/black solid curve, $a_{\mathrm min}$ = 0.005 $\micron$, $a_{\mathrm max}$ = 1.0 $\micron$
($\chi^{2}$ = 619.8).}
\figsetgrpend

\figsetgrpstart
\figsetgrpnum{6.8}
\figsetgrptitle{One-component fits}
\figsetplot{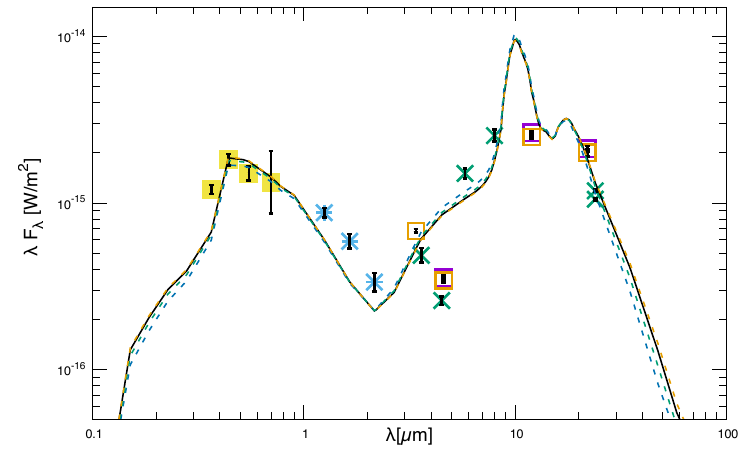}
\figsetgrpnote{One-component model fits to the observed SED for varying shell thickness, $Y$ = 5/blue dashed curve ($\chi^{2}$ = 802.8), $Y$ = 10/green dashed curve ($\chi^{2}$ = 676.7) and $Y$ = 30/orange dashed curve ($\chi^{2}$ = 600.8) along with model \#\,1,s, $Y$ = 20/black solid curve ($\chi^{2}$ = 619.8).}
\label{J045555_thick}
\figsetgrpend

\figsetend

\renewcommand{\thefigure}{6}
\begin{figure}
\plotone{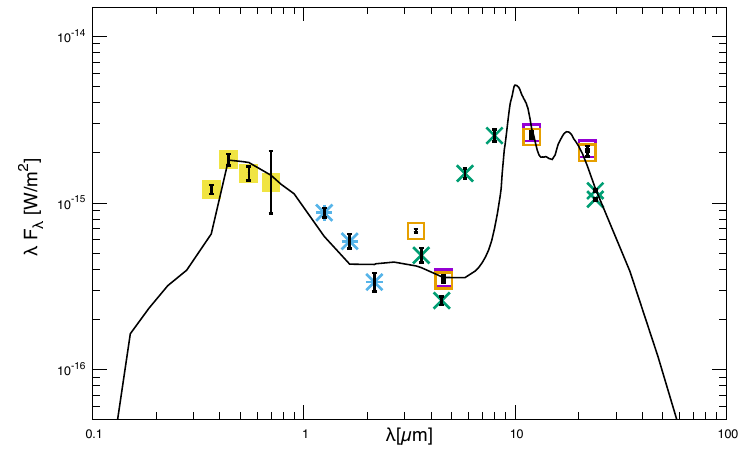}
\caption{The figure shows the adopted two-component model fit (model \#\,4, disk fraction = 0.4, $\chi^{2}$ = 116.6) to the observed SED of the 
post-RGB (disk) source, J045555.15-712112.3. The observed fluxes are de-reddened for Galactic and LMC 
reddening. U,B,V,R (yellow), 2MASS J,H,K (cyan) data are plotted along with WISE (purple) and ALLWISE (orange)
photometry and data from the SAGE-LMC Survey (green) which covers the IRAC and MIPS bands. The error bars are 
indicated in black. All model fits obtained for the source including the adopted best-fit are available in the corresponding Figure Set.}
\end{figure}

\figsetstart
\figsetnum{7}
\figsettitle{J045755.05-681649.2 (post-RGB disk source)}

\figsetgrpstart
\figsetgrpnum{7.1}
\figsetgrptitle{One-component fit}
\figsetplot{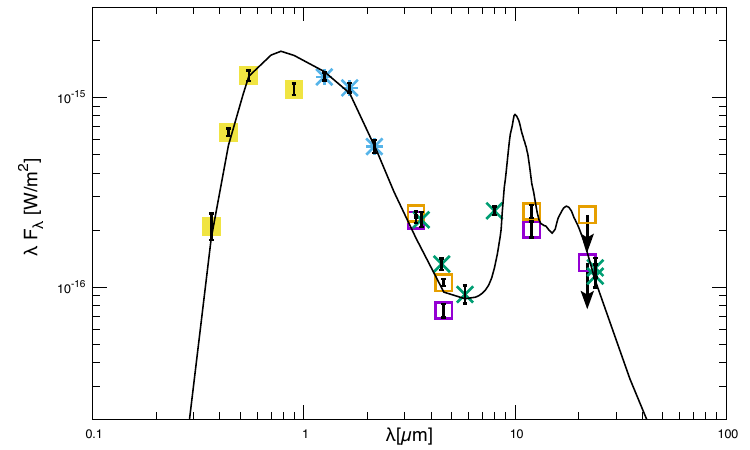}
\figsetgrpnote{One-component model fit (model \#\,1,s, $\chi^{2}$ = 25.2) to the observed SED. The fit is obtained using $T_{\mathrm d}$ (in) = 500 K, Sil-Ow grain type, standard MRN grain size distribution, $\tau$ = 1.0 and $Y$ =  20.}
\figsetgrpend

\figsetgrpstart
\figsetgrpnum{7.2}
\figsetgrptitle{Two-component fit}
\figsetplot{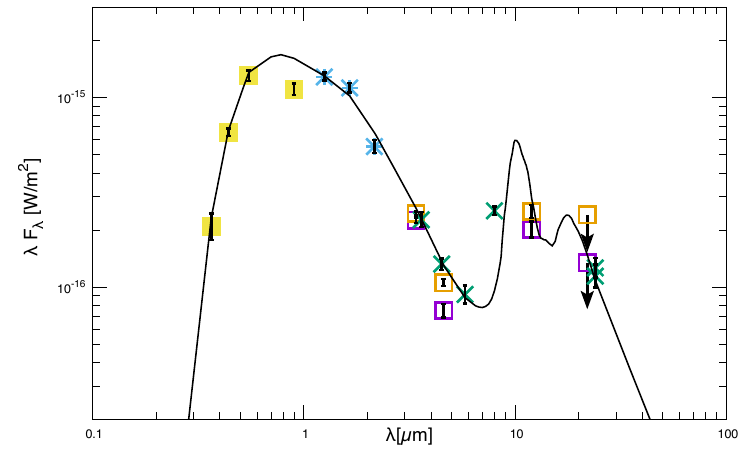}
\figsetgrpnote{Two-component model fit (disk fraction = 0.4, $\chi^{2}$ = 24.7) to the observed SED. The fit (model \#\,2,c) corresponds to the correctly illuminated model \#\,2. 
The warm inner disk/$T_{\mathrm d}$ (in) = 1300 K has Sil-Ow grain type with modified MRN grain size distribution: $a_{\mathrm min}$ = 0.005 $\micron$ and $a_{\mathrm max}$ = 2.0 $\micron$, $\tau$ = 0.5 and $Y$ = 2. The cold outer shell/$T_{\mathrm d}$ (in) = 400 K has Sil-Ow grain type with modified MRN grain size distribution: $a_{\mathrm min}$ = 0.1 $\micron$ and $a_{\mathrm max}$ = 1.0 $\micron$, $\tau$ = 0.9 and $Y$ = 30.} 
\figsetgrpend

\figsetgrpstart
\figsetgrpnum{7.3}
\figsetgrptitle{Comparing fits}
\figsetplot{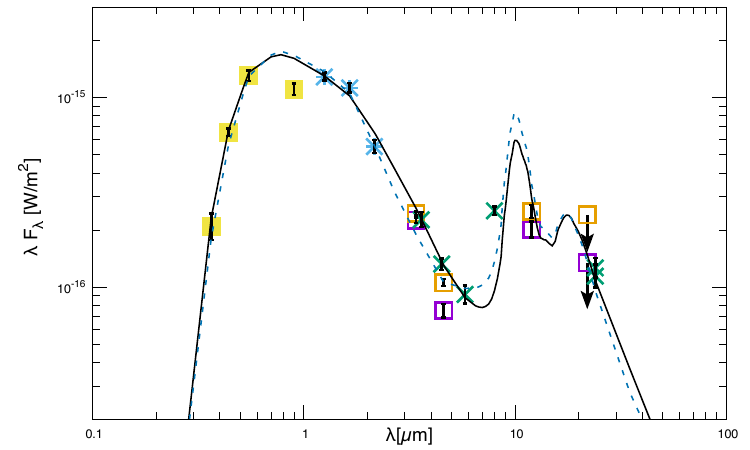}
\figsetgrpnote{Comparing the
one-component model fit (blue dashed curve): $T_{\mathrm d}$ (in) = 550 K, Sil-Ow grain type, standard MRN grain size distribution, $\tau$ = 1.0 and $Y$ =  20, $\chi^{2}$ = 20.7, Table\,\ref{mod-tbl} with the two-component model \#\,2,c (black solid curve), $\chi^{2}$ = 24.7.} 
\label{J045755_comparison}
\figsetgrpend

\figsetgrpstart
\figsetgrpnum{7.4}
\figsetgrptitle{}
\figsetplot{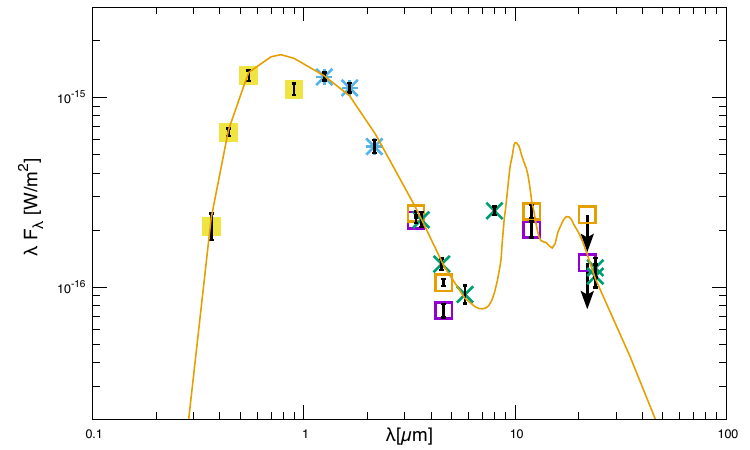}
\figsetgrpnote{Two-component model fit to the observed SED. The fit (orange solid curve, $\chi^{2}$ = 23.9)
is obtained by increasing the thickness of the outer shell
by a factor of 10: $Y$ = 300 in model \#\,2,c (\S\,\ref{thickness}).}
\figsetgrpend

\figsetgrpstart
\figsetgrpnum{7.5}
\figsetgrptitle{One-component fits}
\figsetplot{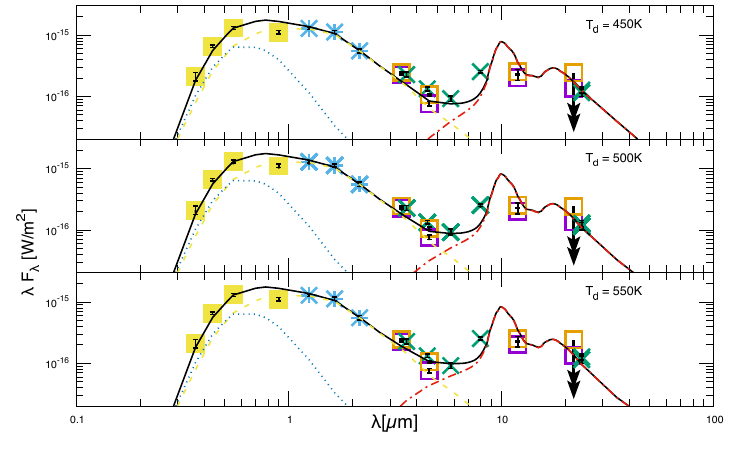}
\figsetgrpnote{One-component model fits to the observed SED for varying
dust temperatures at the inner shell boundary, $T_{\mathrm d}$ (in) = 450 K ($\chi^{2}$ = 31.1), 500 K ($\chi^{2}$ = 25.2) and 550 K ($\chi^{2}$ = 20.7).
The attenuated stellar flux (yellow dashed curve),
scattered light flux (blue dotted curve) and thermal emission (red dash-dotted curve) are also plotted in each case. We
adopted the fit with $T_{\mathrm d}$ (in) = 500 K, model \#\,1,s.}
\label{J045755_temp}
\figsetgrpend

\figsetgrpstart
\figsetgrpnum{7.6}
\figsetgrptitle{One-component fits}
\figsetplot{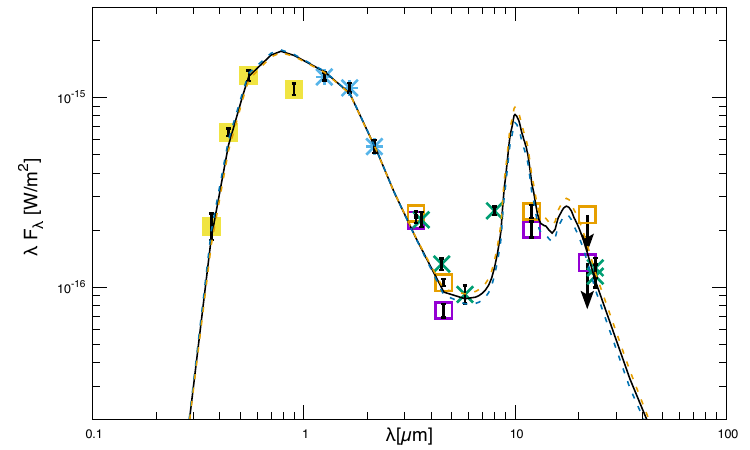}
\figsetgrpnote{One-component model fits to the observed SED for varying shell optical depth
at 0.55 $\micron$: $\tau$ = 0.9/blue dashed curve ($\chi^{2}$ = 25.1) and $\tau$ = 1.1/orange dashed curve ($\chi^{2}$ = 26.9) along with 
model \#\,1,s, $\tau$ = 1.0/black solid curve ($\chi^{2}$ = 25.2).}
\figsetgrpend

\figsetgrpstart
\figsetgrpnum{7.7}
\figsetgrptitle{One-component fit}
\figsetplot{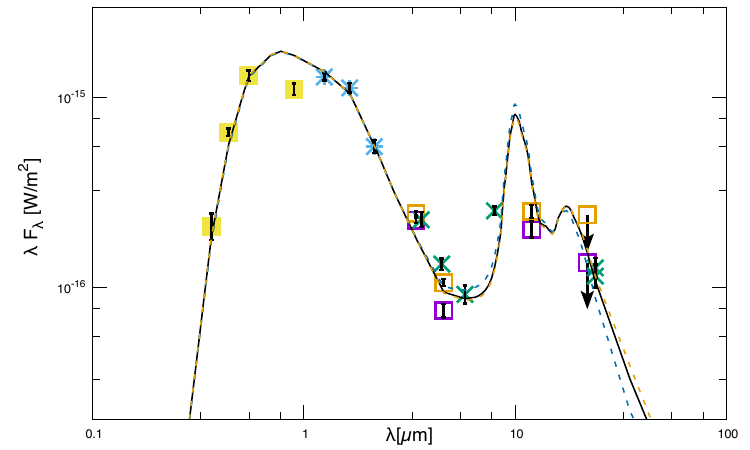}
\figsetgrpnote{One-component model fits to the observed SED of the post-RGB (disk) source, J045755.0 5-681649.2 for varying shell thickness, $Y$ = 5/blue dashed curve ($\chi^{2}$ = 26.7) and $Y$ = 50/orange dashed curve ($\chi^{2}$ = 25.1) along with model \#\,1,s, $Y$ = 20/black solid curve ($\chi^{2}$ = 25.2).}
\figsetgrpend

\figsetend

\renewcommand{\thefigure}{7}
\begin{figure}
\plotone{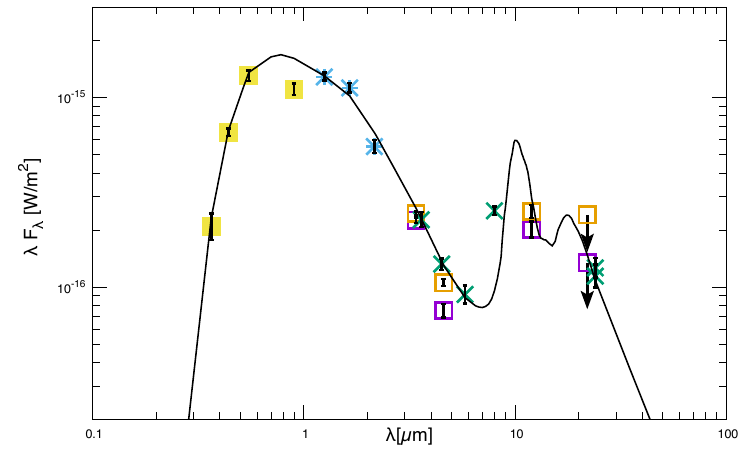}
\caption{The figure shows the adopted two-component model fit (model \#\,2,c, disk fraction = 0.4, $\chi^{2}$ = 24.7) to the observed SED of the post-RGB (disk) source, J045755.05-681649.2. The observed fluxes are de-reddened for Galactic and LMC 
reddening. U,B,V,I (yellow), 2MASS J,H,K (cyan) data are plotted along with WISE (purple) and ALLWISE (orange)
photometry and data from the SAGE-LMC Survey (green) which covers the
IRAC and MIPS bands. The error bars are indicated in black. All model fits obtained for the source including the adopted best-fit are available in the corresponding Figure Set.} 
\end{figure}

\figsetstart
\figsetnum{8}
\figsettitle{J050257.89-665306.3 (post-RGB disk source)}

\figsetgrpstart
\figsetgrpnum{8.1}
\figsetgrptitle{One-component fit}
\figsetplot{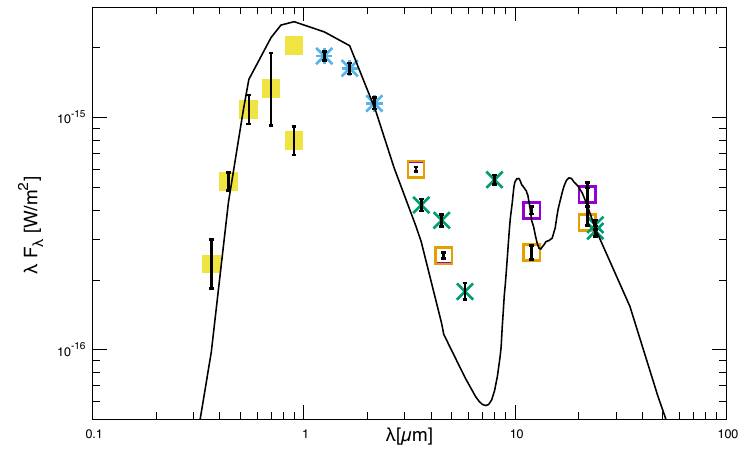}
\figsetgrpnote{One-component model fit (model \#\,1,s, $\chi^{2}$ = 114.3) to the observed SED. I band fluxes have been plotted both from \citet{zaritsky2004} and DENIS \citep{cioni2000} catalog (\S\,\ref{RIdata}).}
\figsetgrpend

\figsetgrpstart
\figsetgrpnum{8.2}
\figsetgrptitle{Two-component fit}
\figsetplot{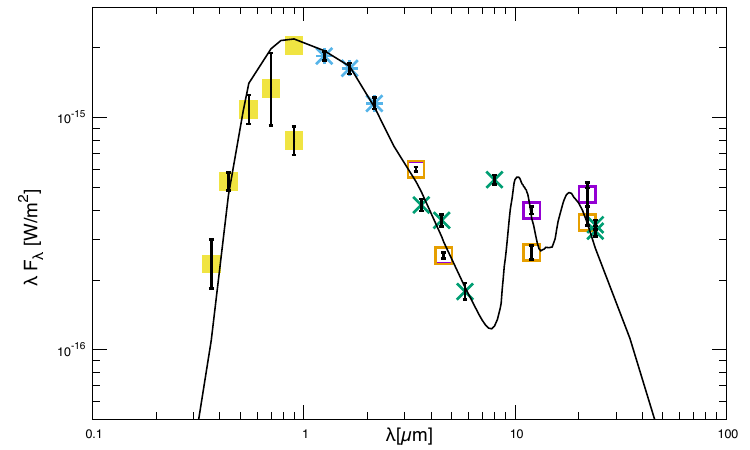}
\figsetgrpnote{Two-component model fit (disk fraction = 0.4, $\chi^{2}$ = 48.4) to the observed SED. The fit (model \#\,2,c) corresponds
to the correctly illuminated model \#\,2.
The disk ($T_{\mathrm d}$ (in) = 1200 K) has Sil-Ow grain type with modified MRN grain size distribution: $a_{\mathrm min}$ = 0.3 $\micron$, $a_{\mathrm max}$ = 5.0 $\micron$, $\tau$ = 0.5 and $Y$ = 3.0. The cold outer shell ($T_{\mathrm d}$ (in) = 240 K) has Sil-Ow grain type with  modified MRN 
grain size distribution: $a_{\mathrm min}$ = 0.005 $\micron$, $a_{\mathrm max}$ = 1.0 $\micron$, $\tau$ = 0.75, $Y$ = 10.}
\figsetgrpend

\figsetgrpstart
\figsetgrpnum{8.3}
\figsetgrptitle{Two-component fit}
\figsetplot{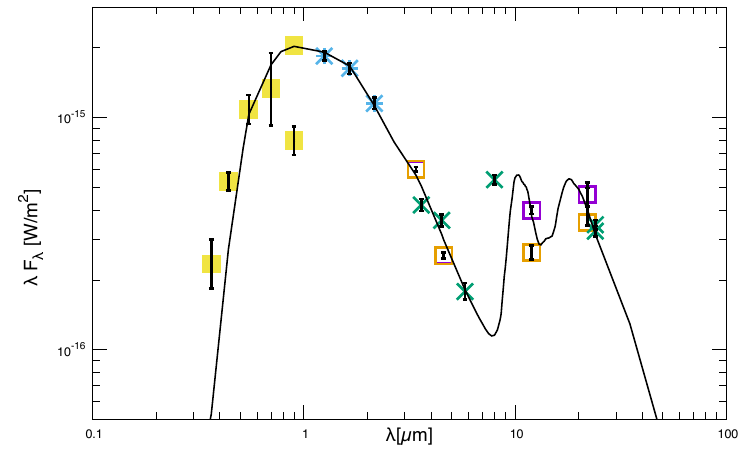}
\figsetgrpnote{Two-component model fit (disk fraction = 0.3, $\chi^{2}$ = 43.9) to the observed SED. The fit corresponds to the correctly illuminated model \#\,3 (model \#\,3,c) wherein the disk ($T_{\mathrm d}$ (in) = 1000 K) is composed of amC-Hn grains with modified MRN grain size distribution: $a_{\mathrm min}$ = 0.005 $\micron$, $a_{\mathrm max}$ = 0.50 $\micron$, $\tau$ = 0.6 and $Y$ = 2. The cold outer shell ($T_{\mathrm d}$ (in) = 250 K) is composed of Sil-Ow grains with standard MRN grain size distribution, $\tau$ = 1.3 and $Y$ = 10.}
\figsetgrpend

\figsetgrpstart
\figsetgrpnum{8.4}
\figsetgrptitle{Comparing fits}
\figsetplot{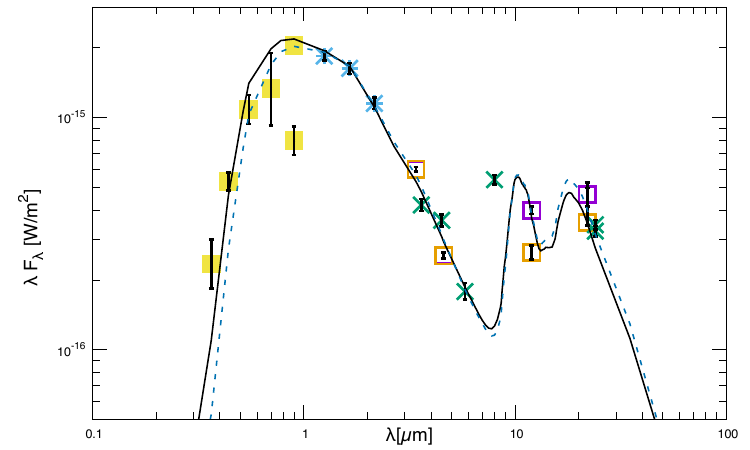}
\figsetgrpnote{Comparing the
two-component model fits, model \#\,2,c (black solid curve,$\chi^{2}$ = 48.4) and model \#\,3,c (blue dashed curve, $\chi^{2}$ = 43.9).} 
\figsetgrpend

\figsetend

\renewcommand{\thefigure}{8}
\begin{figure}
\plotone{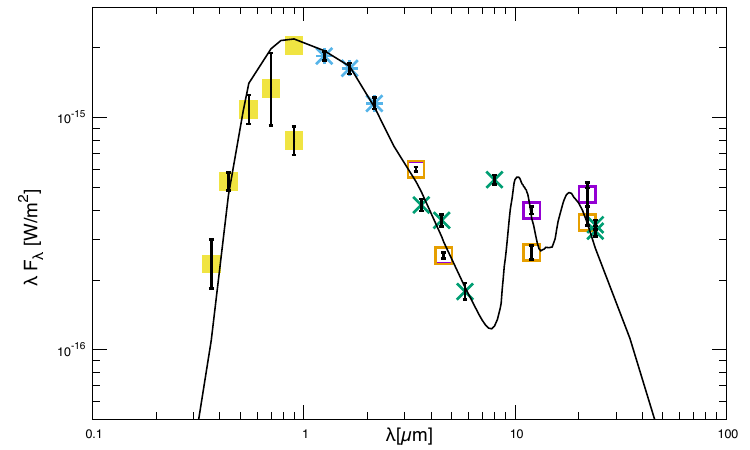}
\caption{The figure shows the adopted two-component model fit (model \#\,2,c, disk fraction = 0.4, $\chi^{2}$ = 48.4) to the observed SED of the post-RGB (disk) source, J050257.89-665306.3. The observed fluxes are de-reddened for Galactic and LMC reddening. U,B,V,R,I (yellow), 2MASS J,H,K (cyan) data are plotted along with WISE (purple) and ALLWISE (orange) photometry and data from the SAGE-LMC Survey (green) which covers the IRAC and MIPS bands. The error bars are indicated in black. All model fits obtained for the source including the adopted best-fit are available in the corresponding Figure Set.}
\end{figure}

\figsetstart
\figsetnum{9}
\figsettitle{J055102.44-685639.1 (post-RGB disk source)}

\figsetgrpstart
\figsetgrpnum{9.1}
\figsetgrptitle{Comparison fits}
\figsetplot{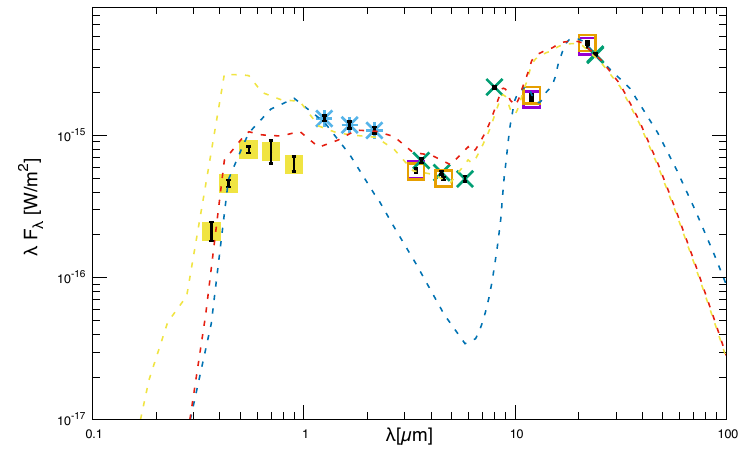}
\figsetgrpnote{One-component model fit to the observed
SED (model \#\,1,s/blue dashed curve, $\chi^{2}$ = 600.2) is plotted along with two-component models, model \#\,2a/yellow dashed curve (disk fraction = 0.4, $\chi^{2}$ = 1800.0) and model \#\,2b/red dashed curve (disk fraction = 0.3, $\chi^{2}$ = 639.7). The model parameters are
given in Table\,\ref{mod-tbl}.}
 \figsetgrpend
 
 \figsetgrpstart
\figsetgrpnum{9.2}
\figsetgrptitle{Two-component fit}
\figsetplot{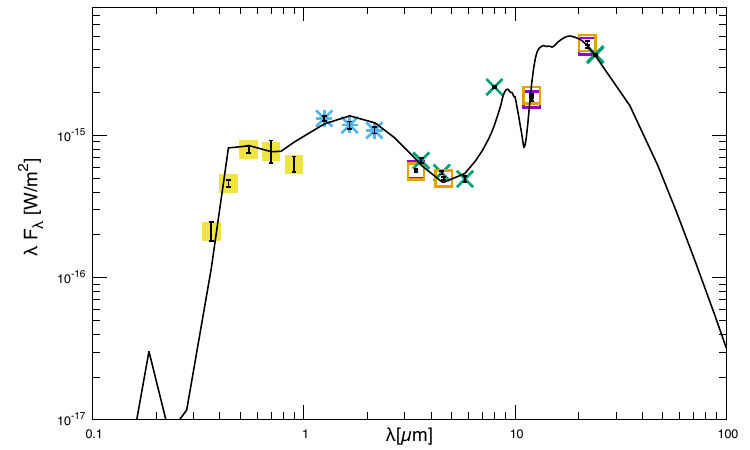}
\figsetgrpnote{Two-component model fit (model \#\,3, disk fraction = 0.3; $\chi^{2}$ = 147.3) to the observed SED. The inner disk  ($T_{\mathrm d}$ (in) = 2000 K) is
made up amC-Hn grains with modified MRN grain size distribution: $a_{\mathrm min}$ = 0.005 $\micron$, 
$a_{\mathrm max}$ = 0.05 $\micron$, $\tau$ = 1.0 and $Y$ = 7. The cool
outer shell ($T_{\mathrm d}$ (in) = 350 K) is made up Sil-Ow/0.4 and SiC-Pg/0.6 grains
with modified MRN grain size distribution: $a_{\mathrm min}$ = 0.005$\micron$,
$a_{\mathrm max}$ = 0.07 $\micron$, $\tau$ = 12.0 and $Y$ = 3.}
 \figsetgrpend

\figsetgrpstart
\figsetgrpnum{9.3}
\figsetgrptitle{Two-component fits}
\figsetplot{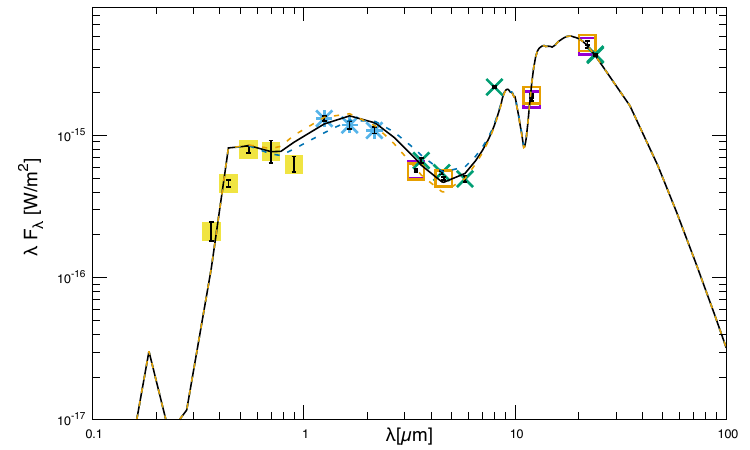}
\figsetgrpnote{Two-component model fits (disk fraction = 0.3) to the observed SED. Fits are obtained by varying the
dust temperature, $T_{\mathrm d}$ at the inner boundary of the dusty disk: $T_{\mathrm d}$ (in) = 1800 K/blue dashed curve ($\chi^{2}$ = 180.1); $T_{\mathrm d}$ (in) = 2200 K/orange dashed curve ($\chi^{2}$ = 159.2) in model \#\,3. The best-fit, \#\,3 corresponds to $T_{\mathrm d}$ (in) = 2000 K/black solid curve ($\chi^{2}$ = 147.3).}
 \figsetgrpend

\figsetgrpstart
\figsetgrpnum{9.4}
\figsetgrptitle{Two-component fits}
\figsetplot{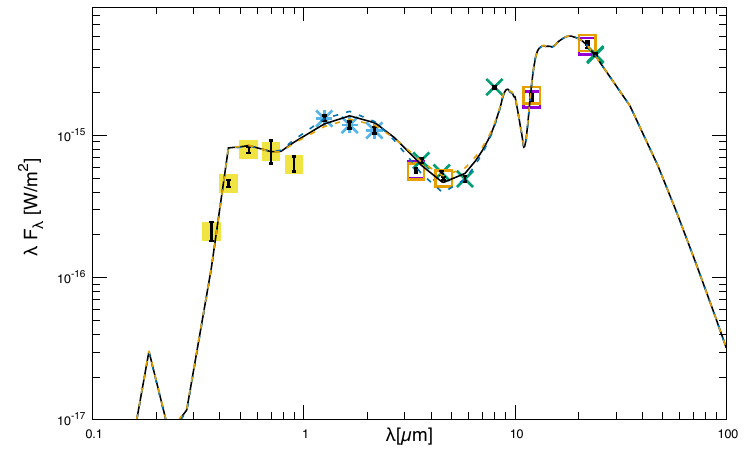}
\figsetgrpnote{Two-component model fits to the observed SED. Fits are obtained by varying the disk thickness, $Y$ = 3/blue dashed curve ($\chi^{2}$ = 159.8) and $Y$ = 20/orange dashed curve ($\chi^{2}$ = 148.1) along with model \#\,3, $Y$ = 7/black solid curve ($\chi^{2}$ = 147.3).} 
\figsetgrpend

\figsetgrpstart
\figsetgrpnum{9.5}
\figsetgrptitle{Two-component fits}
\figsetplot{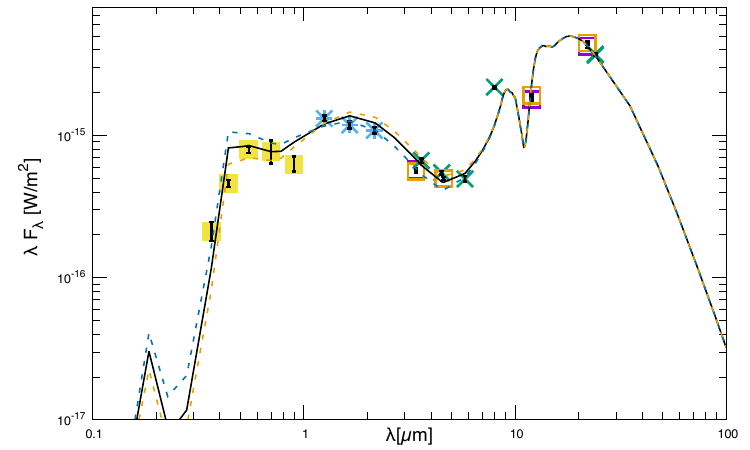}
\figsetgrpnote{Two-component model fits to the observed SED. Fits are obtained by varying the disk optical depth: $\tau$ = 0.8/blue dashed curve ($\chi^{2}$ = 202.5) and 1.2/orange dashed curve ($\chi^{2}$ = 151.3) along with model \#\,3, $\tau$ = 1.0/black solid curve $\chi^{2}$ = 147.3).}
\figsetgrpend

\figsetgrpstart
\figsetgrpnum{9.6}
\figsetgrptitle{Two-component fits}
\figsetplot{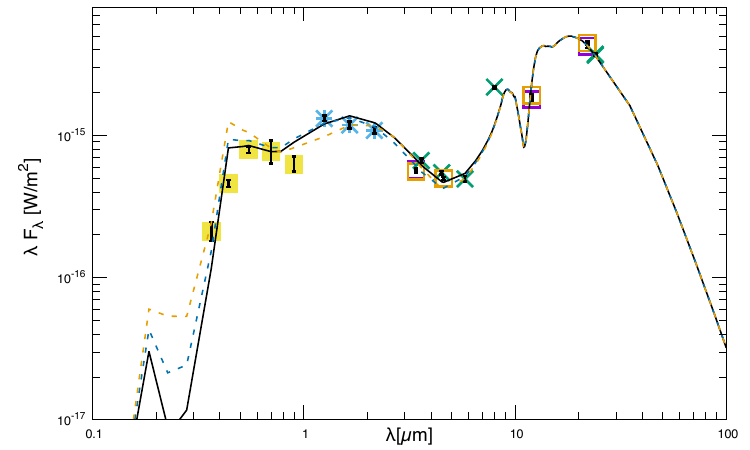}
\figsetgrpnote{Two-component model fits to the observed SED. Fits are obtained by varying the maximum amC-Hn grain
sizes in the inner disk: $a_{\mathrm max}$ = 0.10/blue dashed curve ($\chi^{2}$ = 169.6) and 0.50/orange dashed curve ($\chi^{2}$ = 263.9) along with 
model \#\,3, $a_{\mathrm max}$ = 0.05/black solid curve ($\chi^{2}$ = 147.3).}
\figsetgrpend

\figsetgrpstart
\figsetgrpnum{9.7}
\figsetgrptitle{Two-component fits}
\figsetplot{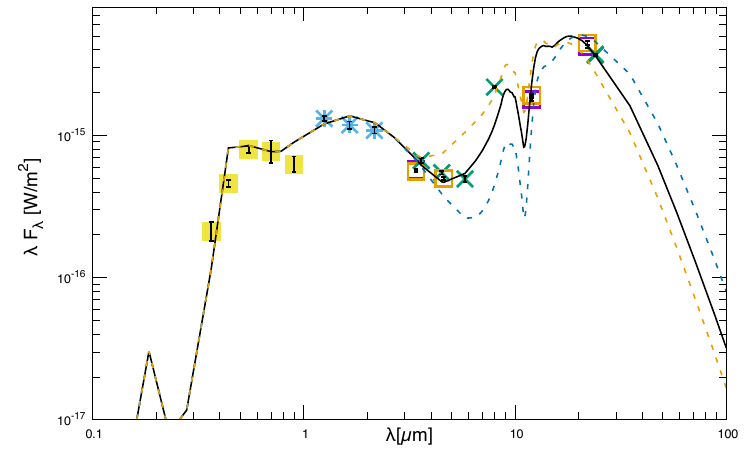}
\figsetgrpnote{Two-component model fits to the observed SED. Fits are obtained by varying the dust temperature at the 
inner boundary of the cold outer shell, $T_{\mathrm d}$ (in) = 250 K/blue dashed curve ($\chi^{2}$ = 458.2) and $T_{\mathrm d}$ (in) = 450 K/orange dashed curve ($\chi^{2}$ = 476.4) along with model \#\,3, $T_{\mathrm d}$ (in) = 350 K/black solid curve ($\chi^{2}$ = 147.3).}
\figsetgrpend

\figsetgrpstart
\figsetgrpnum{9.8}
\figsetgrptitle{Two-component fits}
\figsetplot{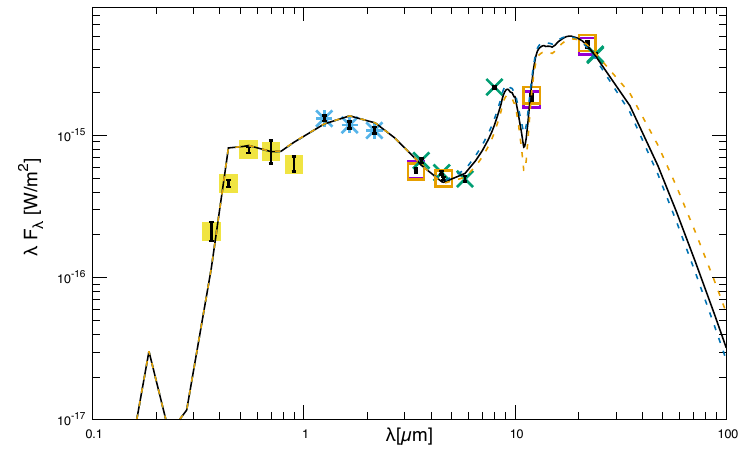}
\figsetgrpnote{Two-component model fits to the observed SED. Fits are obtained by varying the shell thickness, $Y$ = 2/blue dashed curve ($\chi^{2}$ = 151.9) and $Y$ = 10/orange dashed curve  ($\chi^{2}$ = 154.5) along with model \#\,3, $Y$ = 3/black solid curve ($\chi^{2}$ = 147.3).}
\figsetgrpend

\figsetgrpstart
\figsetgrpnum{9.9}
\figsetgrptitle{Two-component fits}
\figsetplot{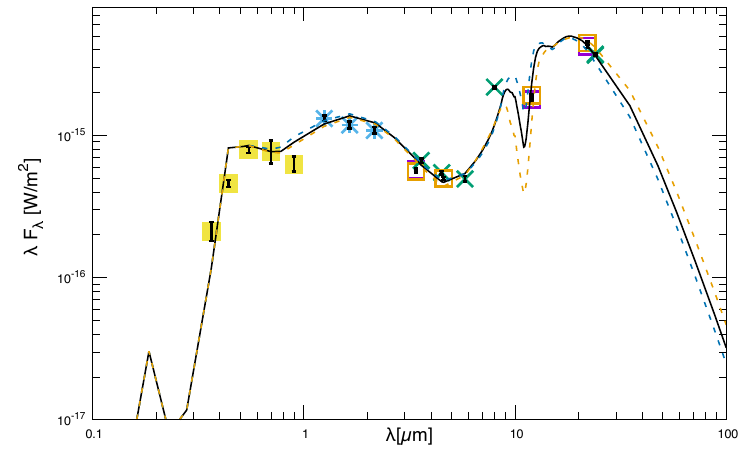}
\figsetgrpnote{Two-component model fits to the observed SED. Fits are obtained by varying the shell optical depth, $\tau$ = 9/blue dashed curve ($\chi^{2}$ = 357.9) and $\tau$ = 18/orange dashed 
curve ($\chi^{2}$ = 189.7) along with model \#\,3, $\tau$ = 12/black solid curve ($\chi^{2}$ = 147.3).}
\figsetgrpend

\figsetgrpstart
\figsetgrpnum{9.10}
\figsetgrptitle{Two-component fits}
\figsetplot{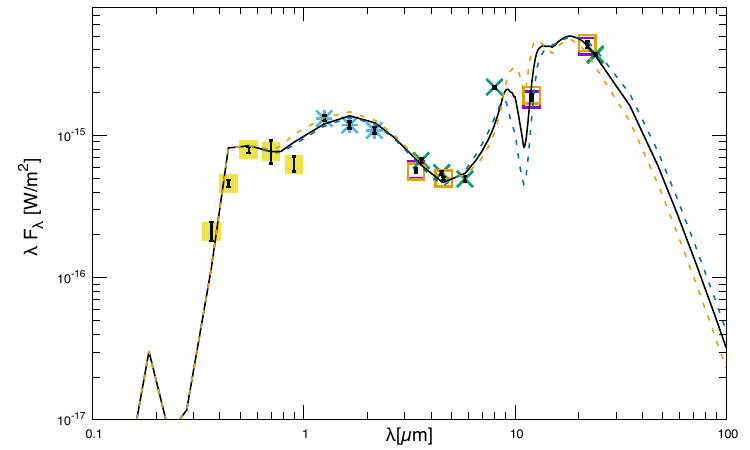}
\figsetgrpnote{Two-component model fits to the observed SED. Fits are obtained by varying the maximum Sil-Ow grain sizes 
in the shell: $a_{\mathrm max}$ = 0.05/blue dashed curve ($\chi^{2}$ = 152.5) and 0.09/orange dashed curve ($\chi^{2}$ = 622.3) along with model \#\,3, $a_{\mathrm max}$ = 0.07/black solid curve $\chi^{2}$ = 147.3). }
\figsetgrpend

\figsetend

\renewcommand{\thefigure}{9}
\begin{figure}
\plotone{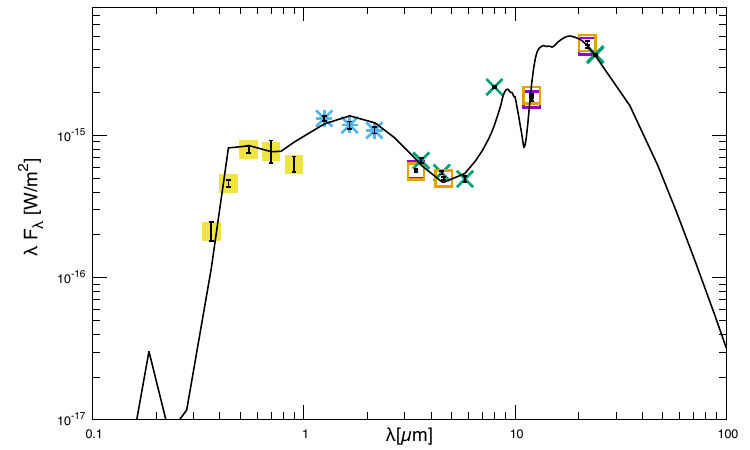}
\caption{The figure shows the adopted two-component model fit (model \#\,3, disk fraction = 0.3; $\chi^{2}$ = 147.3) to the observed SED of the post-RGB (disk) source, J055102.44-685639.1. The observed fluxes are de-reddened for Galactic and LMC
reddening. U,B,V,R,I (yellow), 2MASS J,H,K (cyan) data are plotted along with WISE (purple) and ALLWISE (orange)
photometry and data from the SAGE-LMC Survey (green) which covers the
IRAC and MIPS bands. The error bars are indicated in black. All model fits obtained for the source including the adopted best-fit are available in the corresponding Figure Set.}
\end{figure}

\figsetstart
\figsetnum{10}
\figsettitle{J050632.10-714229.8 (post-AGB shell source)}

\figsetgrpstart
\figsetgrpnum{10.1}
\figsetgrptitle{One-component fit}
\figsetplot{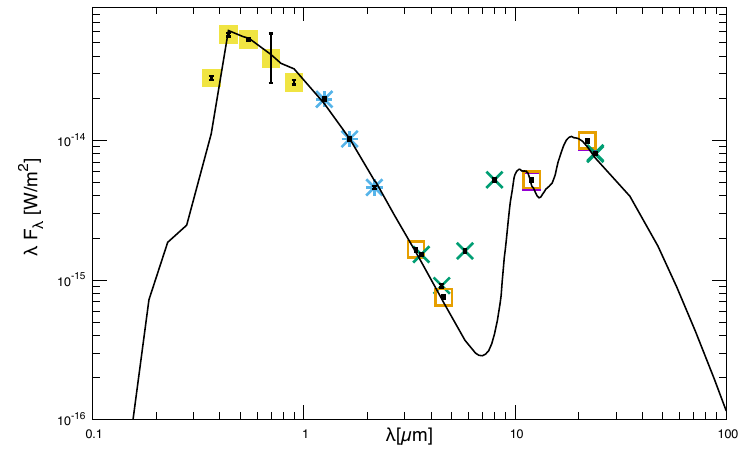}
\figsetgrpnote{One-component model fit (model \#\,1,s, $\chi^{2}$ = 231.1) to the observed SED, 
obtained using Sil-Ow grain type, standard MRN grain size distribution, 
$T_{\mathrm d}$ (in) = 200 K, $\tau$ = 0.6 and $Y$ =  20.}
\label{J050632_model1}
\figsetgrpend

\figsetgrpstart
\figsetgrpnum{10.2}
\figsetgrptitle{Inner shell}
\figsetplot{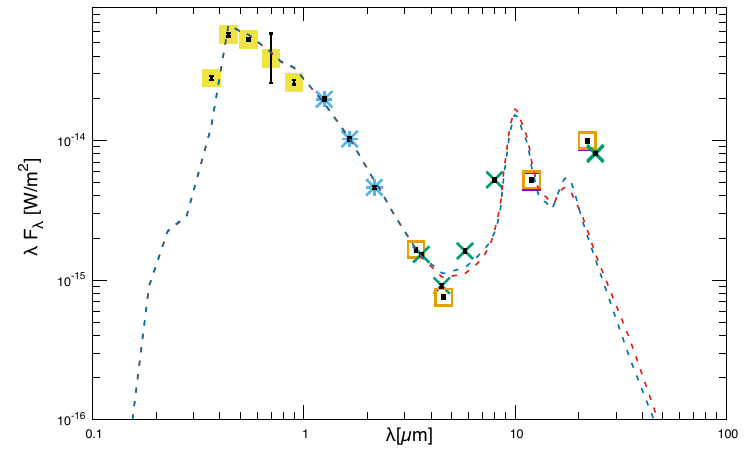}
\figsetgrpnote{Modeling the warm inner shell. Fits are shown to the observed SED in the mid-IR using graintype \#\,1: Sil-Ow (red dashed curve, $\chi^{2}$ = 1528.1) and for comparison,
Sil-Oc (blue dashed curve, $\chi^{2}$ = 1579.5). $T_{\mathrm d}$ (in) = 400 K, MRN grain size distribution, $\tau$ = 0.4 and 
$Y$ = 2.}
\figsetgrpend

\figsetgrpstart
\figsetgrpnum{10.3}
\figsetgrptitle{Inner shell}
\figsetplot{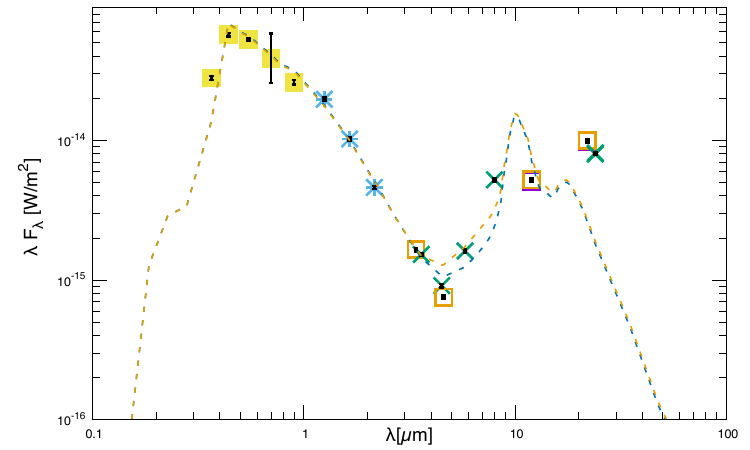}
\figsetgrpnote{Modeling the warm inner shell. 
Fits are shown to 
the observed SED in the mid-IR for amC-Hn and Sil-Ow
grain combinations: Sil-Ow/0.9 and amC/0.1 (blue dashed curve, $\chi^{2}$ = 1412.0) ; Sil-Ow/0.8 and amC/0.2 (orange dashed curve, $\chi^{2}$ = 1445.6).
$T_{\mathrm d}$ (in) = 350 K, $\tau$ = 0.4, $Y$ = 2 and modified grain size distribution:
$a_{\mathrm min}$ = 0.1 $\micron$, $a_{\mathrm max}$ = 0.5 $\micron$ are used.}
\figsetgrpend

\figsetgrpstart
\figsetgrpnum{10.4}
\figsetgrptitle{Inner shell}
\figsetplot{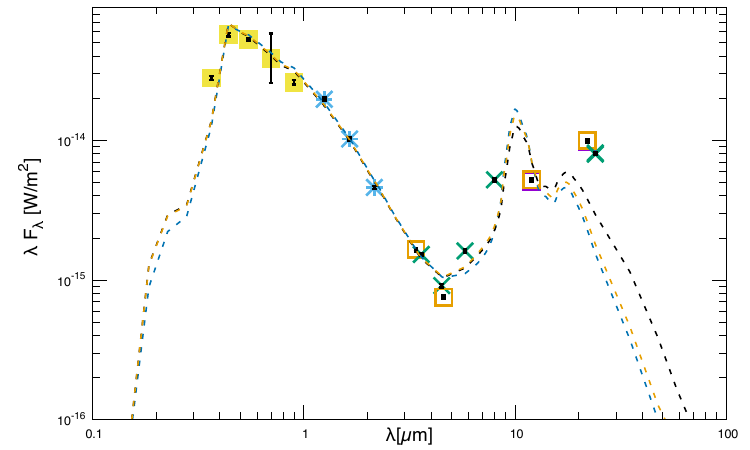}
\figsetgrpnote{Modeling the warm inner shell. Fits are shown to 
the observed SED in the near-IR for different grain compositions: blue dashed curve (\#\,1, $\chi^{2}$ = 1528.1) -- Sil-Ow/1.0, $T_{\mathrm d}$ (in) = 400 K, MRN grain sizes, $\tau$ = 0.4, $Y$ = 2;
orange dashed curve (\#\,2, $\chi^{2}$ = 1412.0) -- amC-Hn/0.1 and Sil-Ow/0.9,
$T_{\mathrm d}$ (in) = 350 K, $a_{\mathrm min}$ = 0.1 $\micron$, $a_{\mathrm max}$ = 0.5 $\micron$, $\tau$ = 0.4, $Y$ = 2;
black dashed curve (\#\,3, $\chi^{2}$ = 988.4)-- grf-DL/0.3, Sil-Ow/0.7, $T_{\mathrm d}$ (in) = 350 K,
$a_{\mathrm min}$ = 0.1 $\micron$, $a_{\mathrm max}$ = 1.0 $\micron$, $\tau$ = 0.4, $Y$ = 5). }
\figsetgrpend

\figsetgrpstart
\figsetgrpnum{10.5}
\figsetgrptitle{Nested shells}
\figsetplot{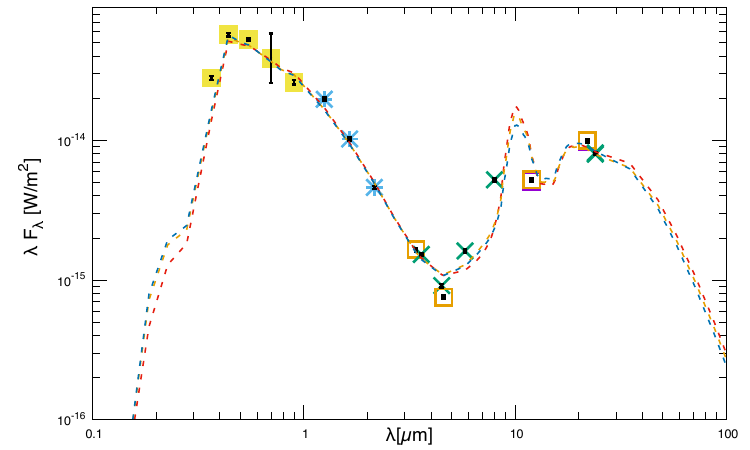}
\figsetgrpnote{The nested shells models. The fits correspond to the correctly illuminated outer shell. Fits are shown 
for different grain compositions in the inner-shell Sil-Ow (\#\,1, red dashed curve, $\chi^{2}$ = 181.6);
amC-Hn/0.1 and Sil-Ow/0.9 (\#\,2, orange dashed, $\chi^{2}$ = 164.2) and grf-DL/0.3 and Sil-Ow/0.7 (\#\,3, blue dashed curve, $\chi^{2}$ = 154.7). The
model parameters for each fit are given in Table\,\ref{mod-tbl-pagb}.}
\figsetgrpend

\figsetgrpstart
\figsetgrpnum{10.6}
\figsetgrptitle{Inner shell}
\figsetplot{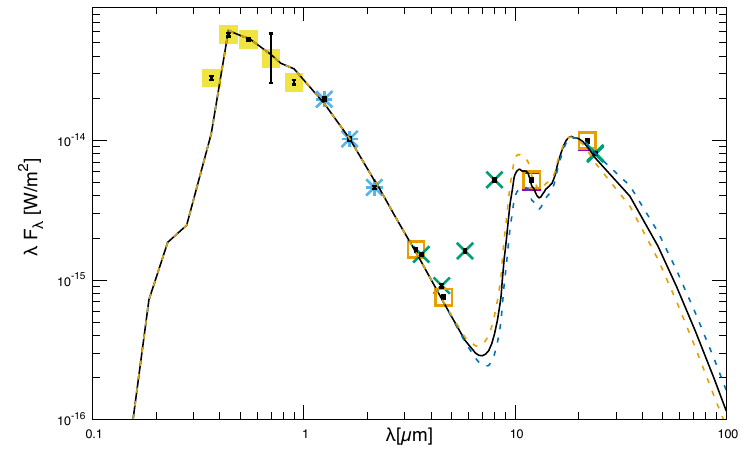}
\figsetgrpnote{One-component model fits to the observed SED. Fits show model \#\,1,s, $T_{\mathrm d}$ (in) = 200 K/black solid curve ($\chi^{2}$ = 231.1) and the effect of varying the dust temperature at the inner shell boundary, $T_{\mathrm d}$ (in) = 220 K/orange dashed curve ($\chi^{2}$ = 253.9) and 180 K/blue dashed curve ($\chi^{2}$ = 248.7).}
\figsetgrpend

\figsetgrpstart
\figsetgrpnum{10.7}
\figsetgrptitle{Inner shell}
\figsetplot{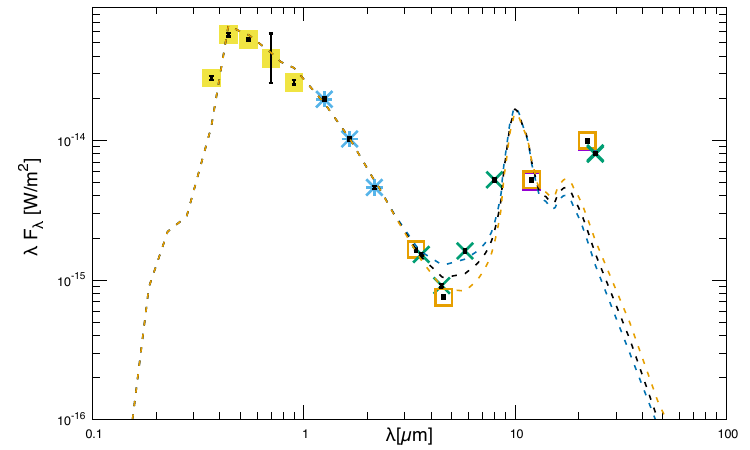}
\figsetgrpnote{Modeling the warm inner shell 
using grain type \#\,1: Sil-Ow. Fits are shown to the observed
SED in the mid-IR using MRN grain size distribution, 
$\tau$ = 0.4, $Y$ = 2 and $T_{\mathrm d}$ (in) = 450 K/blue dashed curve ($\chi^{2}$ = 1690.1), 400 K/black dashed 
curve ($\chi^{2}$ = 1528.2) and 350 K/orange dashed curve ($\chi^{2}$ = 1382.4).}
\figsetgrpend

\figsetgrpstart
\figsetgrpnum{10.8}
\figsetgrptitle{Inner shell}
\figsetplot{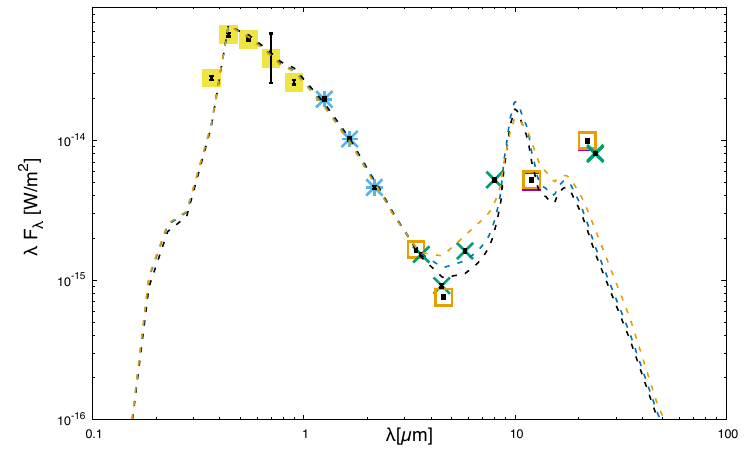}
\figsetgrpnote{Modeling the warm inner shell using grain type \#\,1: Sil-Ow. Fits are shown to the observed
SED in the mid-IR for $T_{\mathrm d}$ (in) = 400 K, $\tau$ = 0.4 and $Y$ = 2  and varying maximum grain sizes: MRN grain sizes/black dashed curve ($\chi^{2}$ = 1528.2); $a_{\mathrm max}$ = 1.0/blue dashed curve ($\chi^{2}$ = 1545.1)
and 5.0/orange dashed curve ($\chi^{2}$ = 1545.0).}
\figsetgrpend

\figsetend

\renewcommand{\thefigure}{10}
\begin{figure}
\plotone{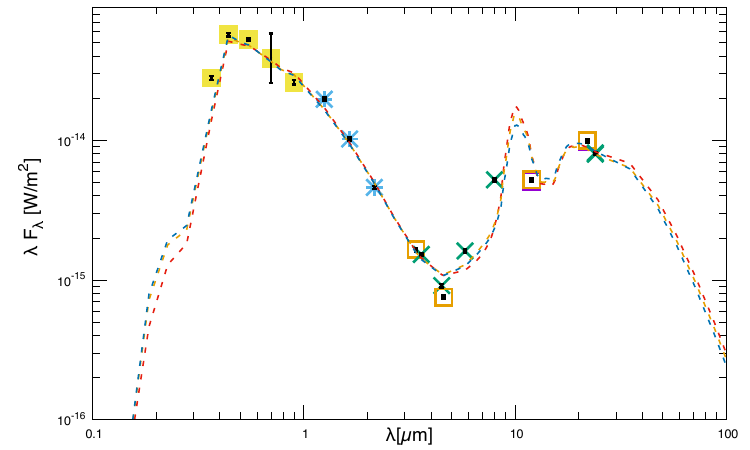}
\caption{The figure shows the nested shells models for the post-AGB (shell) source, J050632.10-714229.8. The fits correspond to the correctly illuminated outer shell, Table\,\ref{mod-tbl-pagb}.
The observed fluxes are de-reddened for Galactic and LMC 
reddening. U,B,V,R,I (yellow), 2MASS J,H,K (cyan) data are plotted along with WISE (purple) and  ALLWISE (orange) photometry and data from the SAGE-LMC Survey (green) which covers the
IRAC and MIPS bands. The error bars are indicated in black. All model fits obtained for the source including the nested shells models are available in the corresponding Figure Set.}
\end{figure}

\figsetstart
\figsetnum{11}
\figsettitle{J051848.84-700247.0 (post-AGB shell source)}

\figsetgrpstart
\figsetgrpnum{11.1}
\figsetgrptitle{One-component fit}
\figsetplot{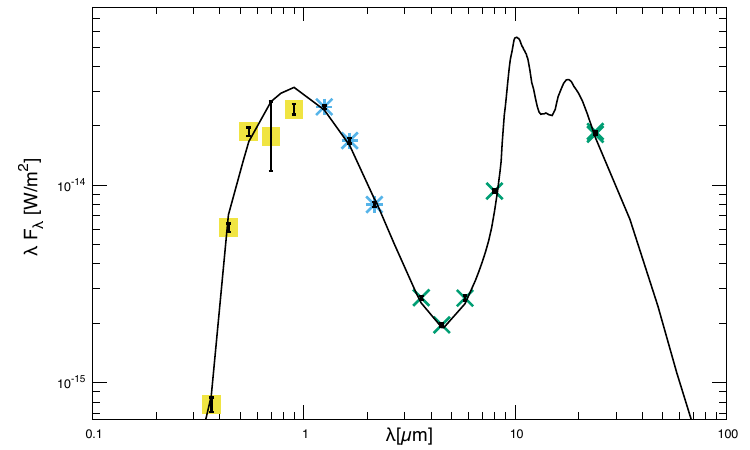}
\figsetgrpnote{One-component model fit ($\chi^{2}$ = 50.4) to the observed SED
is obtained for $T_{\mathrm d}$ (in) = 350 K, using Sil-Ow grain type, standard MRN grain size distribution, $\tau$ = 3.2
and $Y$ =  20.}
\figsetgrpend

\figsetgrpstart
\figsetgrpnum{11.2}
\figsetgrptitle{One-component fits}
\figsetplot{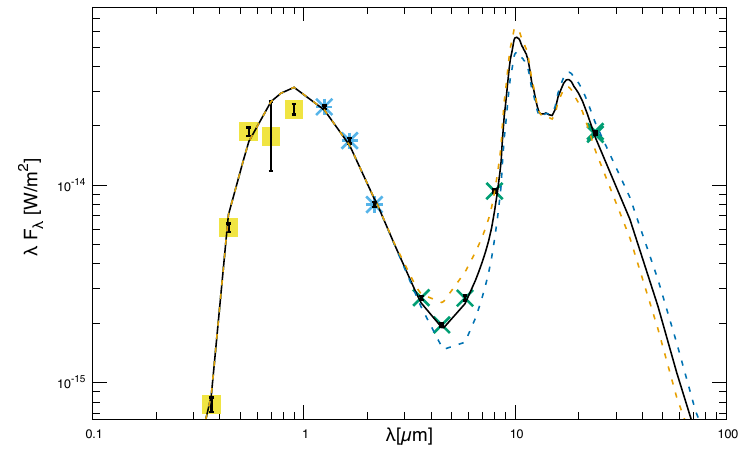}
\figsetgrpnote{One-component model fits to the observed SED for varying dust temperatures at the
inner shell boundary, $T_{\mathrm d}$ (in) = 300 K/blue dashed curve ($\chi^{2}$ = 235.7), 400 K/orange dashed curve ($\chi^{2}$ = 297.4) along with
the best-fit model, $T_{\mathrm d}$ (in) = 350 K/black solid curve ($\chi^{2}$ = 50.4).}
\figsetgrpend

\figsetgrpstart
\figsetgrpnum{11.3}
\figsetgrptitle{One-component fits}
\figsetplot{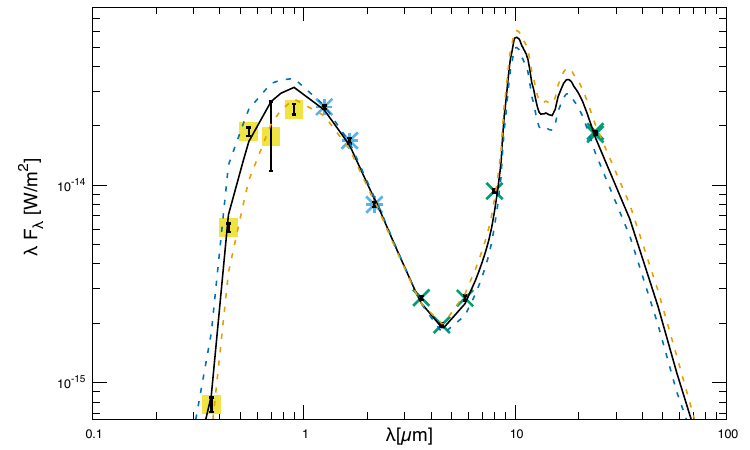}
\figsetgrpnote{One-component model fits to the observed SED for varying shell optical depth
at 0.55 $\micron$, $\tau$ = 2.5/blue dashed curve ($\chi^{2}$ = 470.8) and $\tau$ = 4.0/orange dashed curve ($\chi^{2}$ = 103.0 )along with the best-fit model, $\tau$ = 3.2/black solid curve ($\chi^{2}$ = 50.4).}
\figsetgrpend

\figsetgrpstart
\figsetgrpnum{11.4}
\figsetgrptitle{One-component fits}
\figsetplot{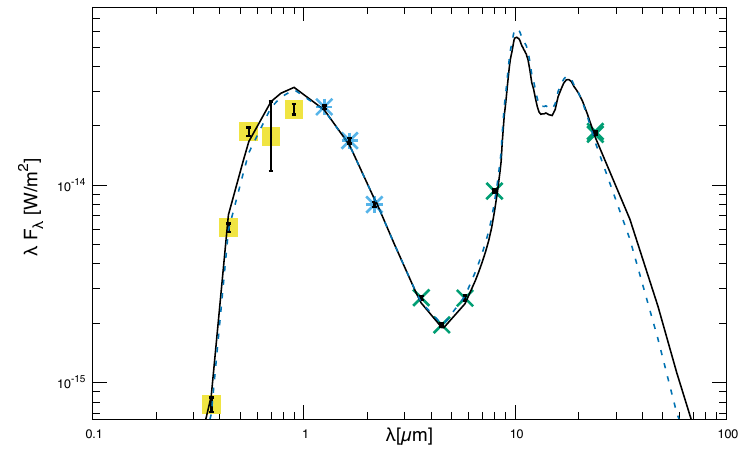}
\figsetgrpnote{One-component model fit to the observed SED for varying shell thickness, $Y$ = 5/blue dashed curve ($\chi^{2}$ = 98.4) along with the best-fit model, $Y$ = 20/black solid curve ($\chi^{2}$ = 50.4).}
\figsetgrpend

\figsetend

\renewcommand{\thefigure}{11}
\begin{figure}
\plotone{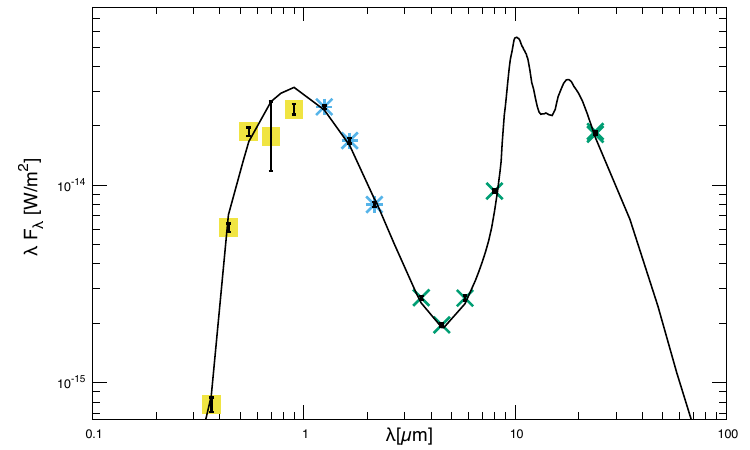}
\caption{The adopted one-component model fit ($\chi^{2}$ = 50.4) to the observed SED of the post-AGB (shell) source, J051848.84-700247.0.
The observed fluxes are de-reddened for Galactic and LMC reddening. U,B,V,R,I (yellow), 2MASS J,H,K (cyan) data are plotted along with data from the SAGE-LMC Survey (green) which covers the
IRAC and MIPS bands. The error bars are indicated in black. All model fits obtained for the source including the adopted best-fit are available in the corresponding Figure Set.}
\end{figure}

\figsetstart
\figsetnum{12}
\figsettitle{J051906.86-694153.9 (post-AGB shell source)}

\figsetgrpstart
\figsetgrpnum{12.1}
\figsetgrptitle{Inner shell}
\figsetplot{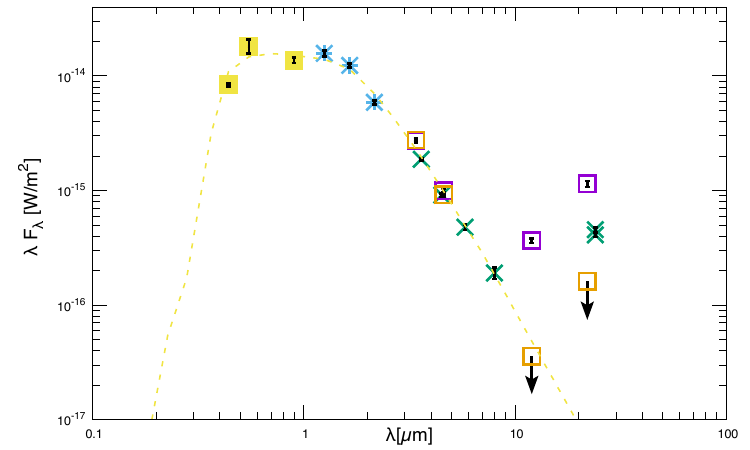}
\figsetgrpnote{Modeling the warm inner shell. Fit to the optical (yellow dashed curve, $\chi^{2}$ = 179.3),
near and mid-IR data is obtained with $T_{\mathrm d}$ (in) = 2000 K, amC-Hn grain type, standard MRN grain size distribution, $\tau$ = 0.35 and $Y$ =  2.}
\figsetgrpend

\figsetgrpstart
\figsetgrpnum{12.2}
\figsetgrptitle{Nested shells}
\figsetplot{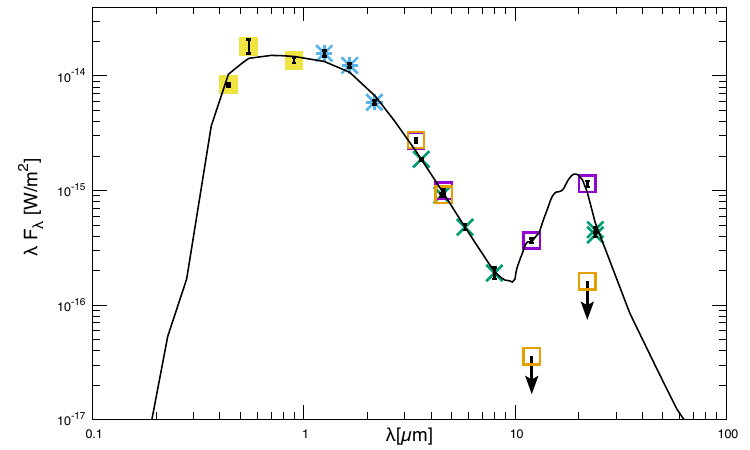}
\figsetgrpnote{The nested shells model ($\chi^{2}$ = 21.3). The
model parameters for the warm inner shell are $T_{\mathrm d}$ (in) = 2000 K,
amC-Hn grains with standard MRN grain size distribution, $\tau$ = 0.35 and $Y$ =  2. The cold outer shell is modeled using
$T_{\mathrm d}$ (in) = 160 K, silicon carbide grains, SiC-Pg with modified MRN grain size distribution:
$a_{\mathrm min}$ = 2.3 $\micron$, $a_{\mathrm max}$ = 3.0 $\micron$, $\tau$ = 0.07 and $Y$ = 2.}
\figsetgrpend

\figsetgrpstart
\figsetgrpnum{12.3}
\figsetgrptitle{Inner shell}
\figsetplot{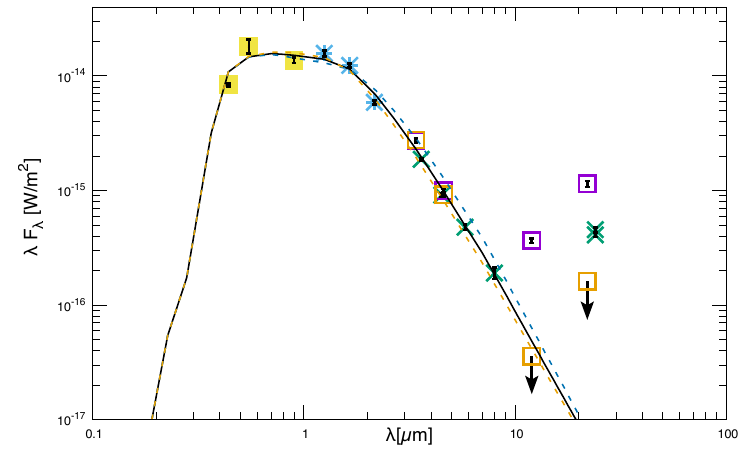}
\figsetgrpnote{Modeling the warm inner shell. Fits are shown 
to the observed SED in the near and mid-IR for varying temperatures at the shell's inner boundary, $T_{\mathrm d}$ = 1700 K/blue dashed curve ($\chi^{2}$ = 225.1) and 2300 K/orange dashed curve ($\chi^{2}$ = 189.0) along with the best-fit model, $T_{\mathrm d}$ = 2000 K/black solid curve, $\chi^{2}$ = 179.3.}
\figsetgrpend

\figsetgrpstart
\figsetgrpnum{12.4}
\figsetgrptitle{Inner shell}
\figsetplot{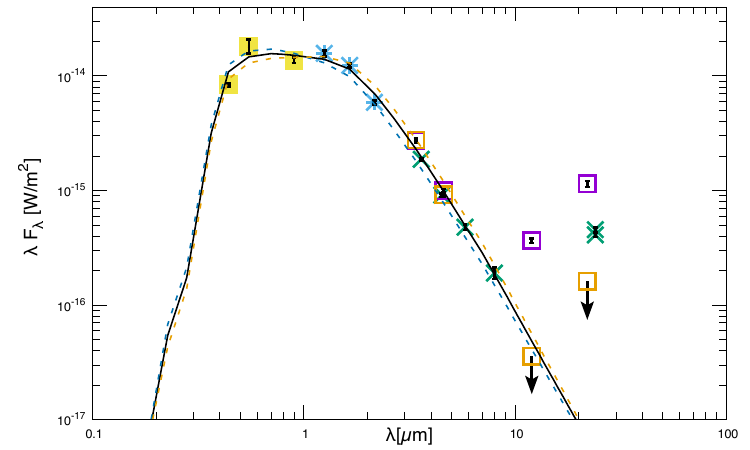}
\figsetgrpnote{Modeling the warm inner shell. Fits are shown 
to the observed SED in the near and mid-IR for varying shell optical depth at 0.55$\micron$,
$\tau$ = 0.20/blue dashed curve ($\chi^{2}$ = 216.4) and 0.50/orange dashed curve ($\chi^{2}$ = 203.5) along with the best-fit model, $\tau$ = 0.35/black solid curve, $\chi^{2}$ = 179.3.}
\figsetgrpend

\figsetgrpstart
\figsetgrpnum{12.5}
\figsetgrptitle{Inner shell}
\figsetplot{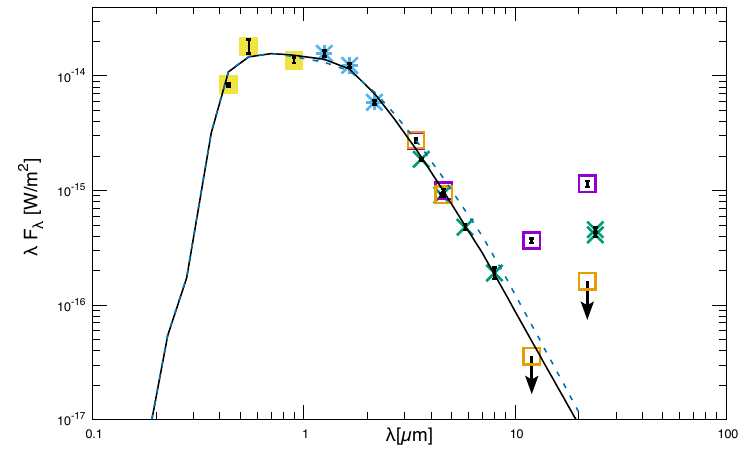}
\figsetgrpnote{Modeling the warm inner shell. Fits are shown 
to the observed SED in the near and mid-IR for varying shell thickness, $Y$ = 7/blue dashed curve ($\chi^{2}$ = 219.2) along with the best-fit model, $Y$ = 2/black solid curve, $\chi^{2}$ = 179.3.}
\figsetgrpend

\figsetgrpstart
\figsetgrpnum{12.6}
\figsetgrptitle{Nested shells}
\figsetplot{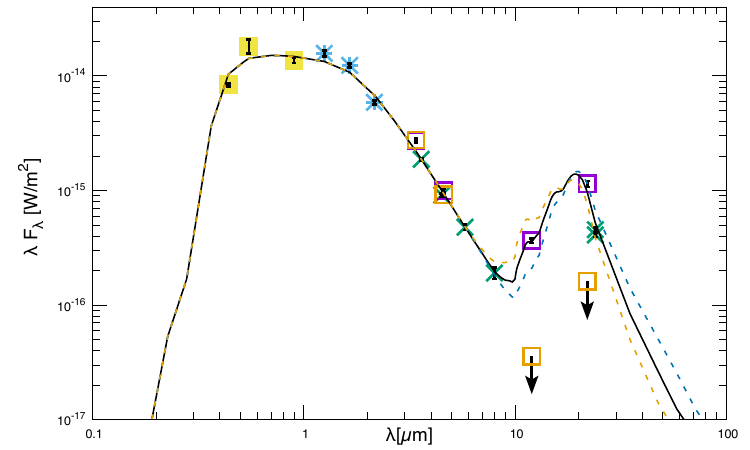}
\figsetgrpnote{Nested shells model fits are shown 
to the observed SED for varying temperature at the inner boundary of the cold outer shell, $T_{\mathrm d}$ (in) = 130 K/blue dashed curve ($\chi^{2}$ = 55.4), 200 K/orange dashed curve ($\chi^{2}$ = 30.5) along with the best-fit model, $T_{\mathrm d}$ (in) = 160 K/black solid curve, $\chi^{2}$ = 21.3.}
\figsetgrpend

\figsetgrpstart
\figsetgrpnum{12.7}
\figsetgrptitle{Nested shells}
\figsetplot{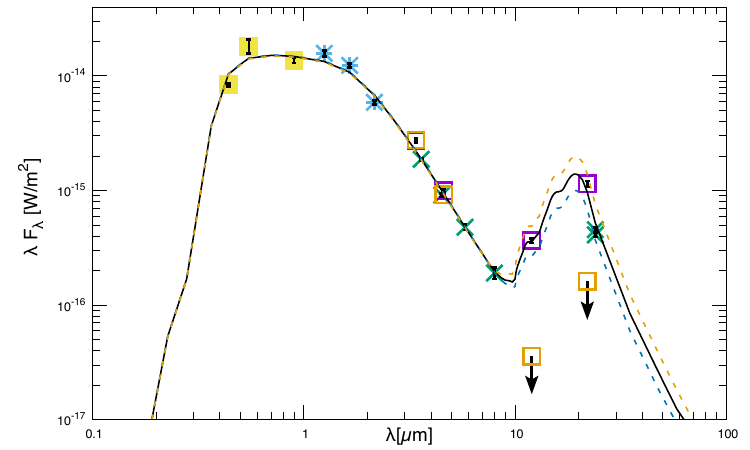}
\figsetgrpnote{Nested shells model fits are shown 
to the observed SED for varying optical depth of the cold outer shell, $\tau$ = 0.05/blue dashed curve ($\chi^{2}$ = 26.5), 0.10/orange dashed curve ($\chi^{2}$ = 80.6) along with the best-fit model, $\tau$ = 0.07/black solid curve ($\chi^{2}$ = 21.3).}
\figsetgrpend

\figsetgrpstart
\figsetgrpnum{12.8}
\figsetgrptitle{Nested shells}
\figsetplot{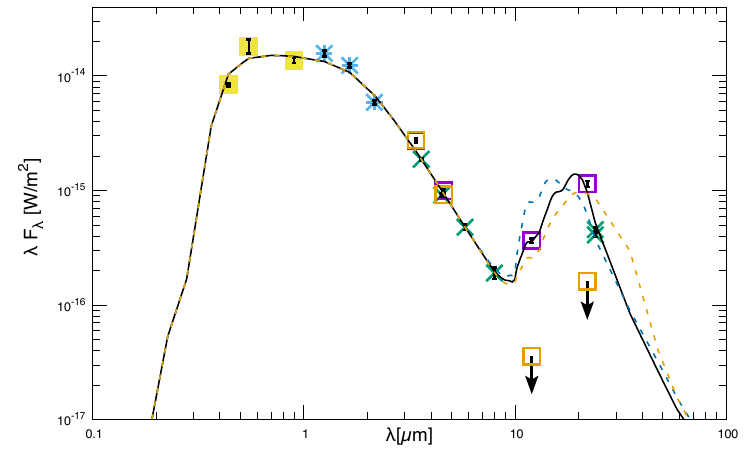}
\figsetgrpnote{Nested shells model fits are shown 
to the observed SED for varying minimum and maximum grain sizes of the cold outer shell,
blue dashed curve ($\chi^{2}$ = 86.0): $a_{\mathrm min}$ = 1.0 $\micron$, $a_{\mathrm max}$ = 3.0 $\micron$, 
orange dashed curve ($\chi^{2}$ = 118.5): $a_{\mathrm min}$ = 2.3 $\micron$, $a_{\mathrm max}$ = 5.0 $\micron$) along with the best-fit 
model, black solid curve ($\chi^{2}$ = 21.3): $a_{\mathrm min}$ = 2.3 $\micron$, $a_{\mathrm max}$ = 3.0 $\micron$.}
\figsetgrpend

\figsetgrpstart
\figsetgrpnum{12.9}
\figsetgrptitle{Nested shells}
\figsetplot{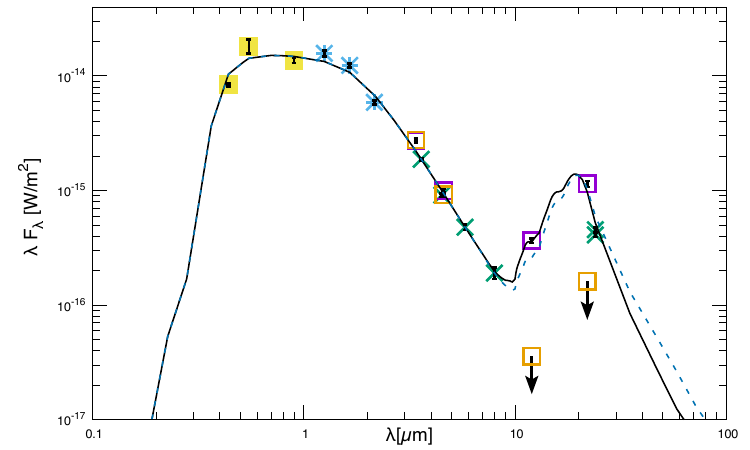}
\figsetgrpnote{Nested shells model fits are shown 
to the observed SED for varying thickness of the cold outer shell, $Y$ = 10/blue dashed curve ($\chi^{2}$ = 35.5) along with the best-fit model, $Y$ = 2/black solid curve ($\chi^{2}$ = 21.3).}
\figsetgrpend

\figsetend

\renewcommand{\thefigure}{12}
\begin{figure}
\plotone{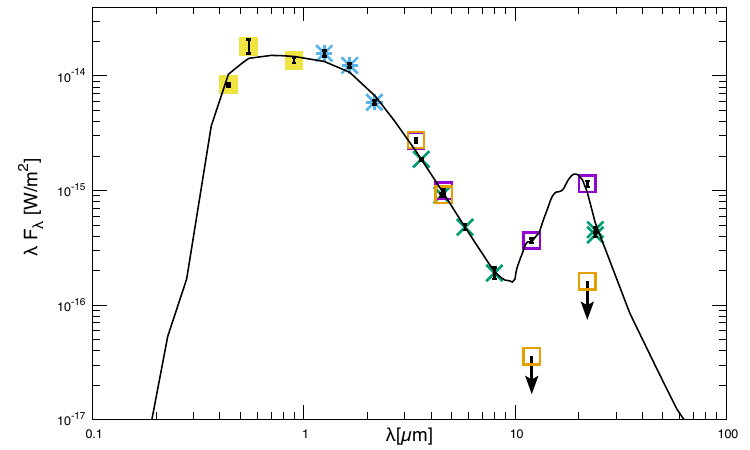}
\caption{The figure shows the adopted nested shells model ($\chi^{2}$ = 21.3) for the post-AGB (shell) source, J051906.86-694153.9. 
The observed fluxes are de-reddened for Galactic and LMC 
reddening. B,V,I (yellow), 2MASS J,H,K (cyan) data are plotted along with 
WISE (purple) and  ALLWISE (orange) photometry
and data from the SAGE-LMC Survey (green) which covers the
IRAC and MIPS bands. The error bars are indicated in black.  All model fits obtained for the source including the adopted best-fit are available in the corresponding Figure Set.}
\end{figure}

\figsetstart
\figsetnum{13}
\figsettitle{J053250.69-713925.8 (post-AGB shell source)}

\figsetgrpstart
\figsetgrpnum{13.1}
\figsetgrptitle{One-component fit}
\figsetplot{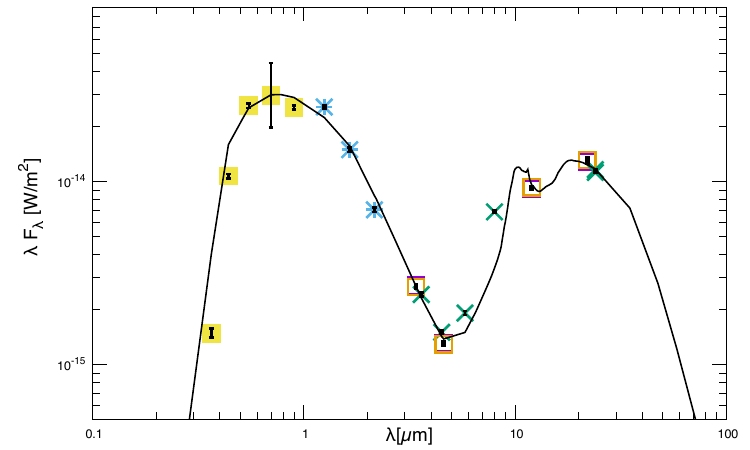}
\figsetgrpnote{One-component model fit ($\chi^{2}$ = 214.4) to the observed 
SED. Fit is obtained using a mix of silicate (Sil-Ow/0.2) and graphite (grf-DL/0.8)  
grains, modified MRN grain size distribution: $a_{\mathrm min}$ = 0.005 $\micron$ and $a_{\mathrm max}$ =  0.50 $\micron$, $T_{\mathrm d}$ (in) = 260 K, $\tau$ = 0.9 and $Y$ = 2.}
\figsetgrpend

\figsetgrpstart
\figsetgrpnum{13.2}
\figsetgrptitle{One-component fits}
\figsetplot{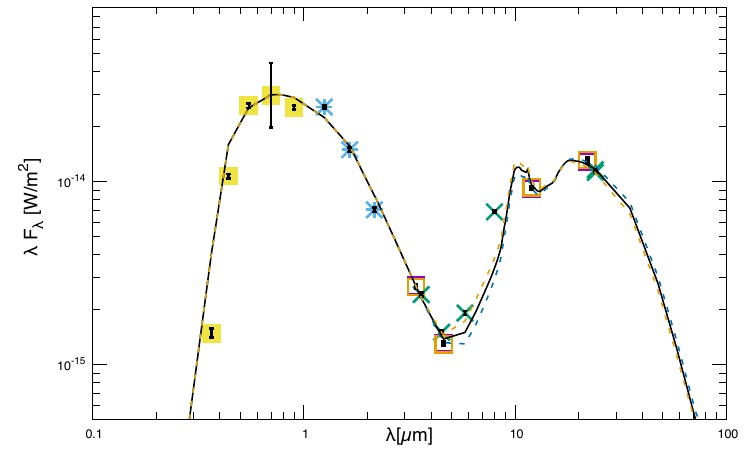}
\figsetgrpnote{One-component model fits to the observed SED for varying dust temperatures
at the inner shell boundary, $T_{\mathrm d}$ (in) = 250 K/blue dashed curve ($\chi^{2}$ = 777.0), $T_{\mathrm d}$ (in) = 270 K/orange dashed curve ($\chi^{2}$ = 460.2) along with the best-fit model,
$T_{\mathrm d}$ (in) = 260 K ($\chi^{2}$ = 214.4).}
\figsetgrpend

\figsetgrpstart
\figsetgrpnum{13.3}
\figsetgrptitle{One-component fits}
\figsetplot{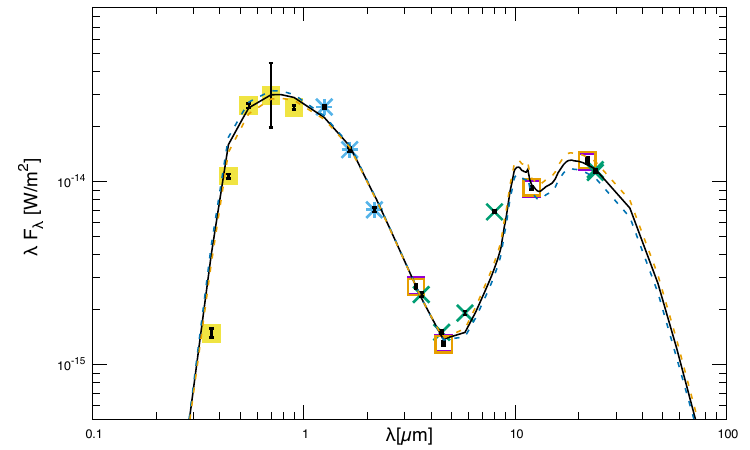}
\figsetgrpnote{One-component model fits to the observed SED 
for varying shell optical depth, $\tau$ = 0.80/blue dashed curve ($\chi^{2}$ = 2220.7), $\tau$ = 1.0/orange dashed curve ($\chi^{2}$ = 2547.2), along with the best-fit model, $\tau$ = 0.90/black solid curve ($\chi^{2}$ = 214.4).}
\figsetgrpend

\figsetgrpstart
\figsetgrpnum{13.4}
\figsetgrptitle{One-component fits}
\figsetplot{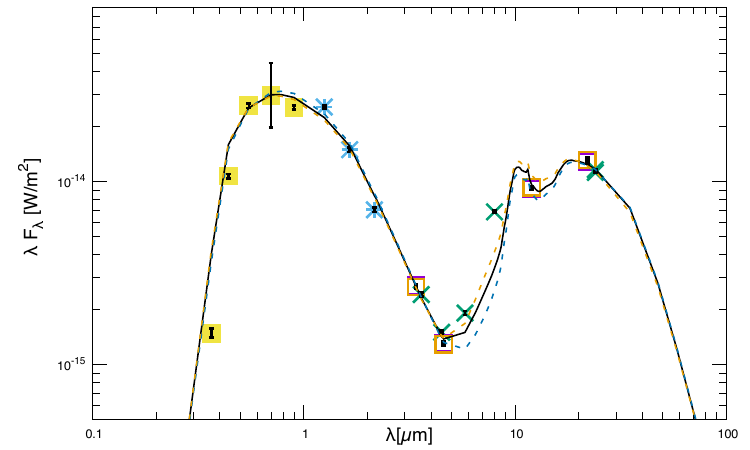}
\figsetgrpnote{One-component model fits to the observed SED 
for varying maximum grain sizes: blue dashed curve ($\chi^{2}$ = 241.4): $a_{\mathrm max}$ = 0.25 $\micron$; orange dashed curve ($\chi^{2}$ = 253.7): $a_{\mathrm max}$ = 1.0 $\micron$ along with the best-fit model, black solid curve ($\chi^{2}$ = 214.4): $a_{\mathrm max}$ = 0.50.}
\figsetgrpend

\figsetgrpstart
\figsetgrpnum{13.5}
\figsetgrptitle{One-component fits}
\figsetplot{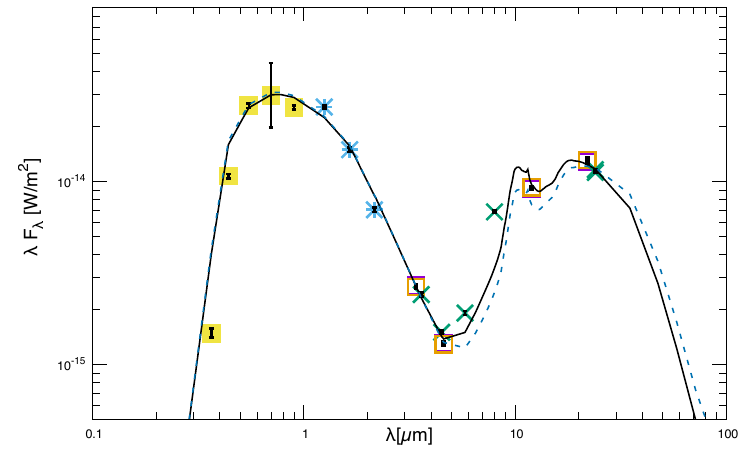}
\figsetgrpnote{One-component model fits to the observed SED 
for varying shell thickness, $Y$ = 5/blue dashed curve ($\chi^{2}$ = 426.7), along with the best-fit model, $Y$ = 2/black solid curve ($\chi^{2}$ = 214.4).}
\figsetgrpend

\figsetend

\renewcommand{\thefigure}{13}
\begin{figure}[!htb]
\plotone{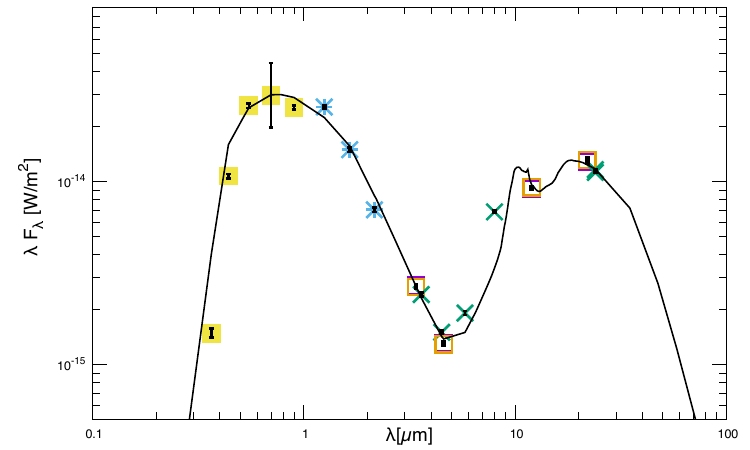}
\caption{The adopted one-component model fit ($\chi^{2}$ = 214.4) to the observed 
SED of the post-AGB (shell) source, J053250.69-713925.8.
The observed fluxes are de-reddened for Galactic and LMC 
reddening. U,B,V,R,I (yellow), 2MASS J,H,K (cyan) data are plotted along 
with WISE (purple) and ALLWISE (orange) photometry
and data from the SAGE-LMC Survey (green) which covers the
IRAC and MIPS bands.
The error bars are indicated in black. All model fits obtained for the source including the adopted best-fit are available in the corresponding Figure Set.}
\label{J053250_model}
\end{figure}

\figsetstart
\figsetnum{14}
\figsettitle{J045623.21-692749.0 (post-AGB disk source)}

\figsetgrpstart
\figsetgrpnum{14.1}
\figsetgrptitle{Nested shells}
\figsetplot{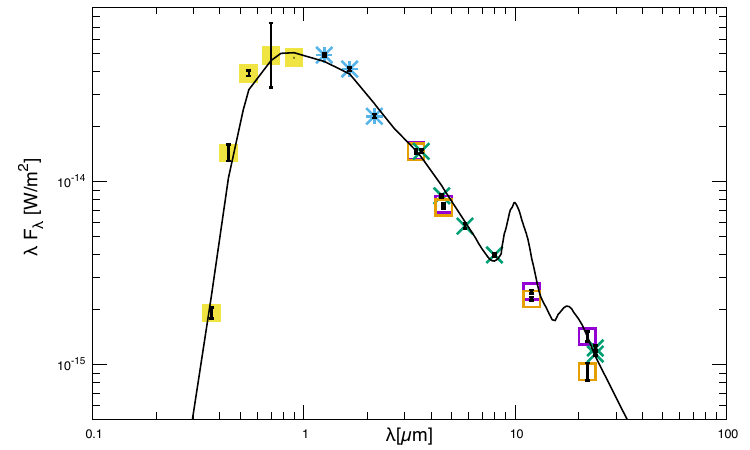}
\figsetgrpnote{The nested shells model (model \#\,1, $\chi^{2}$ = 100.2). The fit corresponds to the correctly illuminated outer shell. 
The inner shell is composed of Sil-Ow grains
with modified MRN grain size distribution: $a_{\mathrm min}$ = 0.005 $\micron$ and $a_{\mathrm max}$ = 5.0 $\micron$, 
$T_{\mathrm d}$ (in) = 1100 K, $\tau$ = 0.6 and $Y$ = 3.0. The outer shell is composed of Sil-Ow grains with
standard MRN grain size distribution, $T_{\mathrm d}$ (in) = 150 K,
$\tau$ = 0.10 and $Y$ = 2.0.}
\figsetgrpend

\figsetgrpstart
\figsetgrpnum{14.2}
\figsetgrptitle{Two-component fit}
\figsetplot{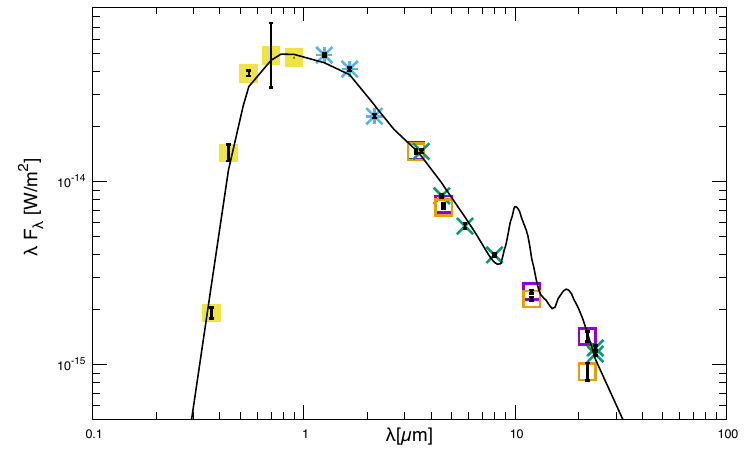}
\figsetgrpnote{Two-component model fit (model\#2,c, disk fraction = 0.5, $\chi^{2}$ = 195.2) to the observed SED. The fit corresponds to the correctly illuminated model\#2,  wherein the 
disk ($T_{\mathrm d}$ (in) = 1100 K) is composed of amorphous carbon (amC-Hn) grains with
modified MRN grain size distribution: $a_{\mathrm min}$ = 0.005 $\micron$, 
$a_{\mathrm max}$ = 2.0 $\micron$, $\tau$ = 0.5 and $Y$ = 2. The cold
outer shell ($T_{\mathrm d}$ (in) = 400 K) is made up of Sil-OW grains and has modified MRN grain size 
distribution: $a_{\mathrm min}$ = 0.40 $\micron$,
$a_{\mathrm max}$ = 1.0 $\micron$, $\tau$ = 0.22 and $Y$ = 15.
}
\figsetgrpend

\figsetgrpstart
\figsetgrpnum{14.3}
\figsetgrptitle{One-component fit}
\figsetplot{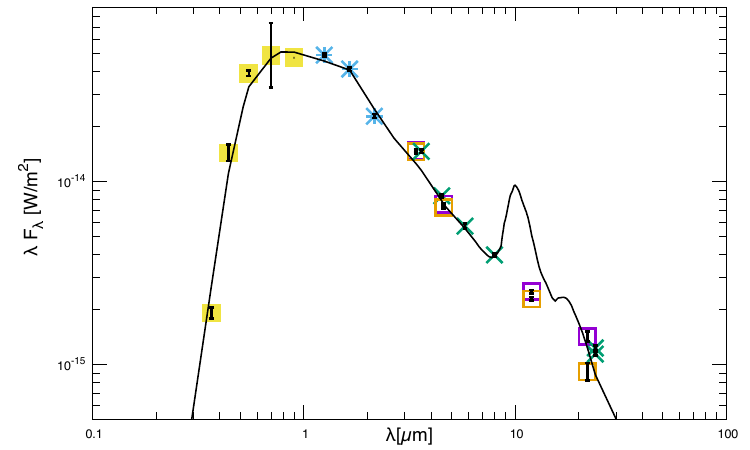}
\figsetgrpnote{One-component model fit ($\chi^{2}$ = 254.2) to the observed SED
is obtained using Sil-Ow grain type, modified MRN grain size distribution: $a_{\mathrm min}$ = 0.005 $\micron$ and $a_{\mathrm max}$ =  5.0 $\micron$, $T_{\mathrm d}$ (in) = 1100 K, $\tau$ = 0.6 and $Y$ = 50.}
\figsetgrpend

\figsetend

\renewcommand{\thefigure}{14}
\begin{figure}
\plotone{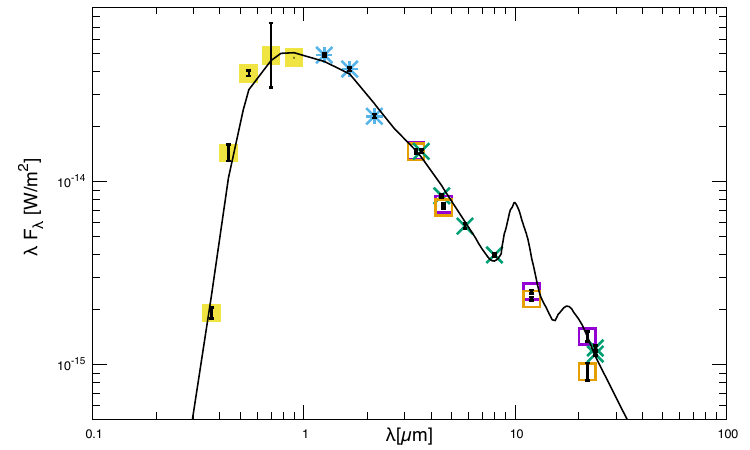}
\caption{
The adopted nested shells model (model \#\,1, $\chi^{2}$ = 100.2) for the post-AGB (disk) source,  J045623.21-692749.0. 
The observed fluxes are de-reddened for Galactic and LMC 
reddening. U,B,V,R,I (yellow), 2MASS J,H,K (cyan) data are plotted along with WISE (purple) and
ALLWISE (orange) photometry and data from the SAGE-LMC Survey (green) which covers the
IRAC and MIPS bands. The error bars are indicated in black. All model fits obtained for the source including the adopted best-fit are available in the corresponding Figure Set.}
\end{figure}

\figsetstart
\figsetnum{15}
\figsettitle{J051418.09-691234.9 (post-AGB disk source)}

\figsetgrpstart
\figsetgrpnum{15.1}
\figsetgrptitle{Two-component fit}
\figsetplot{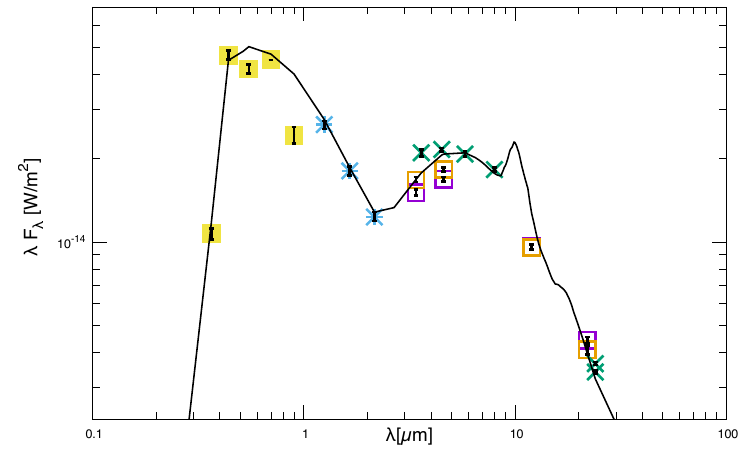}
\figsetgrpnote{Two-component model fit (disk fraction = 0.25, $\chi^{2}$ = 98.0) to the observed SED.
The disk ($T_{\mathrm d}$ (in) = 1100 K) is composed of amorphous carbon (amC-Hn) grains with
modified MRN grain size distribution: $a_{\mathrm min}$ = 0.005 $\micron$, 
$a_{\mathrm max}$ = 2.0 $\micron$, $\tau$ = 5.0 and $Y$ = 15. The cold
outer shell ($T_{\mathrm d}$ (in) = 600 K) is made up of a mix of Sil-Ow/0.4 and grf-DL/0.6 grains with standard MRN grain size distribution, $\tau$ = 0.4 and $Y$ = 30.}
\figsetgrpend

\figsetgrpstart
\figsetgrpnum{15.2}
\figsetgrptitle{Two-component fits}
\figsetplot{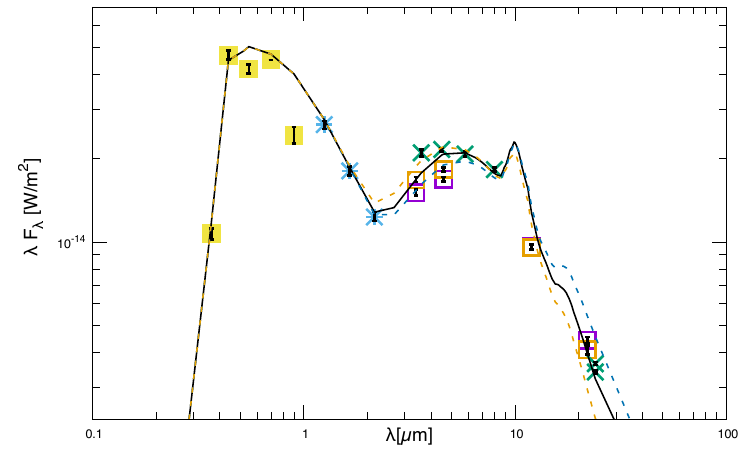}
\figsetgrpnote{Two-component model fits to the observed SED for varying dust temperature at the 
inner boundary of the cold outer shell, $T_{\mathrm d}$ (in) = 450 K/blue dashed curve ($\chi^{2}$ = 325.3) and $T_{\mathrm d}$ (in) = 750 K/orange dashed curve ($\chi^{2}$ = 332.1) along with model \#\,1 ($T_{\mathrm d}$ (in) = 600 K/black solid curve ($\chi^{2}$ = 98.0).}
\figsetgrpend

\figsetgrpstart
\figsetgrpnum{15.3}
\figsetgrptitle{Two-component fits}
\figsetplot{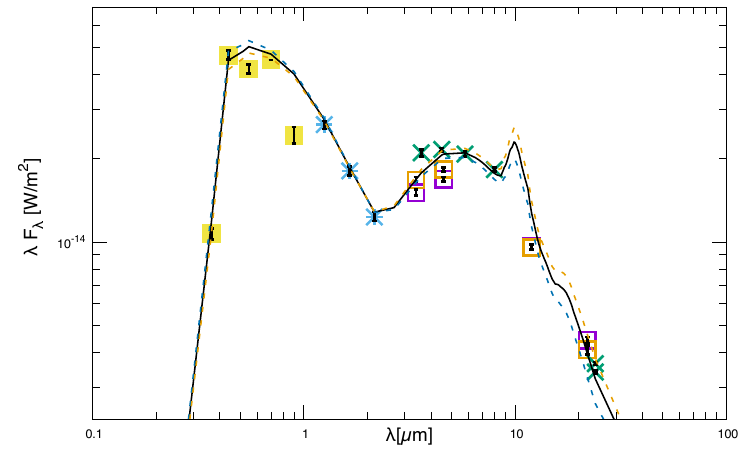}
\figsetgrpnote{Two-component model fits to the observed SED for varying shell optical depth, $\tau$ = 0.3/blue dashed curve ($\chi^{2}$ = 208.8) and $\tau$ = 0.5/orange dashed curve ($\chi^{2}$ = 144.6) along with model \#\,1, $\tau$ = 0.4/black solid curve ($\chi^{2}$ = 98.0).}
\figsetgrpend

\figsetgrpstart
\figsetgrpnum{15.4}
\figsetgrptitle{Two-component fits}
\figsetplot{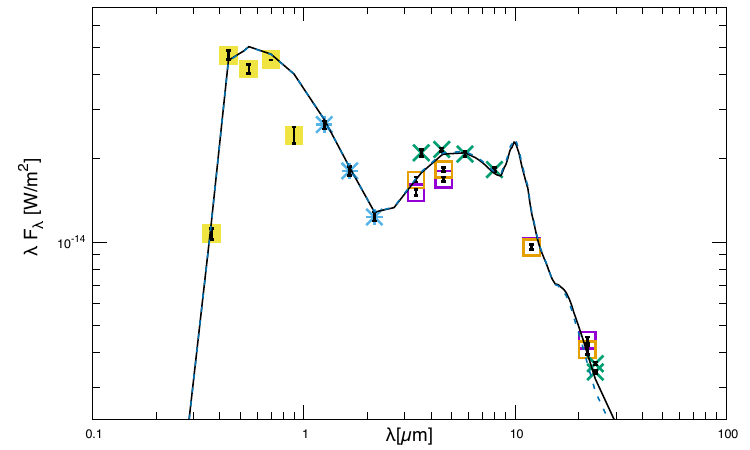}
\figsetgrpnote{Two-component model fits to the observed SED for varying shell thickness, $Y$ = 10/blue dashed curve ($\chi^{2}$ = 155.4) along with model \#\,1, $Y$ = 30/black solid curve
($\chi^{2}$ = 98.0).}
\figsetgrpend

\figsetgrpstart
\figsetgrpnum{15.5}
\figsetgrptitle{Two-component fits}
\figsetplot{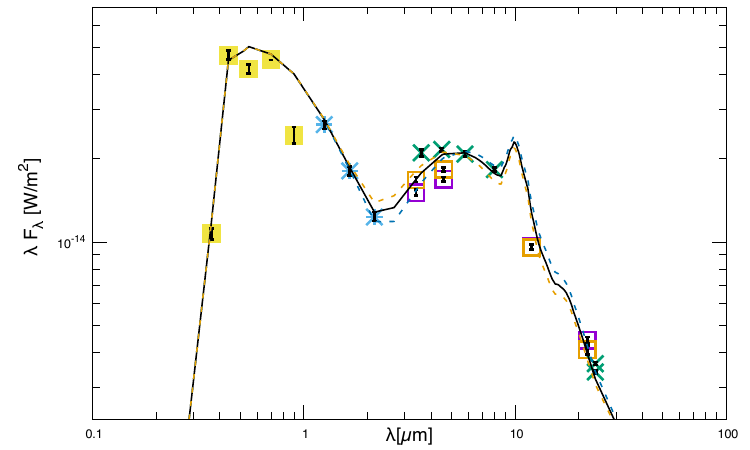}
\figsetgrpnote{Two-component model fits to the observed SED for varying dust temperature at the 
inner boundary of the dusty disk, $T_{\mathrm d}$ (in) = 1000 K/blue dashed curve ($\chi^{2}$ = 126.8) and $T_{\mathrm d}$ (in) = 1200 K/orange dashed curve ($\chi^{2}$ = 104.7) along with model \#\,1 ($T_{\mathrm d}$ (in) = 1100 K/black solid curve ($\chi^{2}$ = 98.0).}
\figsetgrpend

\figsetgrpstart
\figsetgrpnum{15.6}
\figsetgrptitle{Two-component fits}
\figsetplot{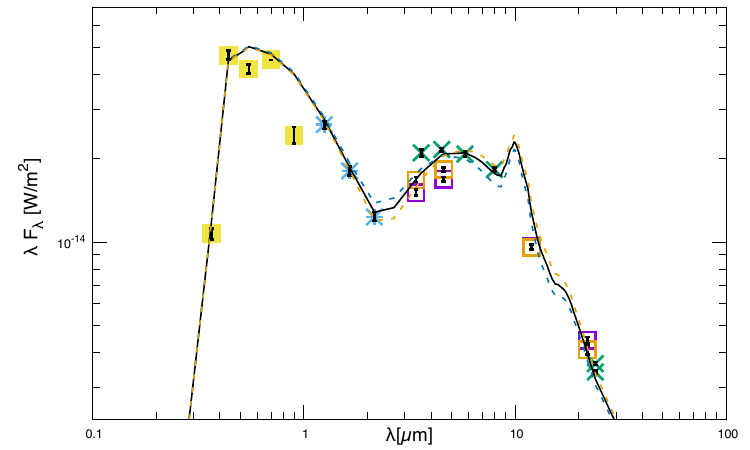}
\figsetgrpnote{Two-component model fits to the observed SED for varying disk optical depth, $\tau$ = 4.0/blue dashed curve ($\chi^{2}$ = 104.4) and $\tau$ = 6.0/orange dashed curve ($\chi^{2}$ = 122.8) along with model \#\,1, $\tau$ = 5.0/black solid curve ($\chi^{2}$ = 98.0).}
\figsetgrpend

\figsetgrpstart
\figsetgrpnum{15.7}
\figsetgrptitle{Two-component fits}
\figsetplot{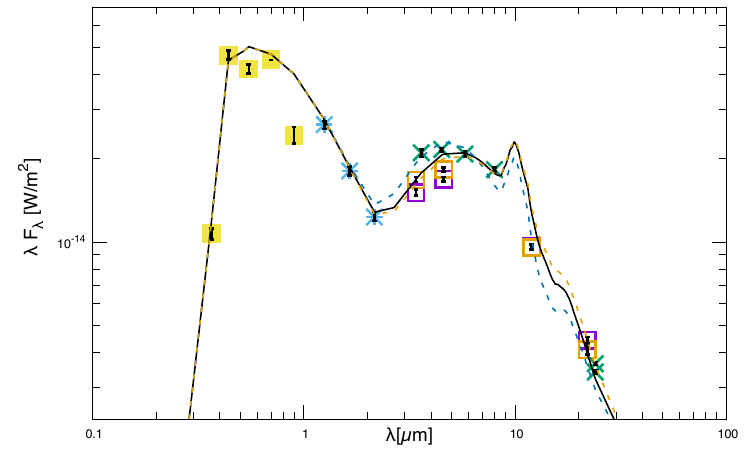}
\figsetgrpnote{Two-component model fits to the observed SED for varying maximum grain sizes in the disk: a$_{max}$ = 1.0 $\micron$/blue dashed curve ($\chi^{2}$ =144.6) and a$_{max}$ = 3.0 $\micron$/orange dashed curve ($\chi^{2}$ = 106.2) along with model \#\,1, a$_{max}$ = 2.0 $\micron$/black solid curve ($\chi^{2}$ = 98.0).}
\figsetgrpend

\figsetgrpstart
\figsetgrpnum{15.8}
\figsetgrptitle{Two-component fits}
\figsetplot{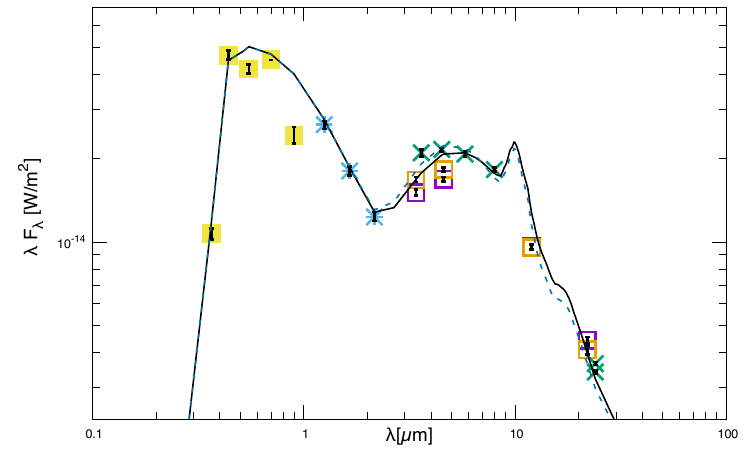}
\figsetgrpnote{Two-component model fits to the observed SED for varying disk thickness, $Y$ = 5/blue dashed curve ($\chi^{2}$ = 125.8) along with model \#\,1, $Y$ = 15/black solid curve
($\chi^{2}$ = 98.0).}
\figsetgrpend

\figsetend

\renewcommand{\thefigure}{15}
\begin{figure}
 \plotone{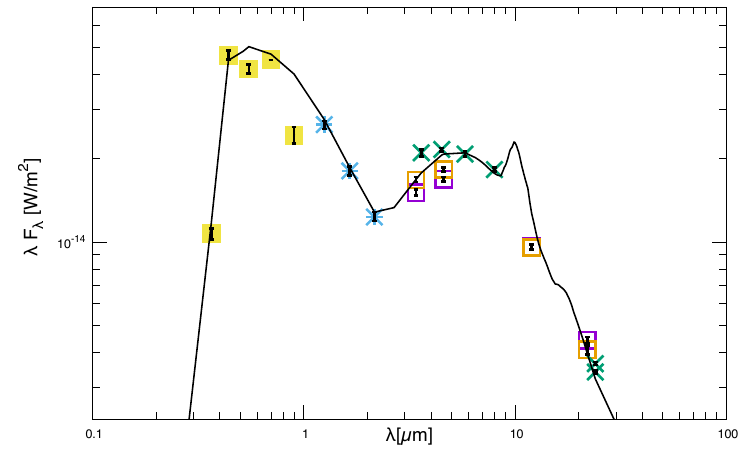}
 \caption{The adopted
two-component model fit (disk fraction = 0.25, $\chi^{2}$ = 98.0) to the observed SED of the post-AGB (disk) source, J051418.09-691234.9.
The observed fluxes are de-reddened for Galactic and LMC 
reddening. U,B,V,R,I (yellow), 2MASS J,H,K (cyan) data are plotted along with WISE (purple) and ALLWISE (orange)
photometry and data from the SAGE-LMC Survey (green) which covers the
IRAC and MIPS bands. The error bars are indicated in black. All model fits obtained for the source including the adopted best-fit are available in the corresponding Figure Set.}
\end{figure}

\figsetstart
\figsetnum{16}
\figsettitle{J055122.52-695351.4 (post-AGB disk source)}

\figsetgrpstart
\figsetgrpnum{16.1}
\figsetgrptitle{One-component fit}
\figsetplot{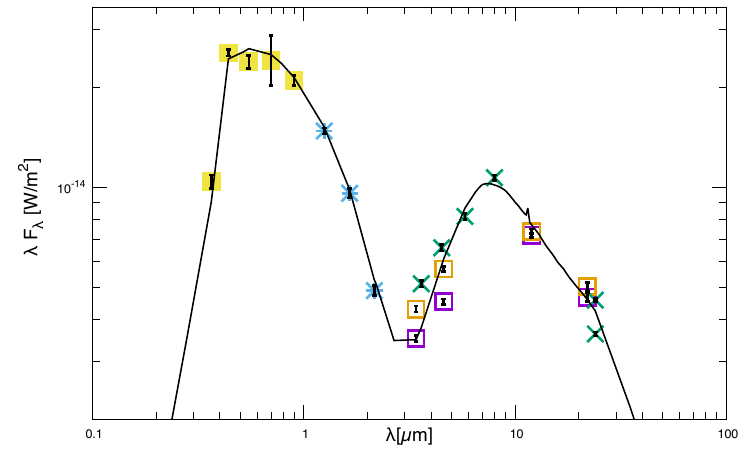}
\figsetgrpnote{One-component model fit ($\chi^{2}$ = 41.8) to the 
observed SED is obtained using a mix of grf-DL/0.7 and amC-Hn/0.3 grains, modified MRN grain size
distribution: $a_{\mathrm min}$ = 0.05 $\micron$, $a_{\mathrm max}$ = 0.3 $\micron$ , $T_{\mathrm d}$ (in) = 450 K, 
$\tau$ = 0.78 and $Y$ =  6.}
\figsetgrpend

\figsetgrpstart
\figsetgrpnum{16.2}
\figsetgrptitle{One-component fits}
\figsetplot{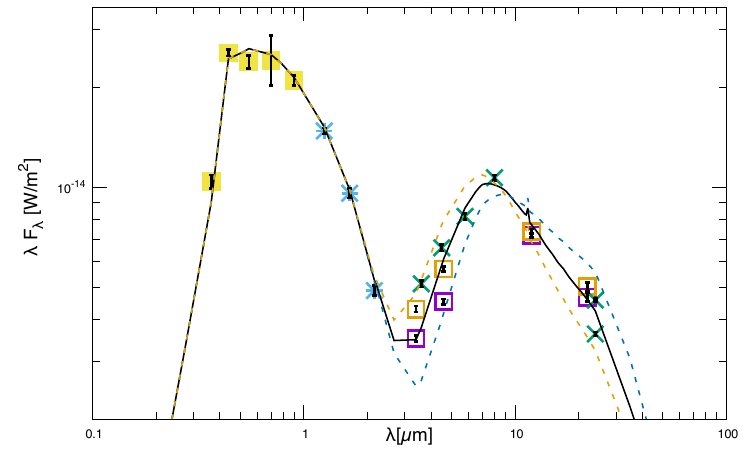}
\figsetgrpnote{One-component model fits to the observed SED for varying dust temperatures at the inner shell boundary, $T_{\mathrm d}$ (in) = 400 K/blue dashed curve ($\chi^{2}$ = 284.3) and 500 K/orange dashed curve ($\chi^{2}$ = 128.5) along with the best-fit model, $T_{\mathrm d}$ = 450 K/black solid curve ($\chi^{2}$ = 41.8).}
\figsetgrpend

\figsetgrpstart
\figsetgrpnum{16.3}
\figsetgrptitle{One-component fits}
\figsetplot{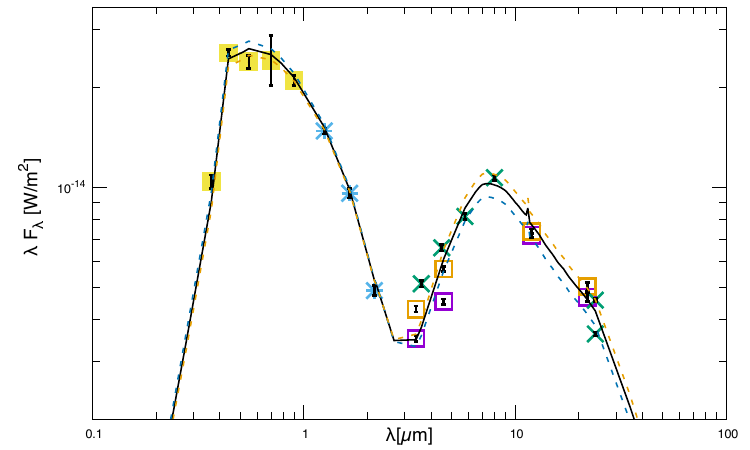}
\figsetgrpnote{One-component model fits to the observed SED 
for varying shell optical depth, $\tau$ = 0.70/blue dashed curve ($\chi^{2}$ = 56.5) and $\tau$ = 0.85/orange dashed curve ($\chi^{2}$ = 63.7) along with the best-fit model, $\tau$ = 0.78/black solid curve ($\chi^{2}$ = 41.8).}
\figsetgrpend

\figsetgrpstart
\figsetgrpnum{16.4}
\figsetgrptitle{One-component fits}
\figsetplot{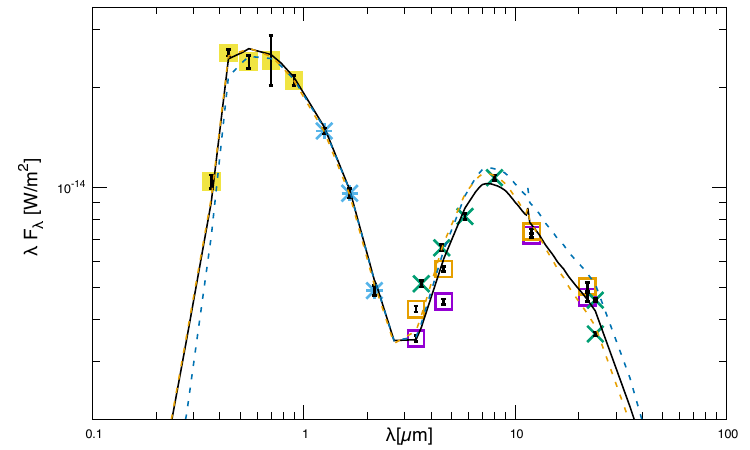}
\figsetgrpnote{One-component model fits to the observed SED 
for varying minimum and maximum grain sizes,
blue/dashed curve ($\chi^{2}$ = 132.8): $a_{\mathrm min}$ = 0.005 $\micron$, $a_{\mathrm max}$ = 0.3 $\micron$; 
orange dashed curve ($\chi^{2}$ = 48.2): $a_{\mathrm min}$ = 0.05 $\micron$, $a_{\mathrm max}$ = 0.5 $\micron$
along with the best-fit model, black solid curve ($\chi^{2}$ = 41.8): $a_{\mathrm min}$ = 0.05 $\micron$, $a_{\mathrm max}$ = 0.3 $\micron$.}
\figsetgrpend

\figsetgrpstart
\figsetgrpnum{16.5}
\figsetgrptitle{One-component fits}
\figsetplot{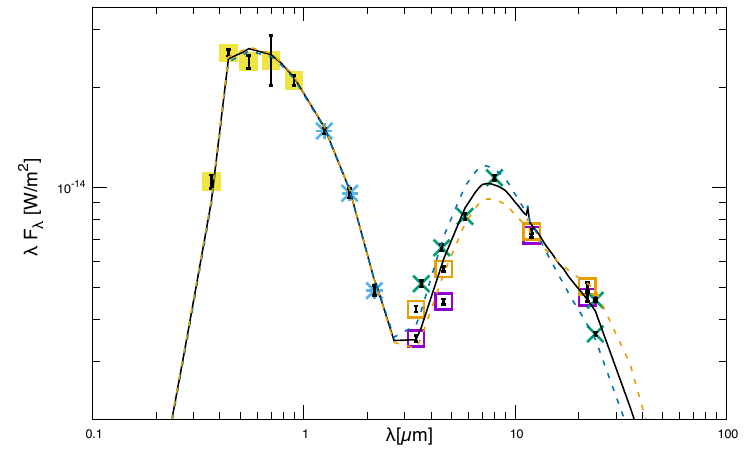}
\figsetgrpnote{One-component model fits to the observed SED for varying shell thickness, $Y$ = 3/blue dashed curve ($\chi^{2}$ = 77.3) and $Y$ = 20/orange dashed curve ($\chi^{2}$= 98.6) along
with the best-fit model, $Y$ = 6/black solid curve ($\chi^{2}$ = 41.8).}
\figsetgrpend

\figsetend

\renewcommand{\thefigure}{16}
\begin{figure}
\plotone{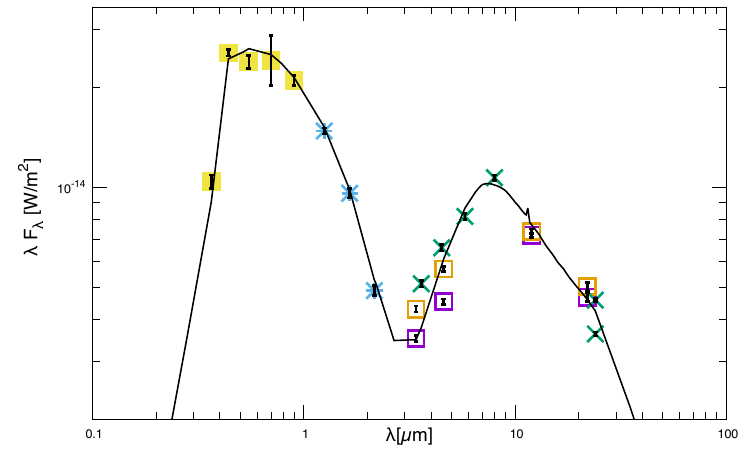}
\caption{The adopted one-component model fit ($\chi^{2}$ = 41.8) to the 
observed SED of the post-AGB (disk) source, J055122.52-695351.4.  The observed fluxes are de-reddened for Galactic and LMC 
reddening. U,B,V,R,I (yellow), 2MASS J,H,K (cyan) data are plotted along with WISE (purple) and ALLWISE (orange)
photometry and data from the SAGE-LMC Survey (green) which covers the
IRAC and MIPS bands. The error bars are indicated in black. All model fits obtained for the source including the adopted best-fit are available in the corresponding Figure Set.
}
\label{J055122_single}
\end{figure}

\figsetstart
\figsetnum{17}
\figsettitle{J052519.48-705410.0 (post-AGB disk source)}

\figsetgrpstart
\figsetgrpnum{17.1}
\figsetgrptitle{One-component fits}
\figsetplot{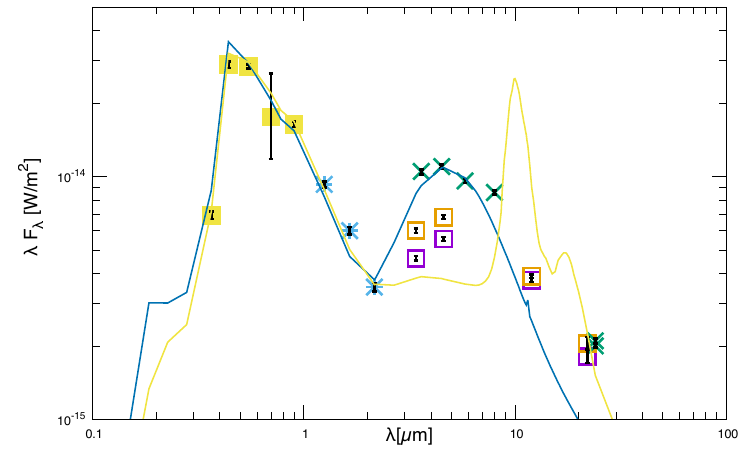}
\figsetgrpnote{One-component model fits to the 
observed SED of the post-AGB (disk) source, J052519.48-705410.0.
Fit to the near and
far-IR data (yellow solid curve/model \#\,1, $\chi^{2}$ = 444.5) was obtained using silicate (Sil-Ow) grains, standard MRN grain size
distribution, $T_{\mathrm d}$ (in) = 800 K, $\tau$ = 1.0 and $Y$ =  10.
Fit to the mid-IR, IRAC data (blue solid curve/model \#\,2, $\chi^{2}$ = 318.3) was obtained using graphite (grf-DL) grains,
modified MRN grain size distribution: $a_{\mathrm min}$ = 0.005 and am$_{\rm max}$ = 1.5,
$T_{\mathrm d}$ (in) = 700 K, $\tau$ = 0.55 and $Y$ =  2.0.}
\figsetgrpend

\figsetgrpstart
\figsetgrpnum{17.2}
\figsetgrptitle{Two-component fit}
\figsetplot{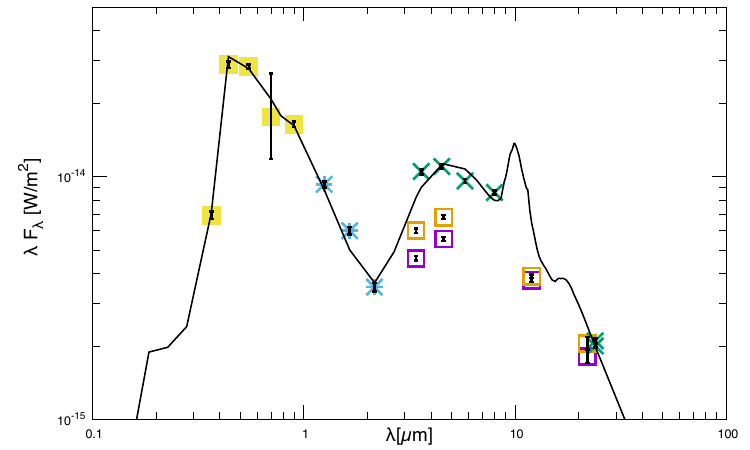}
\figsetgrpnote{Two-component model fit (disk fraction = 0.35, $\chi^{2}$ = 35.9) to the observed SED. The disk ($T_{\mathrm d}$ (in) = 800 K) is composed of Sil-Ow grains  
with MRN grain size distribution: $a_{\mathrm min}$ = 0.005 $\micron$, $a_{\mathrm max}$ = 0.25 $\micron$, $\tau$ = 1.0 and $Y$ = 10.0. The cold outer shell ($T_{\mathrm d}$ (in) = 600 K) is composed of grf-DL grains with modified MRN grain size distribution: $a_{\mathrm min}$ = 0.005 $\micron$, $a_{\mathrm max}$ = 0.5 $\micron$, $\tau$ = 0.8 and $Y$ = 2.0.}
\figsetgrpend

\figsetgrpstart
\figsetgrpnum{17.3}
\figsetgrptitle{Two-component fits}
\figsetplot{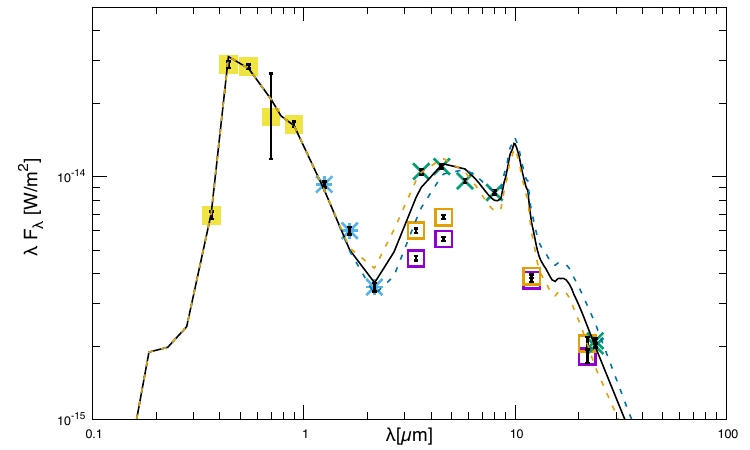}
\figsetgrpnote{Two-component model fits to the observed SED for varying dust temperature at the 
inner boundary of the cold outer shell, $T_{\mathrm d}$ (in) = 550 K/blue dashed curve ($\chi^{2}$ = 148.5) and $T_{\mathrm d}$ (in) = 650 K/orange dashed curve ($\chi^{2}$ = 96.7) along with model \#\,3, $T_{\mathrm d}$ (in) = 600 K/black solid curve ($\chi^{2}$ = 35.9).}
\figsetgrpend

\figsetgrpstart
\figsetgrpnum{17.4}
\figsetgrptitle{Two-component fits}
\figsetplot{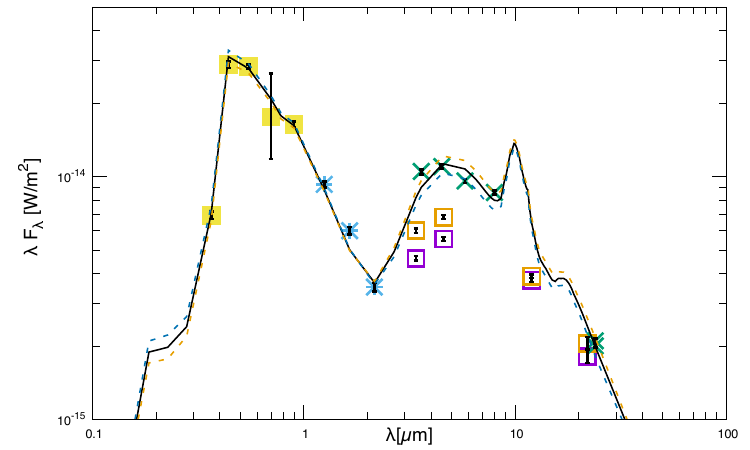}
\figsetgrpnote{Two-component model fits to the observed SED for varying shell optical depth, $\tau$ = 0.7/blue dashed curve ($\chi^{2}$ = 72.3) and $\tau$ = 0.9/orange dashed curve ($\chi^{2}$ = 62.6) along with model \#\,3, $\tau$ = 0.8/black solid curve ($\chi^{2}$ = 35.9).}
\figsetgrpend

\figsetgrpstart
\figsetgrpnum{17.5}
\figsetgrptitle{Two-component fits}
\figsetplot{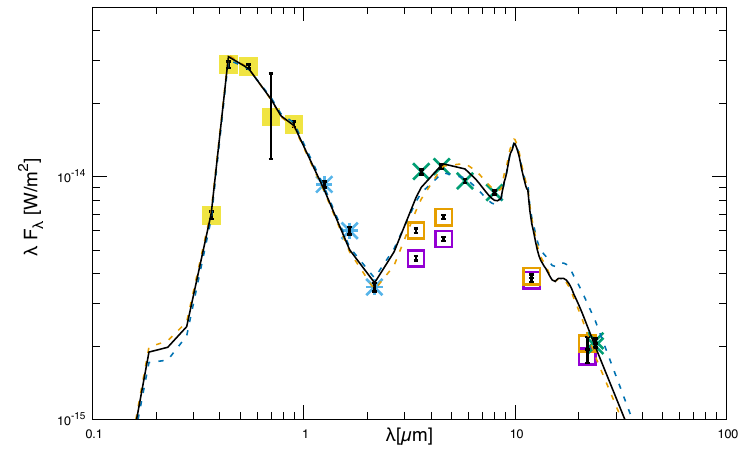}
\figsetgrpnote{Two-component model fits to the observed SED for varying maximum grain sizes in the shell, blue dashed curve ($\chi^{2}$ = 139.3): a$_{max}$ = 0.25 $\micron$; orange dashed curve
($\chi^{2}$ = 80.0): a$_{max}$ = 1.0 $\micron$ along with model \#\,3, a$_{max}$ = 0.5 $\micron$ (black solid curve,
$\chi^{2}$ = 35.9).}
\figsetgrpend

\figsetgrpstart
\figsetgrpnum{17.6}
\figsetgrptitle{Two-component fits}
\figsetplot{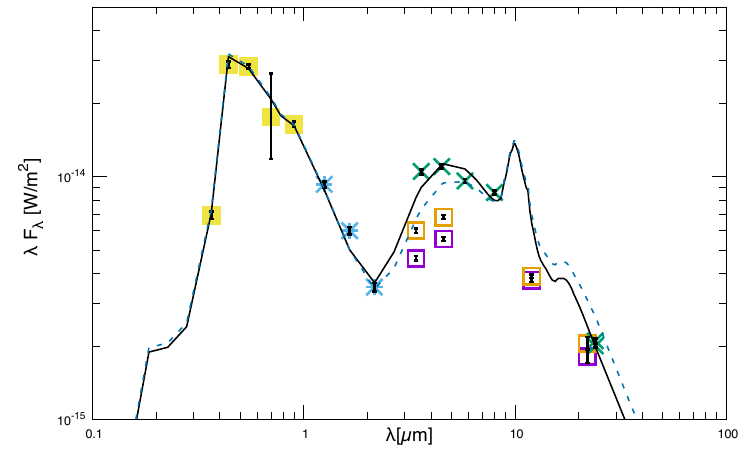}
\figsetgrpnote{Two-component model fits to the observed SED for varying shell thickness, $Y$ = 5/blue dashed curve ($\chi^{2}$ = 227.3) along with model \#\,3, $Y$ = 2/black solid curve ($\chi^{2}$ = 35.9).}
\figsetgrpend

\figsetgrpstart
\figsetgrpnum{17.7}
\figsetgrptitle{Two-component fits}
\figsetplot{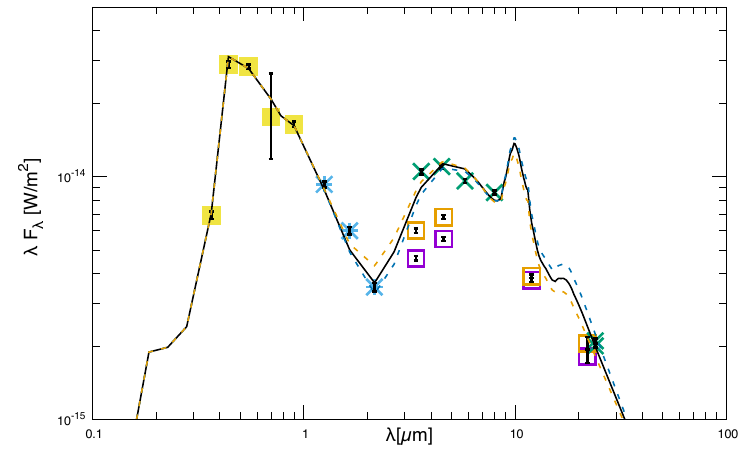}
\figsetgrpnote{Two-component model fits to the observed SED for varying dust temperature at the 
inner boundary of the dusty disk, $T_{\mathrm d}$ (in) = 650 K/blue dashed curve ($\chi^{2}$ = 63.4) and $T_{\mathrm d}$ (in) = 950 K/orange dashed curve ($\chi^{2}$ = 52.6) along with model \#\,3, $T_{\mathrm d}$ (in) = 800 K/black solid curve ($\chi^{2}$ = 35.9).}
\figsetgrpend

\figsetgrpstart
\figsetgrpnum{17.8}
\figsetgrptitle{Two-component fits}
\figsetplot{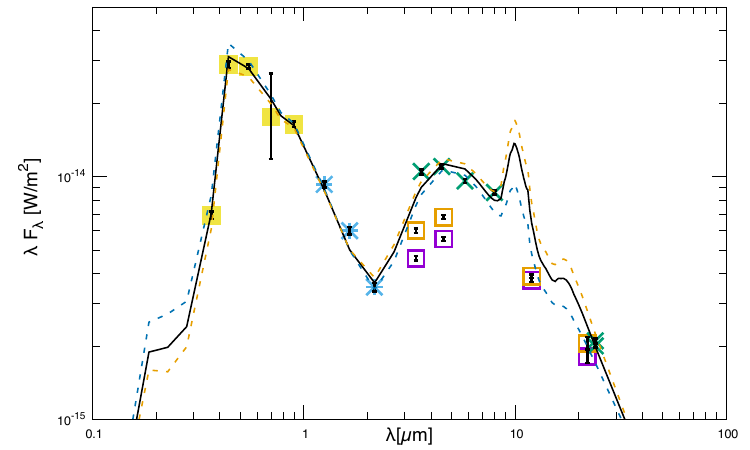}
\figsetgrpnote{Two-component model fits to the observed SED for varying disk optical depth, $\tau$ = 0.5/blue dashed curve ($\chi^{2}$ = 112.1) and $\tau$ = 1.5/orange dashed curve ($\chi^{2}$ = 71.6) along with model \#\,3, $\tau$ = 1.0/black solid curve ($\chi^{2}$ = 35.9).}
\figsetgrpend

\figsetgrpstart
\figsetgrpnum{17.9}
\figsetgrptitle{Two-component fits}
\figsetplot{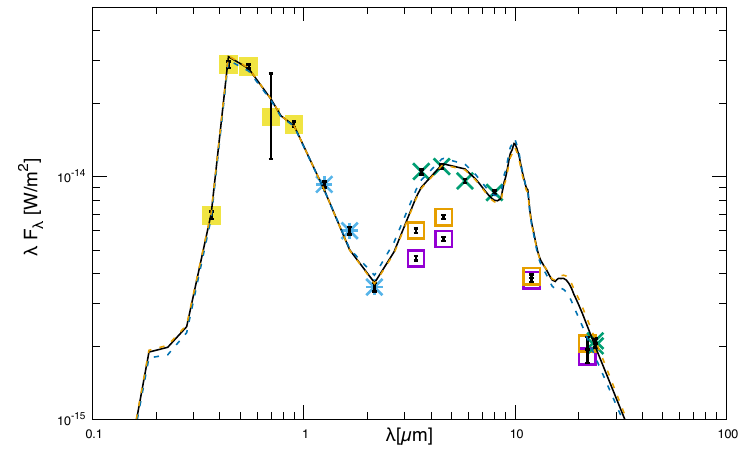}
\figsetgrpnote{Two-component model fits to the observed SED  for varying disk thickness, $Y$ = 2/blue dashed curve ($\chi^{2}$ = 57.7) and $Y$ = 50/orange dashed curve ($\chi^{2}$ = 41.5) along with model \#\,3, shell thickness, $Y$ = 10/black solid curve ($\chi^{2}$ = 35.9).}
\figsetgrpend

\figsetend

\renewcommand{\thefigure}{17}
\begin{figure}
\plotone{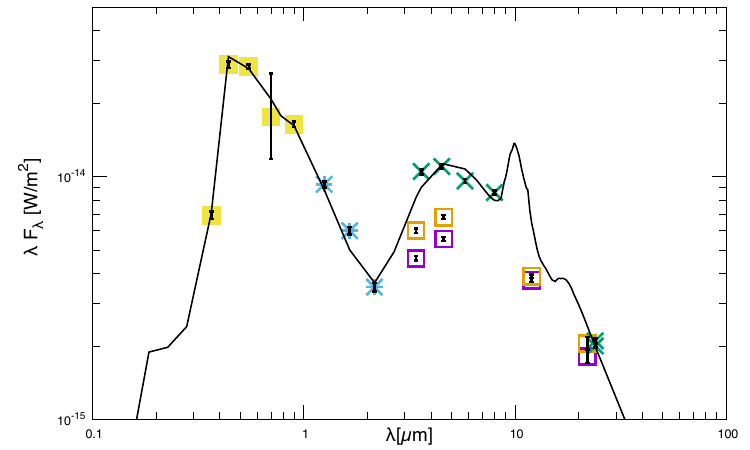}
\caption{The adopted two-component model fit (disk fraction = 0.35, $\chi^{2}$ = 35.9) to the observed SED of the post-AGB (disk) source, J052519.48-705410.0.
The observed fluxes are de-reddened for Galactic and LMC 
reddening. U,B,V,R,I (yellow), 2MASS J,H,K (cyan) data are plotted along with WISE (purple) and ALLWISE (orange) photometry and data from the SAGE-LMC Survey (green) which covers the
IRAC and MIPS bands. The error bars are indicated in black. All model fits obtained for the source including the adopted best-fit are available in the corresponding Figure Set}
\end{figure}

\renewcommand{\thefigure}{18}
\begin{figure*}[!htb]
\plotone{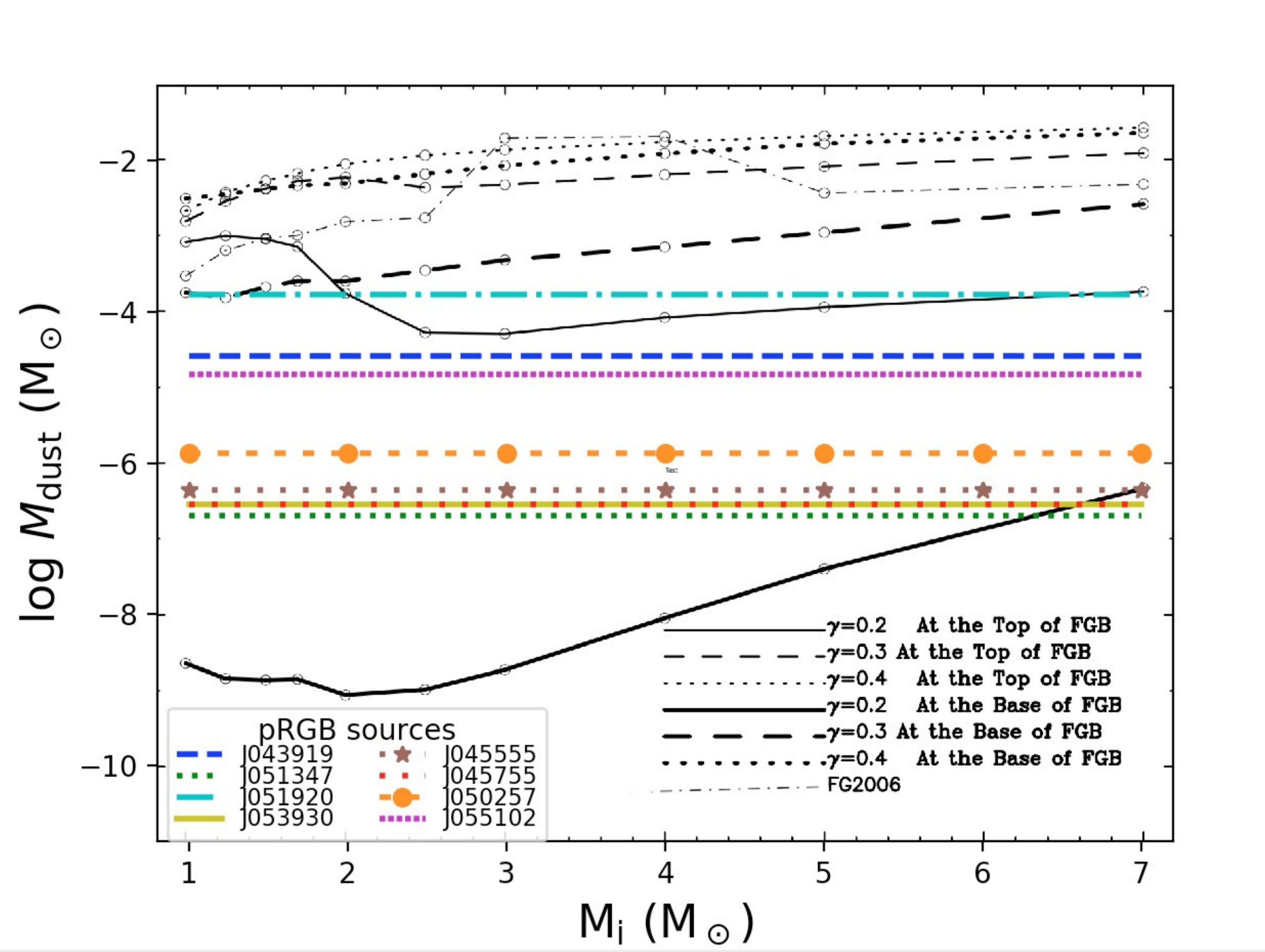}
\caption{The circumstellar dust mass ($M_{\mathrm d}$) of our post-RGB objects are shown (colored horizontal lines) on a plot of theoretically estimated dust masses in the ejecta of common envelope systems versus initial stellar mass, taken from Fig. 2 of \citet{lu2013}. FGB refers to the first red giant branch. FG2006 refers to results by \citet{ferrarotti2006}, showing the dust masses produced in the dust-driven outflows of AGB stars. $M_{\mathrm d}$ scales in proportion to the thickness of
the circumstellar component.  Determination of the latter is  constrained by the long wavelength limit of our data (24 $\micron$). 
Our estimated $M_{\mathrm d}$ values may be treated as lower limits to the actual dust mass in the shells. }
\label{dustmass_lu}
\end{figure*}

\end{document}